\documentclass[12pt]{article}
\usepackage{amsfonts}
\usepackage{epsfig}
\usepackage{amsmath}
\usepackage{amssymb}
\usepackage{color}

\begin{document}

\newcommand{\bq}{\ensuremath{{\bf q}}}
\renewcommand{\cal}{\ensuremath{\mathcal}}
\newcommand{\bqp}{\ensuremath{{\bf q'}}}
\newcommand{\bbq}{\ensuremath{{\bf Q}}}
\newcommand{\bp}{\ensuremath{{\bf p}}}
\newcommand{\bpp}{\ensuremath{{\bf p'}}}
\newcommand{\bk}{\ensuremath{{\bf k}}}
\newcommand{\bx}{\ensuremath{{\bf x}}}
\newcommand{\bxp}{\ensuremath{{\bf x'}}}
\newcommand{\by}{\ensuremath{{\bf y}}}
\newcommand{\byp}{\ensuremath{{\bf y'}}}
\newcommand{\bxpp}{\ensuremath{{\bf x''}}}
\newcommand{\intk}{\ensuremath{{\int \frac{d^3\bk}{(2\pi)^3}}}}
\newcommand{\intq}{\ensuremath{{\int \frac{d^3\bq}{(2\pi)^3}}}}
\newcommand{\intqp}{\ensuremath{{\int \frac{d^3\bqp}{(2\pi)^3}}}}
\newcommand{\intp}{\ensuremath{{\int \frac{d^3\bp}{(2\pi)^3}}}}
\newcommand{\intpp}{\ensuremath{{\int \frac{d^3\bpp}{(2\pi)^3}}}}
\newcommand{\intx}{\ensuremath{{\int d^3\bx}}}
\newcommand{\intxp}{\ensuremath{{\int d^3\bx'}}}
\newcommand{\intxpp}{\ensuremath{{\int d^3\bx''}}}
\newcommand{\drho}{\ensuremath{{\delta\rho}}}
\newcommand{\rhoh}{\ensuremath{{\hat{\rho}}}}
\newcommand{\fh}{\ensuremath{{\hat{f}}}}
\newcommand{\psih}{\ensuremath{{\hat{\psi}}}}
\newcommand{\thetah}{\ensuremath{{\hat{\theta}}}}
\newcommand{\etah}{\ensuremath{{\hat{\eta}}}}
\newcommand{\phih}{\ensuremath{{\hat{\phi}}}}
\newcommand{\0}{\ensuremath{{(\bk,\omega)}}}
\newcommand{\x}{\ensuremath{{(\bx,t)}}}
\newcommand{\xp}{\ensuremath{{(\bx',t)}}}
\newcommand{\xtp}{\ensuremath{{(\bx',t')}}}
\newcommand{\xtpp}{\ensuremath{{(\bx'',t')}}}
\newcommand{\xttpp}{\ensuremath{{(\bx'',t'')}}}
\newcommand{\xtpn}{\ensuremath{{(\bx',-t')}}}
\newcommand{\xtppn}{\ensuremath{{(\bx'',-t')}}}
\newcommand{\xn}{\ensuremath{{(\bx,-t)}}}
\newcommand{\xpn}{\ensuremath{{(\bx',-t)}}}
\newcommand{\xppn}{\ensuremath{{(\bx',-t)}}}
\newcommand{\xpp}{\ensuremath{{(\bx'',t)}}}
\newcommand{\xxp}{\ensuremath{{(\bx,t;\bx',t')}}}
\newcommand{\Crr}{\ensuremath{{C_{\rho\rho}}}}
\newcommand{\Crf}{\ensuremath{{C_{\rho f}}}}
\newcommand{\Crt}{\ensuremath{{C_{\rho\theta}}}}
\newcommand{\Cff}{\ensuremath{{C_{ff}}}}
\newcommand{\Cffh}{\ensuremath{{C_{f\fh}}}}
\newcommand{\Ct}{\ensuremath{{\dot{C}}}}
\newcommand{\Ctt}{\ensuremath{{\ddot{C}}}}
\newcommand{\Crrp}{\ensuremath{{\dot{C}_{\rho\rho}}}}
\newcommand{\Crfp}{\ensuremath{{\dot{C}_{\rho f}}}}
\newcommand{\Crtp}{\ensuremath{{\dot{C}_{\rho\theta}}}}
\newcommand{\Cffp}{\ensuremath{{\dot{C}_{ff}}}}
\newcommand{\Crrpp}{\ensuremath{{\ddot{C}_{\rho\rho}}}}
\newcommand{\thetab}{\ensuremath{{\overline{\theta}}}}

\begin{center}
{\bf\Large
Dynamical field theory for glass-forming liquids, self-consistent
  resummations and time-reversal symmetry}
\newline

{A. Andreanov$^{(1)}$, G. Biroli$^{(1)}$ and A. Lef{\`e}vre$^{(1,2)}$\\}
{\it
(1) Service de Physique Th{\'e}orique, CEA, Gif-sur-Yvette, France\\
(2) Rudolf Peierls Centre for Theoretical Physics, University of
Oxford, England  
 }
\end{center}
\begin{abstract}
We analyze the symmetries and the self-consistent perturbative approaches
of dynamical field theories for glass-forming liquids. In particular we focus
on the time-reversal symmetry which is crucial to obtain fluctuation
dissipation  
relations (FDR). Previous field theoretical treatment violated
this symmetry, whereas others pointed out that constructing
symmetry preserving perturbation theories is a crucial and open issue.
In this work we solve this problem and then apply our results to the
Mode Coupling Theory of the glass transition (MCT).\\
We show that in the context of dynamical field theories for
glass-forming liquids time-reversal symmetry is expressed as a
{\it non-linear} field transformation that leaves the action
invariant. Because of this non-linearity standard perturbation
theories generically do not preserve time-reversal symmetry and in
particular fluctuation-dissipation relations. We show how one can
cure this problem and set up symmetry preserving perturbation
theories by introducing some auxiliary fields. As an outcome we
obtain Schwinger-Dyson dynamical equations that preserve
automatically FDR and that serve as a basis for carrying out
symmetry preserving approximations. We apply our results to the
Mode-Coupling Theory of the glass transition revisiting previous
field theory derivations of MCT equations and showing that they
generically violate FDR. We obtain symmetry preserving
mode-coupling equations and discuss their advantages and
drawbacks. Furthermore we show, contrary to previous works, that
the structure of the dynamic equations is such  that the ideal
glass transition is not cut-off at any finite order of
perturbation theory, even in the presence of coupling between
current and density. The opposite results found in previous field
theoretical works, as the ones based on non-linear fluctuating
hydrodynamics, were only due to an incorrect treatment of
time-reversal symmetry.

\end{abstract}
\begin{center}
\date{today}
\end{center}
\newpage
\tableofcontents
\newpage

 \addcontentsline{toc}{section}{Introduction}
\section*{Introduction}

Liquids, if cooled fast enough in order to avoid crystallization,
enter generically in a metastable phase in which remarkable dynamical phenomena
occur~\cite{ang,DeBenedettiStillinger} upon decreasing the temperature. For instance
the structural relaxation time increases very fast, in many cases faster than
an Arrhenius law, correlation functions have slower than exponential
relaxations, and dynamics become strongly heterogeneous, contrary to
the simple motion which occurs at higher temperature.
Typically at a temperature that is roughly two third of the melting temperature
the relaxation time becomes macroscopic, of the order of
minutes or hours, and the liquid freezes into an amorphous solid called glass.

The equilibrium dynamics of moderately supercooled liquids is rather
well described by
the Mode-Coupling Theory~\cite{reviewDas,reviewDave,kob} (MCT).
MCT is a closure scheme for the correlation function developed twenty
years ago~\cite{ben,leu} leading to a nonlinear integro-differential
equation which has to be
solved self-consistently. It predicts a power law divergence of the relaxation
time and of the viscosity at a finite temperature $T_{MCT}$.
Although it is now clear that this is a spurious transition
several quantitative results compare well with experimental and
numerical findings as for example the wave-vector dependence of the
Debye-Waller
factor (the analog of the Edwards-Anderson parameter for spin-glasses)
see~\cite{reviewDas,reviewDave,kob}. The conventional interpretation is that
at $T_{MCT}$ there is a dynamical crossover: ``hopping or activated''
events not
contained in the theory and which can be roughly neglected for
$T>T_{MCT}$ prevent the existence
of the transition and dominate the slowing down for $T<T_{MCT}$.
Thus, although the transition is strictly speaking avoided, the
dynamics for $T>T_{MCT}$
can be explained in terms of MCT.
This interpretation is based on theoretical extensions of MCT
\cite{extendedGotze,das} and recent developments originating from
mean-field disordered systems~\cite{ReviewMF}.
However, it is important to stress that the MCT dynamical
crossover is not that sharp in real
systems, therefore  many MCT statements have to be considered with great care.
Actually even the existence of a MCT crossover remains a long debated
issue in glass literature, see e.g. \cite{BerthierGarrahan}.
MCT was derived originally using the projection operator
formalism of Mori and Zwanzig~\cite{ben,leu}. A few years later,
MCT was rederived starting from stochastic nonlinear hydrodynamics
equations as a one-loop self-consistent theory~\cite{das}.
This field theory
derivation and subsequent ones have been criticized for two
reasons. First because the mechanism behind the
MCT transition was thought to be a short-scale phenomenon:
locally particles get jammed and jailed in cages due to the interactions
with the nearest neighboring particles in the liquid, resulting in a
fast rattling motion but no structural relaxation.
Stochastic nonlinear hydrodynamics is an effective
theory for moderate and long length scales and therefore it
was thought that many predictions were dependent on the
short-scale cut-off or coincided with the previous one
accidentally \cite{extendedGotze}.

The second criticism is more recent and is related to the
fluctuation-dissipation relation (FDR) between correlation and response
function (see~\cite{mya1} and next sections). The one-loop derivations
of MCT presented in
the literature assume FDR in order to get the correct mode-coupling (MC)
equations but actually they are {\it incompatible} with FDR.
As a consequence they are not consistent and difficult to extend to
more general cases.

Despite these drawbacks the field theory approach is particularly
appealing and is gaining interest for many reasons.
First, within this formalism it has been shown that
MCT is a dynamic critical phenomenon and it leads to strong dynamical
correlations in the four point density correlation function
\cite{BB} (see \cite{KT,FP} for very important early insights based on
the study of mean-field disordered systems). These results, which seem
still out of reach of the projection
operator formalism, have been indeed verified in simulations of model systems
and quantitative MCT predictions~\cite{BB} are in rather good
agreement with numerical results~\cite{TWBBB}(see also \cite{Andersen}).
Furthermore they provide an answer to the first criticism cited above:
the MCT transition is related to growing correlations in the dynamics
and {\it is not a
short scale phenomenon}. The subtle point here is that the
order parameter is the density correlator, a two point function,
whereas in many other cases\footnote{But not all! A famous example is the
BCS superconducting transition.} the order parameter is a one point
function. Therefore the MCT equations have to be interpreted
as mean-field equations on the order parameter. They play the same
role as Weiss mean-field equations for ferromagnets. This makes clear
that, first, they are very different from mode-coupling equations
for critical phenomena. Indeed the latter
are self-consistent equations of the correlations of the order
parameter, and not of the order parameter itself. Second, to obtain
the diverging correlation length
one has to go beyond the mean-field MCT equations and compute
the fluctuations of the order parameter, i.e. a four point
density correlation function. From this perspective
the fact that the order parameter, the density-density correlations,
does not contain any diverging length is natural and does not imply
that the MCT is a short-scale phenomenon.

The second reason which makes a field theory approach appealing is that
it can be used naturally in off-equilibrium regimes. Indeed there is a
strong interest in extending the MCT
equations to regimes in which a liquid is aging or sheared.
There have also been recently very interesting attempts
to do that within the projection operator
formalism~\cite{latz,fuchs,mya}.

 \addcontentsline{toc}{section}{Summary}
\section*{Summary}

The aim of this paper is twofold: first we
analyze field theories for the dynamics of glass-forming liquids in terms of
symmetries of the action.
To our knowledge this has never been done
and is very important in order to preserve {\it physical} symmetries
in approximate treatments. In particular we will focus on
time-reversal
symmetry, which implies important relationships between
observables. A well known example is the fluctuation dissipation
relation (FDR) between correlation and response functions.
Other important examples which have been discovered recently are the Jarzynski
and Crooks equality for non-equilibrium
processes~\cite{Jarzynski,Crooks}.
We will focus on two types of field theories:
the first is obtained from fluctuating nonlinear hydrodynamics (FNH)
equations that describe in a coarse grained way the dynamics
of liquids in which particles obey Newton
equations~\cite{Lubensky,das}.
The second one
is obtained from the stochastic equations derived by Dean~\cite{dean1}.
They are {\it exact} stochastic equations governing the evolution of
the density field of interacting particles evolving with Langevin
dynamics. The field transformations related to time-reversal symmetry
that leave the action invariant are {\it nonlinear} in the
fields. This is at the origin of the violation of FDR (and
time-reversal
symmetry) by self-consistent perturbation theories
that has been already noticed and discussed by Miyazaki and
Reichman~\cite{mya1}.
We shall show that introducing some auxiliary fields this problem
can be solved and one can set up a self-consistent perturbation theory
compatible with FDR. At this stage, we warn the reader that we will use
the words ``perturbation'' and ``perturbative'' loosely throughout
this article, as there will be no small
parameter. In our
context a ``perturbation'' is a formal series expansion, which radius
of convergence may not be known.
The second aim of this work is to analyze the field theory derivation of MCT
using self-consistent one-loop approximations which preserve the fluctuation
dissipation relation between density correlations and response
functions. This will clarify different issues related to the field theory
derivations of MCT and to extended MCT (and the related cut-off of the
transition). In particular we shall show that the two field theories
assoctiated to FNH and Dean equations lead to essentially the same
self-consistent one loop equations.
These equations are very similar to the
usual MCT ones {\it but} but
with a slightly different wavevector-dependence
of the vertex.
This difference is clearly a drawback because the
particular wavevector-dependence of the MCT vertex is
very important for quantitative results.
Furthermore, in the case of the Dean equation, our vertex
leads to an apparently unphysical divergence.
Our results show that some kind of resummation has to be performed in
order to get MCT-like equations that are both quantitatively
succesfull (as the one derived in \cite{ben}) and respect
explicitely time reversal symmetry.
An interesting exact byproduct of our analysis
concerns extended MCT. Indeed, we obtain the exact Schwinger-Dyson
equation for the non-ergodic (Edwards-Anderson) parameter of the
glass phase.
We find that if the time-reversal symmetry is preserved
by the {\it perturbative} self-consistent theory the
MCT transition is not cut-off {\it even when density and currents are
coupled} contrary to the conclusions of previous works
\cite{das,schmitz}. In those cases approximations violating
time reversal symmetry produced a spurious
cut-off of the dynamical transition.
Our conclusion is that from a field theory perspective
the mechanism which cuts the MCT transition off is a
non-perturbative one unrelated to the presence of
density-currents coupling and likely
the same for Brownian or Newtonian dynamics as observed recently
in numerical simulations~\cite{szamel} and suggested also by very
recent works \cite{cates,MMR}.

This manuscript is organized as follows.
In section I we introduce the two field theories for glass-forming liquids
we focus on. From section II to section V, we focus on the field
theory for the Dean equation, which provides a good illustration
of the problems arriving with non-linear symmetries. In section II we
show the fields transformations
which leave the action invariant and are related to time-reversal
symmetry. In section III we discuss how nonlinear field transformations
affect perturbation theory and the origin of violation of FDR in the previous
derivation of MCT. In section IV we show how one can restore FDR in
self-consistent perturbation theories introducing more auxiliary
fields. On the way an exact and compact structure of the Schwinger-Dyson
equation is given, where the consequences of time-reversal symmetry
have been exploited to the maximum. As an inllustration of the general
strategy, we derive in section V
mode-coupling equations which preserves FDR. In section VI, we present
the result of the use of the strategy developped in the previous
sections to fluctuating nonlinear hydrodynamics.
Section VII is devoted to a discussion of MCT related issues in
particular the fact that the transition is not cut-off
even in presence of currents. Finally, appendices deal with the most
technical details.

\section{Dynamical field theory for glass-forming liquids}

The aim of this section is to introduce two field theories
used to study glass-forming liquids such as supercooled liquids
and dense colloidal systems.

\subsection{Brownian Dynamics and the Dean equation}\label{BD}

In the following we derive the field theory corresponding to a system
of $N$ interacting point particles evolving under Langevin dynamics
which provides
a good qualitative description of the dynamics of a dense colloidal suspension
\cite{Pusey}.
Actually one should also take into account hydrodynamic interactions
due to the solvent but they will be neglected for simplicity.
>From a more theoretical point of view this is a
limiting case in which, except for the density, there are no other conserved
variables. We shall consider in the next section the field theory
corresponding to Newtonian dynamics in which energy, momentum and
density are conserved variables.

The starting point is the Langevin equation which describes the
dynamics of $N$ interacting particles,
evolving in Euclidean three dimensional space (coordinates are
labelled by lower
Roman letters, particles labelled by lower Greek letters):
\begin{equation}
\partial_t\bx_\alpha=-\sum_{\alpha<\beta} \nabla
V(\bx_\alpha-\bx_\beta)+{\bf \zeta}_{\alpha},
\label{eqn:lan}
\end{equation}
where $\zeta_\alpha$ is a Gaussian white noise, whose correlations are given
by the Stokes-Einstein relation:
\begin{equation}
\langle
\zeta_{\alpha,i}\x\zeta_{\beta,j}(\bx',t')\rangle
=2T\delta_{\alpha\beta} \delta_{ij}\ \delta(\bx-\bx')\ \delta(t-t').
\end{equation}
Here, $V$ is the pair potential between particles and the time is
expressed in units of the microscopic diffusion constant.

The potential $V$ is defined up to an additive constant. We
shall tune this constant so that $\intx\ V(\bx)=-\frac{T}{\rho_0}$
which will be convenient later\footnote{More precisely, we put the
  system in a box of volume
  $\Omega$, replace the potential $V(\bx)$ by
  $V(\bx)-\frac{1}{\Omega}\left(\frac{T}{\rho_0}-\intx V(\bx)
  \right)$ and take the limit $\Omega\rightarrow\infty$.}.
Observables measured in experiments are functions of the local density
$\rho\x=\sum_\alpha \delta(\bx-\bx_\alpha(t))$.
Using It{\^o} calculus, Dean has shown that the density of a system of
particles obeying (\ref{eqn:lan}) obeys the following Langevin
equation~\cite{dean1}:
\begin{equation}
\partial_t
\rho\x=\nabla\cdot\left(\rho\x\nabla\frac{\delta{\mathcal
    F}}{\delta \rho\x} \right)+\eta\x,
\label{eqn:dean}
\end{equation}
where $\eta$ is a Gaussian random noise, whose correlators are:
\begin{equation}
\langle \eta\x\eta(\bx',t')\rangle=2T\rho\x
\nabla\cdot\nabla' \delta(\bx-\bx')\delta(t-t').
\end{equation}
The prime on the gradient means that it acts on functions of $x'$.
The occurrence of multiplicative noise here is not surprising, as the
local density does not fluctuate in empty regions.
The density functional ${\mathcal F}$ for a fluid of average density
$\rho_0$ can be written as:
\begin{equation}
\begin{split}
{\mathcal F}[\rho]&=T\intx\ \rho(\bx)\left(\ln \frac{\rho(\bx)}{\rho_0}
-1\right)
+\frac{1}{2}\intx\intxp\ \rho(\bx)\rho(\bx')V(\bx-\bx')\\
&=-T{\mathcal S}[\rho]+{\mathcal F}_{int}[\rho].
\label{eqn:df}
\end{split}
\end{equation}
The Langevin equation (\ref{eqn:dean}) is equivalent to the equation
derived heuristically by Kawasaki~\cite{kaw1}, when $-\beta V(\bx)$ is
replaced
by the direct pair correlation function $c(\bx)$, and the functional defined in
(\ref{eqn:df}) becomes the Ramakrishnan-Youssouff (RY) density
functional~\cite{RY}.

The dynamic average of an observable ${\mathcal A}$ over thermal
histories can be expressed as a functional integral over the density
field:
\begin{equation}
\langle{\mathcal A}\rangle=\int{\mathcal
  D}\rho{\mathcal A}[\rho]\langle\delta\left[\partial_t\rho\x-R[\rho,\eta]\x
\right]\rangle_{\eta },
\end{equation}
where $R[\rho,\eta]\x$ is the RHS of (\ref{eqn:dean}), $\langle
\cdot \rangle_{\eta}$ means the average over the Gaussian noise $\eta
$ and $\delta [\cdot]$ is a functional delta function. This is the
standard procedure, see~\cite{zj}, to derive field theories from
stochastic equations. The only subtlety is the absence of a Jacobian.
This is due to the fact that stochastic equations with multiplicative
noise (\ref{eqn:dean}) are defined following the It{\^o} prescription and
therefore the Jacobian is a constant that can be absorbed in the
definition of the functional integral. This is also related to the
Markov property of the stochastic differential equation
(\ref{eqn:dean}) (in the It{\^o} discretization).

By using an integral representation of the functional Dirac distribution
through a conjugated field $\rhoh$ and averaging over the noise
$\eta$, the Martin-Siggia-Rose-de Dominicis-Janssen (MSRJD) action
is obtained~\cite{msrjd}:
\begin{equation}
\langle{\mathcal A}\rangle=\int {\mathcal D}\rho\int {\mathcal D}\rhoh
{\mathcal A}[\rho]\ e^{S[\rho,\rhoh]},
\end{equation}
with
\begin{equation}
\begin{split}
S[\rho,\rhoh]=\intx\int dt\ \Biggl\{&\rhoh\x\left[
-\partial_t\rho\x+\nabla\cdot\left(
\rho\x\nabla\frac{\delta{\mathcal F}[\rho]}{\delta\rho\x}\right)\right]
\\
&+T\rho\x(\nabla \rhoh\x)^2\Biggr\},
\end{split}
\end{equation}
or explicitly:
\begin{eqnarray}
\label{eqn:MSR}
S[\rho,\rhoh]&=&\intx\int dt\ \Biggl\{ \rhoh\x\Biggl[
-\partial_t\rho\x+T\nabla^2\rho\x\\
&+&\nabla\cdot\left(\rho\x
\intxp\ \nabla V(\bx-\bx')\rho(\bx',t)\right)\Biggr]+T\rho\x
(\nabla \rhoh\x)^2\Biggr\}.\nonumber
\end{eqnarray}
For clarity, and as the dynamical action involves one-time quantities,
we shall not write the explicit time dependence of the fields in the
rest of the paper, except for correlation functions involving fields at
different times. For instance $\rho\x$ will be written $\rho_x$, and
$\intx\int dt$ will be replaced by $\int_x$.

\subsection{Fluctuating Nonlinear Hydrodynamics}\label{FH}

We shall now recall the field theory used to investigate the dynamics
of compressible liquids in particular close to the glass transition.
In this case particles evolve under Newtonian dynamics and the
derivation of the corresponding stochastic equations is not from first
principles contrary to Dean's equation. The resulting equations
are meant to be a generalization of hydrodynamic equations to
intermediate time and length scales, and are not expected to lead to
an accurate description
of the physics occuring on short time and length scales. Hence, they
describe the
evolution of slow variables (associated to conserved quantities)
subject to a thermal noise (corresponding to the fast degrees of
freedom that have been integrated out) and they have been called
Fluctuating Nonlinear Hydrodynamics equations (FNH).
A phenomenological derivation and a discussion of the equations can be found
in~\cite{das,Lubensky}. In the following we
will focus on the FNH equations used by Das and Mazenko~\cite{das} to
investigate
the problem of the glass transition. They focused only on density and
momentum as conserved variables but in principle the energy can be
introduced as well. The equations of Das and Mazenko read:
\begin{eqnarray}
\partial_t \rho_x&=&\int d^3\bxp\{ \rho_x, g_{i,x'}\} \frac{\delta
  {\cal F}}{\delta g_{i,x'}}\label{dm1}\\
\partial_t g_{i,x}&=&\int d^3\bxp\{ g_{i,x}, \rho_{x'}\} \frac{\delta
  {\cal F}}{\delta \rho_{x'}}
+\int d^3\bxp\sum_j \{ g_{i,x}, g_{j,x'}\} \frac{\delta {\cal
  F}}{\delta g_{j,x'}}\\\nonumber
&&
+\sum_j \int d^3\bxp\Gamma_{ij}(\bx-\bxp)  \frac{\delta {\cal
  F}}{\delta g_{j,x'}}+\eta_{i,x},\label{dm2}
\end{eqnarray}
where $\rho$ and $g_i$ are respectively the density field and the $ith$
component of the momentum density field and $\eta_i$ is a white Gaussian
noise with variance $\langle \eta_i(\bx,t) \eta_j(\bxp,t')\rangle= 2T
\Gamma_{ij} (\bx-\bxp)\delta(t-t')$. The effective free-energy functional is
${\cal F}={\cal F}_{KIN}+{\cal F}_{U}$, with
\begin{equation}
\begin{split}
{\cal F}_{KIN}[\rho,{\bf g}]&=\frac 1 2 \int d^3 \bx
\frac{\mathbf{g}^2(\bx)}{\rho(\bx)},\\
{\cal F}_{U}[\rho,{\bf g}]&=\frac T m \int d^3 \bx
\rho(\bx) (\log [\rho(\bx)/\rho_0]-1)\\
&-\frac{T}{2m^2}\int d^3 \bx d^3 \bxp c(\bx-\bxp)
(\rho(\bx)-\rho_0)( \rho(\bxp)-\rho_0),
\end{split}
\end{equation}
where $\rho_0$ is the density of the system, $m$ is the particle mass
and $c(\bx)$ is the direct correlation function. Note that the potential
term, ${\cal F}_{U}$, coincides with the Ramakrishnan-Youssouff functional.
The Poisson brackets $\{\cdot,\cdot \}$ and the $\Gamma_{ij}$ are
chosen so that
the continuity equation for the density is verified and in such a way that
the linearized equations coincide with the usual linear hydrodynamic
equations:
\begin{equation}
\begin{split}
\Gamma_{ij}(\bx-\by)&=\left[-\eta_0\left(\frac{1}{3}\nabla_i\nabla_j+\delta_{i,j}\nabla^2\right)-\zeta_0\nabla_i\nabla_j\right]\delta(\bx-\by)\\&
=L_{ij}\delta(\bx-\by)\\
\{\rho(\bx),g_i(\bxp)\}&=-\nabla^i_x\delta(\bx-\bxp)\rho(\bx)\\
\{ g_i(\bx),\rho(\bxp)\}&=-\rho(\bx)\nabla^i_x\delta(\bx-\bxp)\\
\{
g_i(\bx),g_j(\bxp)\}&=-\nabla^j_x\delta(\bx-\bxp)g_i(\bxp)-g_j(\bx)\nabla^i_x\delta(\bx-\bxp),
\end{split}
\end{equation}
and $\eta_0$ and $\zeta_0$ are respectively the bare shear and bulk
viscosity. Using the previous definition one can rewrite the equations
(\ref{dm1},\ref{dm2}) in a more explicit way:
\begin{eqnarray}\label{dm1a}
\partial_t \rho_x&=&-\nabla \cdot{\bf g}_x\\
\partial_t g_{i,x}&=&-\rho_x \nabla_i \frac{\delta {\cal F}_U}{\delta
\rho_x}-\sum_{j}\nabla_j \left(\frac{g_{i,x}g_{j,x}}{\rho_x} \right)
-\sum_{j}L_{ij}\left(\frac{g_{j,x}}{\rho_x} \right)+\eta_{i,x}.\label{dm2b}
\end{eqnarray}
Following the strategy of the previous sections we
find that the corresponding MSRJD action is the integral of:
\begin{eqnarray}
s_x&=&-\rhoh_x\left[\partial_t\rho_x+\nabla_i\left(\rho_x\frac{\delta
F}{\delta g_{i,x}}\right)\right]+T\hat g_{i,x} L_{ij}\hat g_{j,x}\\
&&-\hat g_{i,x}\left[\partial_t g_{i,x}+\rho_x\nabla_i\frac{\delta
    {\cal F}}{\delta\rho_x}+\nabla_j\left(g_{i,x}\frac{\delta {\cal
      F}}{\delta g_{j,x}}\right)+g_{j,x}\nabla_i\frac{\delta {\cal
      F}}{\delta g_{j,x}}+L_{ij}\frac{\delta {\cal F}}{\delta
    g_{j,x}}\right],\nonumber
\label{fh-action}
\end{eqnarray}
where $\hat \rho $ and $\hat {\bf g}$ are respectively the conjugated
field used to express the delta functions corresponding to
Eqs. (\ref{dm1a},\ref{dm2b}), and summation over indices is implicit.
In the next sections, we will focus on BDD only. We shall show later
in section \ref{sec:FNH} how to generalize the results to the case of FNH.

\section{Time-reversal symmetry and fluctuation dissipation relations}
Time-reversal symmetry relates the probabilities of a path and of its
time-reversal
counterpart in configuration space.
This is a very important symmetry obeyed by systems in thermodynamic
equilibrium and it has far reaching consequences. In particular all
physical correlation functions are invariant under time-reversal,
correlation and response function are related by the
fluctuation-dissipation relation. In the context of dynamic field
theory time-reversal symmetry is related to a transformation of the
fields leaving the action invariant. In the following,
we first recall the standard field transformation for one particle evolving
under Langevin dynamics. We will later show how this can be
generalized to the more complex field theories introduced in the previous
section.
\subsection{Langevin dynamics}

Let us now focus on one particle evolving under Langevin
dynamics with additive noise. Denoting the particle
position $X(t)$ and the external potential $V$ the Langevin equation reads:
\begin{equation}
\partial_t X(t)=-\nabla V(X(t))+\eta(t),
\end{equation}
where $\eta$ is a Gaussian white noise with zero mean and variance
$2T$.
After the introduction of a conjugated field $\hat{X}$, the dynamical

action reads
\begin{equation}\label{eqn:msrsimple}
S_0[\phi]=\int_{-\infty}^\infty\, dt\left[
-\hat{X}(t)\cdot\left(\partial_t X(t)+\nabla
V(X(t))\right)+T\hat{X}(t)^2\right],
\end{equation}
where the dynamical field is $\phi=(X,\hat{X})$.
It is easy to check that $S_0$ is invariant under
time-reversal:
\begin{equation}
\mathcal{O}:\left\{
\begin{array}{rl}
t&\rightarrow -t\\
\hat{X}(t)&\rightarrow\hat{X}(t)-\frac{1}{T}\partial_t X(t).
\end{array}\right.
\end{equation}
As a consequence correlation functions of $X (t)$ are invariant
under this fields transformation and therefore they are invariant
under time-reversal.
The application of the fields transformation to the response
function, which can be shown to be equal to
$\langle X(t)\hat{X}(t')\rangle$, leads to
\begin{equation}
  \langle \hat{X}(t')X(t)\rangle=\langle
  \hat{X}(t)X(t')\rangle-\frac{1}{T}\partial_t\langle X(t)X(t')\rangle.
\end{equation}
For $t>t'$, the response vanishes, and we get
\begin{equation}\label{Lfdr}
  \langle \hat{X}(t')X(t)\rangle=-\frac{1}{T}\partial_t\langle
  X(t)X(t')\rangle,
\end{equation}
which is the the FDR between correlation and response functions.

\subsection{Brownian dynamics for the density field}

We will now focus on
the stochastic evolution of the local density field for interacting
Langevin particles (in what follows we use the acronym BDD).
Note that the extension of the ideas in the next paragraphs to generic
multiplicative noise is straightforward.
First we derive the expression of the response function
in terms of fields average and after we show two fields
transformations which lead to a {\it physical} time-reversal symmetry
and, hence, to FDR.

\subsubsection{Response function}

In order to establish FDR,
one has to write the expression for the response of the system to an
external potential. Here, as we probe density fluctuations, this
force is an external potential, which is taken into account by adding
an extra term ${\mathcal F}_{ext}[\rho]=\displaystyle{-\int_x\ \rho_x
  \mu_x}$ to  the free energy
(\ref{eqn:df}). The response $R_{xx'}(t,t')$ at time $t$ and position
$\bx$ to an infinitesimal external force switched on at time $t'$ and
position $\bx'$ is defined by
\begin{equation}
\langle \rho\x\rangle_\mu=\langle\rho\x\rangle_{\mu=0}+\intxpp\int_{t'}^t
dt''\ R_{xx'}(t,t'')\mu(\bx'',t'')+o(\mu),
\end{equation}
where $\langle \cdot\rangle_\mu$ is the average taken with the free
energy functional ${\mathcal F}+{\mathcal F}_{ext}$. Thus, expanding the path
integral to first order in $\mu$, one gets:
\begin{equation}
\begin{split}
\langle
\rho\x\rangle_\mu&=\langle\rho\x\rangle_{\mu=0}\\
&+\intxpp\int_{t'}^\infty
dt''\ \langle \rho\x\rhoh\xttpp\nabla\cdot\left(
\rho\xttpp \nabla\mu\xttpp\right \rangle.
\end{split}
\end{equation}
However, due to causality the time $t''$ in the integral runs until $t$ only,
and integrating twice by parts, one obtains the expression of the
response:
\begin{equation}\label{response}
R_{xx'}(t,t')=-\langle \rho\x \nabla\cdot\left(
\rho\xtp\nabla\rhoh\xtp\right)\rangle.
\end{equation}
This response function is the same as the one studied in~\cite{mya1}.

\subsubsection{First expression of time-reversal symmetry}

The first field transformation related to time-reversal symmetry is
\begin{equation}
{\mathcal T}:\left\{
\begin{array}{rl}
t&\rightarrow -t\\
\rhoh_x&\rightarrow\rhoh_x+f_x,
\end{array}
\right.
\end{equation}
where $f$ verifies
\begin{equation}
\nabla\cdot\left(\rho_x\nabla f_x\right)=-\frac{1}{T}\partial_t \rho_x.
\label{eqn:cond}
\end{equation}
The field $f$ plays a role similar to the longitudinal part of a
current for the density field.
By integrating twice by parts, the variation of (\ref{eqn:MSR}) under
${\mathcal T}$ is found to be
\begin{equation}
-\frac{1}{T}\int_x\ \partial_t {\mathcal F}[\rho_x].
\end{equation}
Hence, at equilibrium
the action (\ref{eqn:MSR}) is invariant under ${\mathcal
T}$. The density field is affected in a simple way by ${\mathcal T}$:
$\rho (\bx,t)\rightarrow \rho (\bx,-t)$. This implies that
any average of the type $\langle\prod_i
\rho(\bx_i,t_i)\rangle$ is invariant under time-reversal.
By making the change ${\mathcal T}$ in the expression (\ref{response})
of the response function and using the
condition (\ref{eqn:cond}), we get:
\begin{equation}
\frac{1}{T}\partial_{t'} C_{xx'}(t-t')=R_{xx'}(t-t')-R_{xx'}(t'-t),
\label{eqn:FDT}
\end{equation}
where $C_{xx'}(t-t')=\langle \drho\x\drho\xtp\rangle$.
This is the FDR, which can be easily generalized to
correlators of more density fields:
\begin{equation}
\begin{array}{rcl}
\displaystyle{\frac{1}{T}\partial_{t'}C_{x_1\cdots x_n
    x'}(t_1\cdots,t_n,t')}&=&\displaystyle{R_{x_1\cdots x_n
    x'}(t_1\cdots,t_n,t')}\\
&-&\displaystyle{R_{x_1\cdots x_n x'}(-t_1\cdots,-t_n,-t')},
\end{array}
\end{equation}
where
\begin{equation}
R_{x_1\cdots x_n x'}(t_1\cdots,t_n,t')=-\langle \nabla\cdot\left(
\rho\xtp\nabla\rhoh\xtp\right)\prod_{i=1}^n
\rho(\bx_i,t_i) \rangle
\end{equation}
and
\begin{equation}
C_{x_1\cdots x_n x'}(t_1\cdots,t_n,t')=\langle\rho\xtp \prod_{i=1}^n
\rho(\bx_i,t_i)\rangle.
\end{equation}

\subsubsection{Second expression of time-reversal symmetry}

As said above, the time-reversal symmetry may be expressed through
another transformation of the fields. Indeed, consider the following
change:
\begin{equation}
{\mathcal U}:\left\{
\begin{array}{rl}
t&\rightarrow -t\\
\rhoh_x&\rightarrow-\rhoh_x+\beta\displaystyle{\frac{\delta {\mathcal
    F}}{\delta\rho_x}}.
\end{array}
\right.
\end{equation}
We remark that this transformation involves only the non-physical
field $\rhoh$, as does ${\cal T}$. Thus, both transforms clearly give rise
to the same relations between correlation functions.

As above the dynamical action can be easily proved to be invariant
under ${\mathcal U}$.
The time-reversal symmetry of any density
correlation functions follows directly.
FDR can also be derived using the identity:
\begin{equation}
\langle \rho\x\frac{\delta S}{\delta\rhoh\xtp}\rangle=0,
\end{equation}
which implies
\begin{equation}
\begin{split}
\langle \rho\x\partial_{t'}\rho\xtp\rangle=&\langle
\rho\x\nabla\cdot\left(\rho\xtp\nabla\frac{\delta {\cal F}}{\delta\rho\xtp}
\right)\rangle\\
&-2T\langle\rho\x\nabla\cdot\left(\rho\xtp\nabla\rhoh\xtp\right)\rangle.
\end{split}
\end{equation}
Splitting the last term in two identical parts and applying the
transformation ${\cal U}$ to one of them leads us to
\begin{equation}
\begin{split}
\langle \rho\x\partial_{t'}\rho\xtp\rangle=&\langle
\rho\x\nabla\cdot\left(\rho\xtp\nabla\frac{\delta {\cal F}}{\delta\rho\xtp}
\right)\rangle\\
&-T\langle\rho\x\nabla\cdot\left(\rho\xtp\nabla\rhoh\xtp\right)\rangle\\
&+T\langle\rho\xn\nabla\cdot\left(\rho\xpn\nabla\rhoh\xtpn\right)\rangle\\
&-\langle
\rho\xn\nabla\cdot\left(\rho\xtpn\nabla\frac{\delta {\cal F}}{\delta\rho\xtpn}
\right)\rangle.
\end{split}
\end{equation}
The two terms containing no $\rhoh$ in the RHS cancel
and one gets (\ref{eqn:FDT}).

>From ${\cal U}$ one can derive another useful identity as
follows. Applying the transformation to ${\cal G}_{xx'}(t-t')=\langle
\rho(\bx,t)\rhoh(\bxp,t')\rangle$ one gets:
\begin{equation}\label{eqn:idU}
\beta\langle \rho(\bx,t) \displaystyle{\frac{\delta {\cal
    F}[\rho,\rhoh]}{\delta\rho(\bxp,t')}}\rangle=\Theta(t-t')\langle
    \rho(\bx,t)\rhoh(\bxp,t')\rangle+\Theta(t'-t)\langle
    \rho(\bxp,t')\rhoh(\bx,t)\rangle.
\end{equation}
As we shall show in paragraph \ref{section:quad}, this identity allows
substantial
simplifications when ${\cal F}$ is quadratic.
Indeed in that case ${\cal U}$ is a linear
transformation and (\ref{eqn:idU}) then becomes a linear
relation between the density-density correlator $\langle
\delta\rho(\bx,t) \delta\rho(\bxp,t')\rangle $ and the naive response
${\cal G}_{xx'}(t-t')$, which are both two-points functions.

\section{Nonlinear symmetries and perturbation theories}

As will be emphasized all along this article, a physical symmetry is
related to fields transformations that leave the action invariant.
In the following we shall show from a general point
of view why the linearity of these
transformations is essential to construct symmetry preserving
perturbative expansions. None of BDD and FNH field theories has this
property, namely the fields transformations related to time-reversal
symmetry are nonlinear. In the following we shall highlight all the
complications which arise in perturbation theories
because of the nonlinearity, focusing
on BDD. In particular this will make clear why
mode-coupling approximations (MCA) generically violate time-reversal
symmetry and hence FDR.
\subsection{General discussion}
For the sake of generality, let us consider a field $\phi_{\alpha }$
(where $\alpha=(x,t)$ in our case) and a generic field theory:
\begin{equation}\label{genericfieldtheory}
\int \prod_{\alpha }d\phi_{\alpha }\exp \left[
-\frac{1}{2}\phi_{\alpha } (G_0^{-1})_{\alpha ,\beta }\phi_{\beta }
+gV (\phi )\right],
\end{equation}
where $g$ is the coupling constant controlling the perturbative
expansion, $V$ is the interaction and the field $\phi$ is
assumed to have a zero mean. The bare propagator is
$G_0$ and the bare action $S_0[\phi]=-\frac{1}{2}\phi_{\alpha }
(G_0^{-1})_{\alpha,\beta}\phi_\beta +gV (\phi)$.

Now consider a linear transformation ${\mathcal O}$ of the field
$\phi_{\alpha}\rightarrow \phi_\alpha'=O_{\alpha,\beta}\phi_\beta$
which leaves $S_0$ invariant. We require that the integration
measure is left invariant (the contrary giving rise to so-called
anomalies in quantum field theory), which means that ${\mathcal O}$ must have
determinant of modulus equal to one (or at least to a constant). Then
we have the following Ward-Takahashi (WT)
identities:
\begin{equation}\label{eqn:WT}
\langle\phi_{\alpha_1}\cdots\phi_{\alpha_n}\rangle=O_{\alpha_1,\beta_1}\cdots
O_{\alpha_n,\beta_n}\langle\phi_{\beta_1}\cdots \phi_{\beta_n}\rangle.
\end{equation}
One immediately sees the problem when the symmetry is not
linear. Consider for
instance a transformation of the form $\phi_{\alpha}\rightarrow
\phi_\alpha'=O_{\alpha,\beta}\phi_\beta^2$. The WT identities are
\begin{equation}
\langle\phi_{\alpha_1}\cdots\phi_{\alpha_n}\rangle=O_{\alpha_1,\beta_1}\cdots
O_{\alpha_n,\beta_n}\langle\phi_{\beta_1}^2\cdots \phi_{\beta_n}^2\rangle.
\end{equation}
Clearly, a nonlinear symmetry induces relations between correlation
functions of different orders in the fields.
A concrete example is the proof of FDR obtained in the previous sections for
BDD field theory.
The general idea behind mode-coupling approximation
and other self-consistent approaches is to provide closed
equations for two point functions. Hence, because the transformation
mixes correlation functions of different order in the fields
it seems difficult, if not impossible, to construct MC-like approximations
which preserve the symmetry.
In order to understand where perturbation theories fail in the case of non
linear symmetries it is useful to recall what are the key ingredients that
make them work when the symmetry is linear.

Let us first focus on standard (non self-consistent) perturbation theory.
A linear symmetry implies a relation between two point functions, see Eq.
(\ref{eqn:WT}). Why is this relation preserved order by order in perturbation
theory in $g$?
The reason is that the potential part $V(\phi)$ is itself invariant under
${\mathcal O}$ and thus (\ref{eqn:WT}) is also true when both LHS and RHS
are computed at any order in $g$.
For concreteness, let us focus on the MSRJD field theory for one particle
evolving under Langevin dynamics.
The dynamical action
can be split into two parts: $S_{QUAD}+S_{INT}$, with
\begin{equation}
S_{QUAD}[\phi]=\int_{-\infty}^\infty\, dt\left[
-\hat{X}(t)\cdot\partial_t X(t)+T\hat{X}(t)^2\right]
\end{equation}
and
\begin{equation}
S_{INT}[\phi]=-g\int_{-\infty}^\infty\, dt\, \hat{X}(t)\cdot\nabla V(X(t)).
\end{equation}
Both parts of the action are invariant under time-reversal and one can
expand in powers of $g$:
\begin{equation}
\langle \hat{X}(t')X(t)\rangle=\sum_{n=0}^\infty
\frac{{(-g)}^n}{n!} \langle\hat{X}(t')X(t)
S_{INT}[\phi]^n\rangle_{GAUSS},
\end{equation}
where $\langle\cdot\rangle_{GAUSS}$ stands for the Gaussian average with
action $S_{QUAD}$. However $S_{INT}$ is itself invariant under the
action of ${\cal O}$, thus the identity (\ref{Lfdr}), FDR,
is true at any order in perturbation theory.

Let us now focus on self-consistent perturbation theory.
This approximation scheme consists in cutting the perturbation series
at a given order and replacing the bare propagator by the
dressed one in the formal expression of the self-energy (some diagrams
present in bare perturbation theory have to be neglected in order to avoid
double counting). Doing so one
obtains self-consistent equations for the evolution of the dressed
propagator which in principle can be improved considering more diagrams.
Note also that these self-consistent equations can be often shown to be exact
in some large $N$ limit, where the field has become a $N$
component field~\cite{bckm}.

The formal way to obtain these self-consistent perturbation theory
is through the Legendre transform $\Gamma(G)=\log F (\Sigma (G),G)$
of two points functions which generates all the two particles
irreducible diagrams~\cite{DeDom}:
\begin{equation}
\begin{split}
\exp \Gamma (G)=\int \prod_\alpha d\phi_{\alpha }\exp
\Biggl[&-\frac{1}{2}\sum_{\alpha,\beta}
\phi_\alpha (G_0^{-1})_{\alpha,\beta}\phi_\beta
+g V (\phi)\\
&-\frac{1}{2} \sum_{\alpha\beta}\Sigma_{\alpha
  ,\beta}(\phi_\beta\phi_{\alpha}-G_{\beta,\alpha})\Biggr],
\end{split}
\end{equation}
where $\Sigma $ is determined by the condition $\frac{\partial F
}{\partial\Sigma }=0$, i.e. $\Sigma $ such that the propagator of the
theory equals $G$. De Dominicis and Martin have shown that $\Gamma $
is equal to
\begin{equation}\label{gamma}
-\frac{1}{2}Tr \left( G^{-1}_0\
G\right)+\frac{1}{2}Tr\ln G+\phi_{2PI}( G),
\end{equation}
where $\phi_{2PI}(G)$ is the sum of all 2PI diagrams with
the full propagator $G$ used as internal lines~\cite{DeDom}.
The physical propagator of the theory is
obtained by finding the $G$ which makes $\Gamma $ stationary, i.e. solving
$\frac{\partial
\Gamma }{\partial G}=0$. Hence, it is immediate to check that the
self-energy as a function of the propagator is $\Sigma
(G)=2\frac{\partial \phi_{2PI}}{\partial G}$. Practically the usual
self-consistent or mode-coupling approximations keep the lowest non-trivial
diagrams, the bubbles, in $\phi_{2PI}$ .

Let us now prove that the symmetry is preserved if the field transformation
$\phi_{\alpha }\rightarrow \phi_{\alpha }'=O_{\alpha
,\beta }\phi_{\beta }$ that leaves the action
invariant is linear.
For the same reason as before we assume $|\det O|=1$.
Under this transformation the propagator transforms into
$G'_{\alpha ,\beta }=M_{\alpha ,\beta ;\alpha ',\beta '}G_{\alpha '
,\beta '}$ where $M_{\alpha ,\beta ;\alpha ',\beta '}=O_{\alpha
  ,\alpha '}O_{\beta ,\beta '}$ ($|\det M|=1$). We shall prove that
this relation is preserved in the self-consistent perturbative expansion.

The first thing we want to prove is that $\Gamma (MG)=\Gamma (G)$ which
in a certain sense is expected since a physical symmetry should not change
the value of the functional. This is easy to prove starting from the
expression
for $\Gamma (MG)$ and rewriting $\phi_{\alpha }=O_{\alpha ,\beta }\phi '
_{\beta }$ and changing variable in the functional integral. The
measure as well as the action is invariant so that one finds
\begin{eqnarray}
\Gamma (MG)=\log\int \prod_{\alpha }d\phi_{\alpha }\exp &\Biggl[
-\frac{1}{2}\sum_{\alpha\beta}\phi_{\alpha } (G^{0})^{-1}_{\alpha ,\beta }\phi_{\beta }
+gV (\phi)\\ \nonumber
&\ \,-\frac{1}{2} (\Sigma M)_{\alpha ,\beta }(\phi_{\alpha
}\phi_{\beta }-G_{\alpha ,\beta })\Biggr].
\end{eqnarray}
Calling $\Sigma '=\Sigma M$ and noticing that $\frac{\partial
\Gamma (\Sigma,MG)}{\partial\Sigma }=0$ is equivalent to
$\frac{\partial \Gamma  (\Sigma',G)}{\partial\Sigma'}=0$ we obtain
that indeed $\Gamma (MG)=\Gamma (G)$. Furthermore since this is true
for any value of $g$, the coupling constant, then this identity is true
for $\Gamma $ expanded to any given order in $g$.
The consequences of this identity are particularly useful. First if
$G$ is a solution of $\frac{\partial \Gamma }{\partial G}=0$ so is $MG$.
Thus, if there is a unique solution - the symmetry is unbroken - then
$G=MG$. In the case of time-reversal symmetry
this identity leads to FDR. The fact that the solution is unique is
expected in our case since time-reversal symmetry is
certainly not broken at equilibrium.
Since this is true for any value of $g$ then
it is true when $\Gamma $ is
expanded to any finite order in g, i.e. for the self-consistent
equations written
to any finite order in g.
Finally the last remark following from $\Gamma (MG)=\Gamma (G)$ is
the symmetry relation verified by the self energy: $\Sigma (MG)=\Sigma (G)M $.

To recap, we have shown that a symmetry related to a linear
transformation in the field is a symmetry of the self-consistent
equations written to any finite order in g.
That is the reason why FDR is preserved for standard MSRJD Langevin field
theory. Note that a very different proof can be found in~\cite{Ma}.

\subsection{Subtleties in perturbation theory}

In this section we focus on BDD field theory and we highlight the difficulties
and the failures of perturbation theories {\it vis {\`a} vis} FDR.
The case of FNH is conceptually identical but practically more clumsy because
of the larger number of fields.
Let us start with bare perturbation theory.
In order to do a perturbative analysis it is convenient to separate
the Gaussian zero mean part of the local density field from the
interacting one by introducing density fluctuations in (\ref{eqn:MSR})
$\drho_x=\rho_x-\rho_0$. This gives:
\begin{equation}
S=\int_x \left(s_{0,x} +s_{INT,x} \right),
\end{equation}
with:
\begin{eqnarray}
s_{0,x}&=&\rhoh_x\left(-\partial_t\drho_x+T\nabla^2\drho_x
+\rho_0\int_{x'}\,
V(\bx-\bx')\nabla^2\drho_{x'}\right)+T\rho_0\left(\nabla\rhoh_x\right)^2
\nonumber \\
s_{INT,x}&=&T\drho_x(\nabla \rhoh_x)^2 +\rhoh_x\ \nabla\cdot
\left(\drho_x\int_{x'}\, \nabla V(\bx-\bx')\,\drho_{x'}\right).
\end{eqnarray}
This may be written in a more compact form through the
bidimensional vector field $(\drho,\rhoh)^\dagger$.
In Fourier space, the inverse of the propagator of this field is:
\begin{equation}
\tilde G^{-1}_0=\left(
\begin{array}{cc}
0 & i\omega+T\bk^2\left(1+\beta\rho_0 V(\bk)\right) \\
-i\omega+T\bk^2\left(1+\beta\rho_0 V(\bk)\right) & -2 T\rho_0\bk^2
\end{array}
\right).
\label{eqn:prop1}
\end{equation}
However, including the potential in the form (\ref{eqn:prop1}) makes
it practically impossible to write weak coupling expansions that preserve time-reversal symmetry. Hence we
shall drop it out of the inverse of the propagator:
\begin{eqnarray}
G^{-1}_0&=
\left(\begin{array}{cc}
C_0(\bk,\omega)& G_0(\bk,\omega)\\
G_0^*(\bk,\omega)& 0
\end{array}
\right)^{-1}
\\ \nonumber
&=\left(
\begin{array}{cc}
0 & i\omega+T \bk^2\\
-i\omega+T\bk^2 & -2 T\rho_0\bk^2,
\end{array}
\right),
\label{eqn:prop}
\end{eqnarray}
and we shall treat the quadratic term as an insertion in a line. This
gives the following Feynman rules:
\begin{itemize}
\item bare density correlator:\\
\includegraphics[width=.2\textwidth]{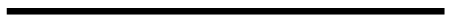}
$\ \ \ \ C_0(\bk,\omega)$
\item bare naive response: \\
\includegraphics[width=.2\textwidth]{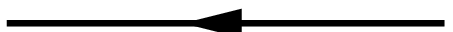}
$\ \ \ \ {\cal G}_0(\bk,\omega)$
\item line insertion:\\
\includegraphics[width=.09\textwidth]{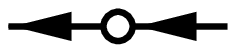}
 $\ \ \ \ -\rho_0 \bk^2 V(\bk)$
\item potential vertex:\\
\includegraphics[width=.14\textwidth]{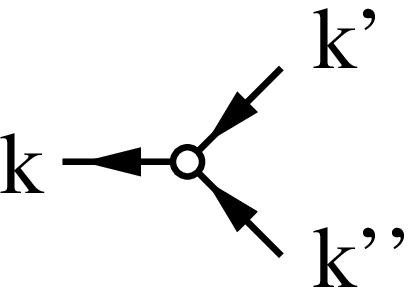}
$\ \ \ \ \ \ \  \frac{1}{2}\left(\bk\cdot\bk' V(\bk')+\bk\cdot\bk'' V(\bk'') \right)=\Gamma(\bk,\bk',\bk'')$
\vskip 4mm
\item noise vertex:\\
\vskip 1mm
\includegraphics[width=.14\textwidth]{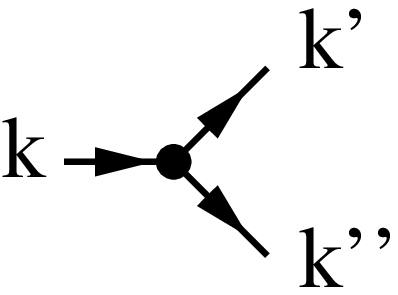}
$\ \ \ \ -T\bk'\cdot\bk'$
\end{itemize}
\vskip 2mm
The bare density correlator is
\begin{equation}
C_0(\bk,\omega)=\frac{2T\rho_0\bk^2}{\omega^2+ \left(T\bk^2\right)^2},
\label{eqn:barc}
\end{equation}
and the bare naive response
\begin{equation}
{\cal G}_0(\bk,\omega)=\frac{1}{T\bk^2-i\omega}.
\label{eqn:barg}
\end{equation}
Due to the form of the vertices, diagrams with tadpoles and hence
corrections to the average density vanish to all orders (the
momentum at the entrance into the tadpole is zero).

In terms of density fluctuations, the response is
\begin{eqnarray}
R_{xx'}(t,t')&=&-\langle \drho\x \nabla\cdot\left(
\drho\xtp\nabla\rhoh\xtp\right)\rangle
-\rho_0\nabla^2\langle \drho\x \rhoh\xtp\rangle\nonumber\\
&=&\chi_{xx'}(t,t')-\rho_0\ \nabla^2 {\cal G}_{xx'}(t,t'),
\label{eqn:rep}
\end{eqnarray}
where ${\cal G}$ is the naive response and $\chi$ is an ``anomalous'' response.
Having a look at (\ref{eqn:rep}), one sees that part of the non-triviality of
the FDR arises from the anomalous response $\chi$, which itself comes
from the multiplicative aspect of the noise, or equivalently from the
nonlinearity of the transformation of the fields associated with
time-reversal.

A perturbative expansion in powers of the potential preserves the symmetry.
Indeed the dynamical average of any function ${\cal
A}[\rho,\rhoh]$ may be written as
\begin{equation}
\begin{split}
\langle{\cal A}[\rho,\rhoh]\rangle&=\int{\cal D}\rho\int{\cal D}\rhoh
\ {\cal A}[\rho,\rhoh]\ e^{S_{FREE}[\rho,\rhoh]+S_V[\rho,\rhoh]}\\
&=\sum_{p=1}^\infty\int{\cal D}\rho\int{\cal D}\rhoh
\ {\cal A}[\rho,\rhoh] \frac{(S_V[\rho,\rhoh])^p}{p!}
e^{S_{FREE}[\rho,\rhoh]},
\end{split}
\end{equation}
where $S_{FREE}$ contains the terms of the action which do not contain
the potential.
The key point is that both $S_{FREE}$ and $S_V$ are invariant under
${\cal T}$, and thus FDR is verified to any finite order
in the expansion in power series of $S_V$.
However $S_{FREE}$ {\it is not quadratic in the fields}. The nonlinearity
of the fields transformation relate the noise cubic term to quadratic terms.
Thus this vertex must be taken into account
non-perturbatively. In other words, to compute the correlators at a
given order in the power
series expansion in $S_V$, one has to include contributions at all
orders in the noise vertex. This is a difficult but not impossible task
because, at a given order $p$ in
$S_V$, the diagrammatic expansion contains a finite number of
diagrams, due to the absence of the propagator
connecting two $\rhoh$'s. Indeed, if one considers a diagram with $p$
potential vertices and $q$ noise vertices contributing to a correlation
function of
$r$ $\rho$'s and $s$ $\rhoh$'s, one must have $q+s\leq p+r$. Hence
such a diagram must have less than $p+r-s$ noise
vertices

Thus bare perturbation theory which preserves the nonlinear symmetry can be
set up but is considerably more complicated than the usual one.
Finally, this discussion makes it clear that
any approximation which drops a part of the diagrams
of the full expansion in terms of the noise vertex is expected to be
in contradiction with the FDR. This is indeed what happens in
self-consistent approximations as we shall show below.

\subsection{Violation of fluctuation dissipation relations in self-consistent perturbation theory}
\begin{figure}[htb]
\begin{center}
\includegraphics[width=.2\textwidth]{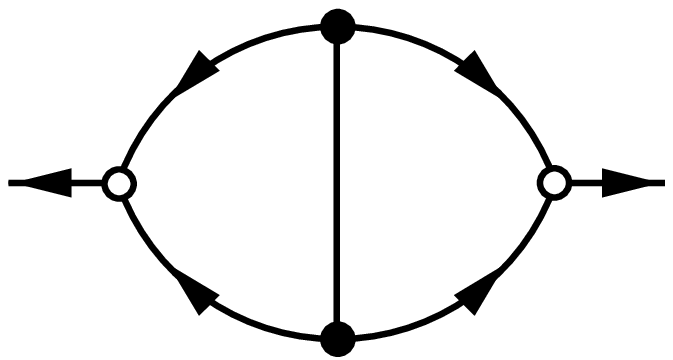}
\end{center}
\caption{Example of diagram which contribute at order $V^2$ to FDR but
is absent of the MC equations.}\label{fig:FDT}
\end{figure}

Let us now focus on self-consistent perturbation theory in particular
on the mode-coupling approximation introduced by Kawasaki~\cite{kaw2}
that consists in neglecting vertex renormalization.
First, we write the Schwinger-Dyson (SD) equations
\begin{equation}\label{eqn:SD}
G^{-1}_0\cdot G=1+\Sigma\cdot G,
\end{equation}
where $\Sigma$ is the self energy:
\begin{equation}
\Sigma\0=\left(
\begin{array}{cc}
\Sigma_{\rho\rho}\0 & \Sigma_{\rho\rhoh}\0\\
\Sigma_{\rhoh\rho}\0 & \Sigma_{\rhoh\rhoh}\0
\end{array}
\right),
\end{equation}
and the associative product $\cdot$ is defined as follows:
\begin{equation}
(A\cdot B)(\bk,t)=\int_{-\infty}^\infty dt'\,A(\bk,t-t')B(\bk,t').
\end{equation}
Causality and reality of the density auto-correlator imply that the
self-energies verify:
\begin{eqnarray}
\Sigma_{\rho\rhoh}(\bk,t)&=&\Sigma_{\rhoh\rho}(\bk,-t)\\
\Sigma_{\rho\rho}(\bk,t)&=&0.
\label{eqn:van}
\end{eqnarray}
The first diagrams contributing to the self-energies are\\
\begin{tabular}{ccc}
& & \\
$\Sigma_{\rhoh\rho}^{(2)}$&$=$&\includegraphics[width=.6\textwidth,clip]{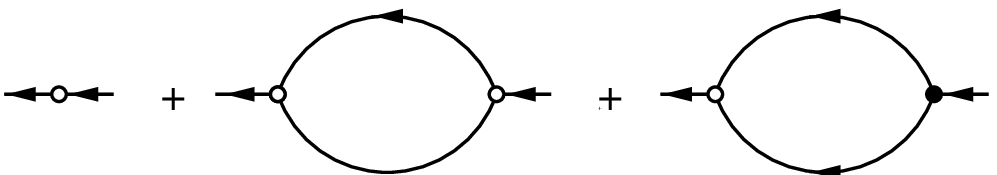}\\
$\Sigma_{\rhoh\rhoh}^{(2)}$&$=$&\includegraphics[width=.76\textwidth,clip]{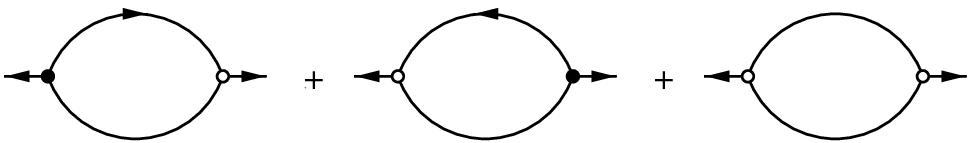}\\
& &
\end{tabular}\\
Diagrams of higher orders all contain vertex renormalization.
Hence, if one neglects renormalization of both vertices, the
SD equations (\ref{eqn:SD}) become MC equations for
(\ref{eqn:dean}) (for $t>0$):
\begin{eqnarray}
\partial_t {\cal G}(\bk,t)&=-\rho_0 T\bk^2\ \left(1+\beta\
V(\bk)\right) {\cal G}(\bk,t)+\int_{-\infty}^t dt'\
\Sigma_{\rhoh\rho}(\bk,t-t')\ C(\bk,t')\nonumber \\
\partial_tC(\bk,t)&=-\rho_0 T\bk^2\ \left(1+\beta\
V(\bk)\right)C(\bk,t)+\int_0^\infty dt'\
\Sigma_{\rhoh\rhoh}(\bk,t-t')\  {\cal G}(\bk,t')\nonumber\\
&+\int_{-\infty}^t dt'\ \Sigma_{\rhoh\rho}(\bk,t-t')\ C(\bk,t'),\label{eqn:naiveSD}
\end{eqnarray}
with:
\begin{eqnarray}
\Sigma_{\rhoh\rho}(\bk,t)&=&4\intq\  {\cal G}(\bk,t)\
C(\bk-\bq,t)\Gamma(\bq,\bk,\bk-\bq)\Gamma(\bk,\bq,\bk-\bq)\nonumber\\
&&-2\intq\  {\cal G}(\bk,t)\  {\cal G}(\bk-\bq,t)\bq\cdot(\bk-\bq)\Gamma(\bq,\bk,\bk-\bq)\nonumber\\
\Sigma_{\rhoh\rhoh}(\bk,t)&=&2\intq\ C(\bk,t)\
C(\bk-\bq,t)^2\\\nonumber
&&-8\intq\ \Re  {\cal G}(\bk,t)\ C(\bk-\bq,t)\bq\cdot\bk\Gamma(\bq,\bk,\bk-\bq).
\end{eqnarray}
These equations are not compatible with FDR, as can be seen from the
solution of SD equations at low orders in the potential.
FDR is trivially verified at order zero. At order one the MC equations are exact
and hence they are automatically compatible with FDR.
Incompatibilities appear at order two, where diagrams such as
those shown in Fig.~\ref{fig:FDT}, the first diagrams contributing to vertex
renormalization, have to be taken into account in the non
self-consistent perturbation theory and therefore also in the
self-consistent one (as discussed previously,
in order to preserve FDR one has always to take into account
the contributions at all orders in the noise vertex contribute to FDR).
This suggests that if one wants to improve the approximation by keeping for instance
the first vertex corrections, one has to include at least all the
diagrams of order two (in powers of the potential) in the self-energy. In
that case, the incompatibility with FDR would be an effect of order
three. However, nothing guaranties that the violation of FDR by the
self-consistent approximation is attenuated when the order
of the approximation is increased.
Another consequence of practical importance is violation of causality
in (\ref{eqn:naiveSD}), where times in the integrals are not
restricted in $[0,t]$. On the contrary, when time-reversal - and thus
FDR - is preserved, times in the integrals run from $0$ to $t$.
To conclude this paragraph, we remark that incompatibilities with FDR
have arisen from an explicit breaking of the time-reversal
symmetry which is due to the nonlinearity of the field transformation
related to time-reversal symmetry.

\section{Restoration of time-reversal symmetry in perturbative
expansions}

In this section, by introducing some extra fields, we construct a
generalization of the field theory described previously in which
time-reversal symmetry corresponds
to a field transformation which is {\it linear}.
As a consequence all the problems described in the previous section
are eliminated. In particular this allows to set up
self-consistent perturbative equations which preserve FDR.
We explain in detail our procedure for the BDD field theory ; the result
of this strategy for the FNH field theory will be given in section \ref{sec:FNH}.

\subsection{Introducing extra fields}\label{sec:morefields}

As already discussed in order to overcome the violation of
time-reversal
symmetry in perturbative expansions we need to make
the related fields transformation linear.
This can be achieved introducing some auxiliary fields.
Furthermore we also introduce
a {\it response field} such that the response function becomes
explicitly a two-fields correlation function.
We have found two different fields transformation related to the same
physical symmetry (time-reversal). As a consequence we can make each
one or both of the symmetries linear which leads to three
different field theories. In this
section we shall treat the case in which only either the $\cal U$ or
 $\cal T$
transformation
is linearized. The case of the
completely linearized theory (both symmetries are made linear)
is considered in Appendix E.

Let's consider the time-reversal transformation ${\cal T}$ first: we
start from the identity:
\begin{equation}
\langle{\cal A}\rangle=\int{\cal D}\rho{\cal A}[\rho]\langle
\delta\left(\partial_t\rho\x-R[\rho,\eta]\x\right)\rangle,
\end{equation}
where $R[\rho,\eta]\x$ is the RHS of (\ref{eqn:dean}). We now plug
into the functional integral the representation of the identity
\begin{equation}
\int {\cal D}f\ \prod_{x,t} \delta \left(\nabla \cdot(\rho_x\nabla
f_x)
+\frac{1}{T}\partial_{t} \rho_x \right)\left(\det \left[-\nabla\cdot(
  \rho_x \nabla) \right]\right)=1.
\end{equation}
We put a minus sign in such a way that the operator inside
the determinant is positive definite. Thus we don't need to take
the absolute value of the determinant.
We exponentiate the delta function using an auxiliary
field $\fh$ and we introduce fermionic fields $\phi$ and $\overline{\phi}$ to
exponentiate the determinant.
As a consequence there is a new term to add to
the previous action that reads:
\begin{equation}
\int_x \hat f_x \left(\nabla (\rho_x\nabla
f_x)+\frac{1}{T}\partial_{t} \rho_x \right)
-\int_x\ \rho_x\nabla \phi_x\cdot\nabla \overline{\phi}_x.
\end{equation}
Furthermore we introduce also the field $\psi=\nabla \cdot(\rho
\nabla \rhoh)$ which allows for the usual two-fields correlator
representation of
the response function. This leads to introduce a conjugated field
$\psih$ for
the Fourier representation of the delta function related to $\psi$. The final action is
the integral of
\begin{equation}
\begin{split}
s_x=& -\rhoh_x\partial_t\rho_x+\psi_x\frac{\delta{\cal
    F}}{\delta\rho_x}-T\rhoh_x\psi_x
+\psih_x(\psi_x-\nabla_x\cdot(\rho_x\nabla_x\rhoh_x))\\
& +\fh_x\left(\frac{1}{T}\partial_t\rho_x+\nabla_x\cdot(\rho_x\nabla_x
    f_x)\right)-\rho_x\nabla_x\cdot\left(\phi_x\nabla_x\overline\phi_x\right).
\label{msrd-action-1}
\end{split}
\end{equation}
This action now remains invariant up to the boundary terms under the
following linear
transformation ${\cal T}_1$: first invert the time $t\rightarrow -t$
then change the
fields in sequential order as follows:
\begin{equation}
{\cal T}_1:\left\{
\begin{array}{rcl}
\rhoh_x & \rightarrow & \rhoh_x +f_x\\
\psi_x &\rightarrow& \psi_x+\frac{1}{T}\partial_{t} \rho_x \\
\psih_x &\rightarrow& \psih_x+T f_x\\
\fh_x &\rightarrow& -\fh_x+T f_x+\psih_x +T\rhoh_x\\
f_x&\rightarrow& -f_x
\end{array}
\right.
\end{equation}
Formally we write this as $\tilde\phi=O\cdot\phi$, where
$\phi=(\rho,\rhoh,\psi,\psih,f,\fh,\phi,\overline\phi)^T$ and
$\tilde\phi(\bx,t)=\phi(-\bx,t)$. This implies an identity for
correlators:
\begin{equation}
\label{eqn:mgm}
\tilde{G}=O\cdot G\cdot O^T.
\end{equation}
The transformation has a determinant of modulus one, as a product of
simple transformations with this property.

Let us now show that this transformation implies FDR. Consider
$R_{xx'}(t'-t)=-\langle\rho(\bx,-t)\psi(\bx',-t')\rangle$. Under the
transformation ${\cal T}_1$ this transforms into
$R_{xx'}(t-t')-\frac{1}{T}\partial_{t'}C_{xx'}(t-t')$. Thus the
equality (\ref{eqn:mgm}) implies in particular
$R_{xx'}(t-t')=R_{xx'}(t'-t) +\frac{1}{T}\partial_{t'}C_{xx'}(t-t')$
which is the fluctuation-dissipation relation.

We now show how to linearize
the second transformation ${\cal U}$. We introduce
the field $\theta=\delta{\cal F}/\delta\rho$ and the conjugate one
$\thetah$ to exponentiate the delta function \footnote{The usefulness
of introducing these two fields when dealing with the BDD field
theory was noticed by C. Chamon and L.F. Cugliandolo from a
sligthly different perspective \cite{CC}.}.

The action $S$ is then transformed into an integral of
\begin{equation}
\label{eqn:minimal-action}
s_x=-\rhoh_x\partial_t\rho_x+T\rho_x(\nabla_x\rhoh_x)^2
+\thetah_x\left(\theta_x-\frac{\delta{\cal F}}{\delta\rho_x}\right)
-\rho_x(\nabla_x\rhoh_x)(\nabla_x\theta_x).
\end{equation}
The associated linear transformation ${\cal U}_1$ is
\begin{equation}
{\cal U}_1:\left\{
\begin{array}{rcl}
\rhoh_x&\rightarrow& -\rhoh_x+\frac{1}{T}\theta_x\\
\thetah_x&\rightarrow& \thetah_x-\frac{1}{T}\partial_t\rho_x,
\end{array}
\right.
\end{equation}
As before we write it as $\tilde\phi=O\cdot\phi$, where
$\phi=(\rho,\rhoh,\theta,\thetah)^T$ and
$\tilde\phi(\bx,t)=\phi(-\bx,t)$. The same
identity (\ref{eqn:mgm}) holds for correlators. Again the
transformation has a  determinant of modulus one.
The response function can be written as
$R_{xx'}(t-t')=\langle\rho(\bx,t)\thetah(\bx',t')\rangle$.
Using the transformation ${\cal U}_1$ we find again that
$\langle\rho(\bx,-t)\thetah(\bx',-t')\rangle$ equals
$R_{xx'}(t-t')-\frac{1}{T}\partial_{t'}C_{xx'}(t-t')$, hence FDR.
Note that this second field theory is considerably simpler than the previous
one because it has less fields.

One might wonder how the final result depends on the choice of the
linearized field theory. As far as self-consistent perturbation theory
and MCT is concerned we have written the dynamical equations obtained by
\begin{itemize}
\item a) using the completely linearized theory where both symmetries
  are rendered linear.
\item b) only the fields involved in the transformation ${\cal T}_1$
  in addition to $\rho$.
\item c) only the fields involved in the transformation ${\cal U}_1$
  in addition to $\rho$.
\item d) writing the terms of the action such that the potential $V$ is in one of the vertices (in the goal of making the link with standard MCT).
\end{itemize}
At the order of one loop, we have found the same sets of equations for
correlation and response function at long times in all cases. This is
not surprising, since the different transformations do not affect the
physical fields and change the response fields in the same way. We
thus expect this to be valid at all orders. Indeed, as we will show
below, one gets closed equations
for the dynamical evolutions of correlators involving only the fields
$\rho$ and $\theta$.
In addition, this tends to confirm that FDR
makes the results robust with respect to the choice of extra dynamical
variables.  Thus in the following, we will focus on the
simplest theory written above in terms of $\rho$, $\rhoh$, $\theta$ and
$\thetah$ only, the choice of the fields to work with being merely a
matter of taste. We refer the reader interested in the theory with all
fields introduced above to Appendix E.

\subsection{Minimal theory preserving fluctuation dissipation
  relations and its basic properties}

In order to avoid cumbersome calculations, we shall from now on
describe a minimal (in the sense of the number of fields) theory for
which the symmetry associated to FDR is linear. It is the theory
produced by ${\cal U}_1$.
In the following we shall describe the WT relations
for correlation functions and self-energies due to time-reversal
symmetry. These are particularly useful because they make clear that
there are only three independent correlation functions or
self-energies out of sixteen.
In the next section we shall write down the general form
of the Schwinger-Dyson equations. Using the WT relations
we find that these reduces only to three
independent integro-differential equations.

Henceforth we use the density of action (\ref{eqn:minimal-action}) and
the symmetry ${\cal U}_1$. We split the density of action into a Gaussian
part $s_0$ and the interaction part $s_{INT}$:
\begin{eqnarray}
\label{action-gauss}
s_{0,x}&=&-\rhoh_x\partial_t\delta\rho_x-T\rho_0\rhoh_x\nabla_x^2\rhoh_x
+\thetah_x\theta_x-\thetah_x
(W\star \delta\rho)_x+\rho_0\rhoh_x\nabla^2\theta_x\\
\nonumber\\
\label{action-int}
s_{INT,x}&=&T\delta\rho_x(\nabla\rhoh_x)^2
-\delta\rho_x(\nabla_x\rhoh_x)(\nabla_x\theta_x)
+T\thetah_x\sum_{n>1}\frac{1}{n}\left[-\frac{\delta\rho_x}{\rho_0}\right]^n,
\end{eqnarray}
where
\begin{equation}
(W\star \rho)_x=\int_y W(\bx-\by)\rho_y
\end{equation}
and
\begin{equation}
W(\bx)=V(\bx)+\frac{T}{\rho_0}\delta(\bx).
\end{equation}
It is useful to remark here that each term of the expansion in $n$ is
independently invariant under ${\cal U}_1$.

One gets the following relations for correlators from identity
(\ref{eqn:mgm}) applied to the case of ${\cal U}_1$ (the limit
$t,t'\to\infty, \tau=t-t'$ is taken):
\begin{eqnarray}
C_{\rho\rhoh,xx'}(\tau)&=&\frac{\Theta(\tau)}{T}C_{\rho\theta,xx'}(\tau)\label{eqn:idcorr1}\\
C_{\theta\rhoh,xx}(\tau)&=&\frac{\Theta(\tau)}{T}C_{\theta\theta,xx'}(\tau)\\
\label{eqn:idcorr2}
C_{\theta\thetah,xx'}(\tau)&=&-\frac{1}{T}\frac{\partial}{\partial\tau}\left(\Theta(\tau)\,C_{\rho\theta,xx'}(\tau)\right)\label{eqn:idcorr3}\\
C_{\rho\thetah,xx'}(\tau)&=&-\frac{\Theta(\tau)}{T}\frac{\partial}{\partial\tau}C_{\rho\rho,xx'}(\tau)=R_{xx'}(\tau),\label{eqn:idcorr4}
\end{eqnarray}
with the obvious notation $C_{ab,xx'}(t-t')=\langle a_x(t)
b_{x'}(t')\rangle$.  These identities imply only three independent
two-fields correlators, namely $C_{\rho\rho,xx'}$,$C_{\rho\theta,xx'}$
and $C_{\theta\theta,xx'}$. Moreover, the causality of the theory,
which is insured by the It{\^o} discretization, is explicit. A
perturbative proof of causality can be found in Appendix B and comes
from the causality of the bare propagator, which in Fourier space is:
\begin{equation}
\label{eqn:bare-corr}
\left(
\begin{array}{cccc}
\frac{2T\rho_0 k^2}{(\omega^2+\rho_0^2 k^4 W(\bk)^2)} &
-\frac{1}{i\omega+\rho_0k^2W(\bk)} & \frac{2T\rho_0 k^2
W(\bk)}{(\omega^2+\rho_0^2 k^4 W(\bk)^2)}& \frac{\rho_0
  k^2}{i\omega+\rho_0k^2W(\bk)}\\
-\frac{1}{-i\omega+\rho_0k^2W(\bk)} & 0 &
-\frac{W(\bk)}{-i\omega+\rho_0k^2W(\bk)} & 0\\
-\frac{2T\rho_0 k^2W(\bk)}{(\omega^2+\rho_0^2 k^4 W(\bk)^2)} &
-\frac{W(\bk)}{i\omega+\rho_0 k^2W(\bk)}& -\frac{2T\rho_0
k^2W(\bk)^2}{(\omega^2+\rho_0^2 k^4
W(\bk)^2)}&-\frac{i\omega}{i\omega+\rho_0k^2W(\bk)}\\
\frac{\rho_0k^2}{-i\omega+\rho_0k^2W(\bk)} &
0&\frac{i\omega}{i\omega+\rho_0k^2W(\bk)} & 0
\end{array}
\right).
\end{equation}
The bare propagator helps also to understand the anomaly in
(\ref{eqn:idcorr2}) where the time derivative of the Heaviside
function is present. One can easily see that the bare propagator
(\ref{eqn:bare-corr}) has the form:
\begin{equation}
C_{0,\theta\thetah}(\bk,\omega)=-1+\mbox{function of}(\bk,\omega).
\end{equation}
It is easy to prove using (\ref{action-gauss},\ref{action-int})
that there are no diagrammatic corrections to the constant part of the above
correlator, that is in the above equation the constant part persists
perturbatively and we can write:
\begin{equation}
\label{eqn:delta-anomaly}
C_{\theta\thetah,xx'}(\tau)=-\frac{\Theta(\tau)}{T}
\frac{\partial}{\partial\tau}
C_{\rho\theta,xx'}(\tau)-\delta(\tau).
\end{equation}
One can find (see appendix C) that
$C_{\rho\theta,xx'}(0)=T$. The latter together with (\ref{eqn:delta-anomaly})
allows to write (\ref{eqn:idcorr2}).

Recall that the Schwinger-Dyson equations have the form $G_0^{-1}\cdot
G=1+D$, with $D=\Sigma\cdot G$.
Thus, $\Sigma$ transforms under ${\cal U}_1$ in the following way:
\begin{equation}\label{eqn:vdv}
\tilde\Sigma=O^{-T}\cdot\Sigma\cdot O^{-1},
\end{equation}
enforcing the following constraints on self-energies:
\begin{eqnarray}
\Sigma_{\rhoh\rho}(\bk,\tau)&=&\frac{\partial}{\partial\tau}\Sigma_{\thetah\theta}(\bk,\tau)\\
\Sigma_{\thetah\rho}(\bk,\tau)&=&-\frac{\Theta(\tau)}{T}\frac{\partial}{\partial\tau}\Sigma_{\thetah\thetah}(\bk,\tau)\\
\Sigma_{\rhoh\theta}(\bk,\tau)&=&-\frac{\Theta(\tau)}{T}\Sigma_{\rhoh\rhoh}(\bk,\tau)\\
\Sigma_{\rhoh\thetah}(\bk,\tau)&=&T\Theta(-\tau)\Sigma_{\thetah\theta}(\bk,-\tau)-T\Theta(\tau)\Sigma_{\thetah\theta}(\bk,\tau).
\end{eqnarray}
The other elements ($\Sigma_{\rho\rho}$, $\Sigma_{\theta\rho}$ and $\Sigma_{\theta\theta}$) vanish. One can write the equation for $\Sigma_{\rhoh\rho}(\bk,\tau)$ in the following form:
\begin{equation}
\label{s-energy-anomaly}
\Sigma_{\rhoh\rho}(\bk,\tau)=-\frac{1}{T}\frac{\partial}{\partial\tau}\left[\Theta(\tau)\Sigma_{\rhoh\thetah}(\bk,\tau)\right].
\end{equation}
As before we start with the diagrammatic analysis recovering a set of
diagrams which contribute to the $\delta$-function term. Their
direct resummation is cumbersome. We used the Schwinger-Dyson
equations to identify the $\delta$-function term.
Note that as for correlation functions there are only three
independent self-energy terms. All the others are either zero or
related through WT identities.


\subsection{Dynamical equations}

In the following section we transcribe the full dynamical equations
and leave their derivation for Appendix C. There are only three independent equations. Using
the transformation laws under $\cal U_{1}$ it is easy to see that there
are not more than four independent equations. The proof that
one of this equations is trivially verified once the other three are
verified is more tricky and done in Appendix D.

Let us first choose $\tau>0$ and write the equations which will give
the time evolution of the correlators at strictly positive
time-difference.
The values at $\tau=0$ of the correlators will be
obtained from the study of the singularities (remember there is a
$\delta(\tau)$ in the RHS of (\ref{eqn:SD})).

We first consider $(G_0^{-1}\cdot G-\Sigma\cdot
  G)_{\rhoh\rho}(\bk,\tau)=0$. The corresponding equation is
\begin{eqnarray}\label{eqn:SDrhr}
\partial_\tau C_{\rho\rho}(\bk,\tau)+\rho_0 k^2
C_{\rho\theta}(\bk,\tau)=
\int_0^\tau dt\,\Sigma_{\rhoh\theta}(\bk,\tau-t)
C_{\rho \theta}(\bk,t)
\\ \nonumber
+\int_0^\tau dt\,\Sigma_{\thetah\theta}(\bk,\tau-t)\partial_t
C_{\rho\rho}(\bk,t).
\end{eqnarray}
Now consider $(G_0^{-1}\cdot G-\Sigma\cdot
  G)_{\rhoh\theta}(\bk,\tau)=0$. The corresponding equation is
\begin{eqnarray}\label{eqn:pp}
\partial_\tau C_{\rho\theta}(\bk,\tau)+\rho_0 k^2
C_{\theta\theta}(\bk,\tau)=-\Sigma_{\thetah\theta}(\bk,\tau)
C_{\rho\theta}(\bk,0)\\ \nonumber
+\int_0^\tau dt\,\Sigma_{\rhoh\theta}(\bk,\tau-t)
C_{\theta\theta}(\bk,t)+
\int_0^\tau dt\,\Sigma_{\thetah\theta}(\bk,\tau-t)\partial_t C_{\rho
  \theta}(\bk,t).
\end{eqnarray}
Finally consider $(G_0^{-1}\cdot G-\Sigma\cdot
  G)_{\thetah\rho}(\bk,\tau)=0$. The corresponding equation is
\begin{eqnarray}\label{eqn:Ct}
W(\bk)C_{\rho\rho}(\bk,\tau)-C_{\rho
  \theta}(\bk,\tau)=\frac{1}{T}\Sigma_{\thetah\thetah}(\bk,0)
C_{\rho \rho}(\bk,\tau)\\ \nonumber
+\int_0^\tau dt\,\Sigma_{\thetah\theta}(\bk,\tau-t)
C_{\rho \theta}(\bk,t)-
\frac{1}{T}\int_0^\tau dt\,\Sigma_{\thetah\thetah}(\bk,\tau-t)
\partial_t C_{\rho\rho}(\bk,t).
\end{eqnarray}
As discussed at the beginning of this section, there is an extra
equation that at first sight might seem independent from the first
three (but in fact is not, see Appendix D). It comes from
$(G_0^{-1}\cdot G-\Sigma\cdot
  G)_{\thetah\theta}(\bk,\tau)=0$ and reads:
\begin{eqnarray}\label{eqn:Ct2}
&&W(\bk)C_{\rho\theta}(\bk,\tau)-C_{\theta\theta}(\bk,\tau)=\frac{1}{T}\Sigma_{\thetah\thetah}(\bk,0)
C_{\rho \theta}(\bk,\tau)-\frac{1}{T}\Sigma_{\thetah\thetah}(\bk,\tau)
C_{\rho \theta}(\bk,0)\nonumber\\
&+&\int_0^\tau dt\,\Sigma_{\thetah\theta}(\bk,\tau-t)
C_{\theta\theta}(\bk,t)-
\frac{1}{T}\int_0^\tau dt\,\Sigma_{\thetah\thetah}(\bk,\tau-t)
\partial_t C_{\rho\theta}(\bk,t).
\end{eqnarray}

We stress here that obtaining such consistent structure -
  which is the {\em exact one} - for the dynamical equations is far from
trivial. Approximations which violate FDR generally lead to a
different structure. In such cases, the resulting conclusions about
the existence of the glass transition are very suspicious because they
might be due to the violation of this fundamental structure.

\subsection{Static limit}

The static equations are obtained by taking $\tau=0$ in the SD
equations. This is an advantage of the present field theory
approach, in which the statics is included in the dynamics, compared
to derivations of dynamical equations based on
Mori-Zwanzig formalism.
We remark that the closed set of static equations cannot be obtained from the
evolution equations as written above, as the singularities at
$\tau=0$ have been excluded from the latter.
The correct derivation is given in Appendix C.
The correlators and their derivatives at initial time are:
\begin{eqnarray}
C_{\rho\theta}(\bk,0)&=&T\\
\dot{C}_{\rho\rho}(\bk,0^+)&=&-\rho_0\bk^2C_{\rho
\theta}(\bk,0)=-T\rho_0\bk^2\\
C_{\theta\theta}(\bk,0)&=&W(\bk) C_{\rho\theta}(\bk,0)=TW(\bk)\\
\dot{C}_{\rho \theta}(\bk,0^+)
&=&-W(\bk)T\rho_0\bk^2-T\Sigma_{\thetah\theta}(\bk,0)\label{eqn:Cff0}\\
C_{\rho\rho}(\bk,0)&=&\frac{T}{W(\bk)}+\frac{1}{TW(\bk)}
\Sigma_{\thetah\thetah}(\bk,0)C_{\rho\rho}(\bk,0)\label{eqn:Ct0},
\end{eqnarray}
where $\dot{C}$ stands for $\partial_\tau C$. The equation
(\ref{eqn:Ct0}) is obtained from (\ref{eqn:Ct}) and
(\ref{eqn:stt}).

\subsection{Equation for the non-ergodicity parameter}

In MCT, the glass phase is characterized by a non-zero value of the
so-called non-ergodicity parameter~\cite{got,got1}, which signals the
existence of
infinite-time correlations. The non-ergodicity parameter $f(\bk)$ is defined
as follows:
\begin{equation}
f(\bk)=\lim_{t\rightarrow\infty}\frac{C_{\rho\rho}(\bk,t)}{S(\bk)}.
\end{equation}
In MCT, $f(\bk)$ is zero in liquid phase and jumps to a non-zero value
at the glass transition. Here the equation for the non-ergodicity
parameter can be obtained by taking the limit $\tau\rightarrow\infty$
in (\ref{eqn:SDrhr}-\ref{eqn:Ct2}).

We first give an argument in support of the vanishing of
$C_{\rho\theta}(\bk,\infty)$ and $C_{\theta\theta}(\bk,\infty)$.
The physical interpretation
is the following. We label by $\alpha$ each ergodic component into
which the system may break (if no ergodic to non-ergodic transition
occurs then there is only one such component). Then in each component the
system decorrelates completely at long times, and we can write:
\begin{equation}
\langle\rho(\bk,0)\frac{\delta{\cal
    F}}{\delta\rho(-\bk,\infty)}\rangle=\sum_\alpha \ {\cal W}_\alpha
\langle\rho\rangle_\alpha \langle\frac{\delta{\cal
    F}}{\delta\rho}\rangle_\alpha,
\end{equation}
where each component is characterized by its weight ${\cal W}_\alpha$
and the static average $\langle\cdot\rangle_\alpha$ inside it.
However, the average force $\langle\frac{\delta{\cal
    F}}{\delta\rho}\rangle_\alpha$ vanishes, and hence
$C_{\rho\theta}(\bk,\infty)$ too. This also applies to
$C_{\theta\theta}(\bk,\infty)$.
We stress here that the possibility to express infinite time
averages in term of static quantities is intimately related to the
fact that the dynamical equations are consistent with the dynamical
symetries at equilibrium. Indeed this guarranties that the asymptotic
measure is Gibbsian. Let us add that
$C_{\rho\theta}(\bk,\infty)=0$ may also be seen as a direct
consequence of $C_{\rho\rhoh}(\bk,\infty)=0$.
Furthermore, Eq. (\ref{eqn:SDrhr}) gives
$\Sigma_{\thetah\theta}(\bk,\infty)=0$, and putting (\ref{eqn:Ct0})
into (\ref{eqn:Ct0}) leads to the {\it exact equation}:
\begin{equation}
\frac{f(\bk)}{1-f(\bk)}=\frac{1}{T^2}S(\bk)\
\Sigma_{\thetah\thetah}(\bk,\infty).\label{eqn:nonerg}
\end{equation}
Of course this equation is too difficult to be analysed directly
because $\Sigma_{\thetah\thetah}$ contains an infinite number of terms
that cannot be resummed. The same occurs in the Mori-Zwanzig formalism,
where $\Sigma_{\thetah\thetah}$ is related to the long time limit of
the memory kernel. One has to resort to approximation schemes, the
simplest one being replace $\Sigma_{\thetah\thetah}$ by its one loop
expression. That will be discussed in the following.


\subsection{Quadratic density functionals}\label{section:quad}

In this section we discuss the particular case in which the
density functional is taken to be quadratic in the fields.
This corresponds to make a simple quadratic expansion of
the entropic part ${\cal S}$ and to replace in the FNH case the
kinetic term by ${\bf  g}^{2}/2\rho_{0}$ where $\rho_{0}$ is the
average density. The former case has been already studied
in~\cite{mya1} and the latter in~\cite{schmitz}.
In these works, especially in~\cite{mya1}, the issue of preserving FDR
has been already investigated and it has been shown that
one-loop mode-coupling equations preserve FDR in this case (contrary
to the general case).

>From our perspective the reason of this preservation is simple:
in these cases the fields transformations become linear even without
introducing extra fields. Let us consider in detail the BDD case
studied by Miyazaki and Reichman~\cite{mya1}.

When expanded at order two, the entropic part is written as
\begin{equation}
{\cal S}[\rho]\approx {\cal S}[\rho_0]-T\int_x\
\frac{\delta\rho_x^2}{2\rho_0}.
\label{eqn:exp}
\end{equation}
Thus the free energy functional ${\cal F}[\rho]$ becomes quadratic:
\begin{equation}
\begin{split}
{\cal F}[\rho]&=\frac{1}{2}\int_{x,x'}\ \rho_x\rho_{x'}\
W(\bx-\bxp).
\label{eqn:quad}
\end{split}
\end{equation}
The equilibrium measure from (\ref{eqn:quad}) is Gaussian:
\begin{equation}
\begin{split}
{\cal P}[\rho]&\propto e^{\displaystyle{-\frac{1}{2} \intk\
  \frac{\rho(\bk)\rho(-\bk)}{S(\bk)}}}\\
S(\bk)&=\langle \delta\rho(\bk)\delta\rho(-\bk)\rangle=\frac{T}{W(\bk)},
\end{split}
\end{equation}
where we have used $\intx\ W(\bx)=0$.
This
Gaussian form considerably simplifies the structure of the resulting
field theory. Indeed, the transformation ${\cal U}$ defined earlier
becomes linear:
\begin{equation}
{\cal U}:\left\{
\begin{array}{rl}
t&\rightarrow -t\\
\rhoh_x&\rightarrow-\rhoh_x+\beta\int_{x'}\ W(\bx-\bxp)\ \rho_{x'}
\end{array}
\right.
\end{equation}
As stated previously, the use of ${\cal U}$ makes it possible to
derive an identity
between the density-density correlators and the naive
response. One can write
\begin{equation}
\begin{split}\label{eqn:idUquad}
\langle\rho(\bx,t)\rhoh(\bxp,t')\rangle=&-\langle\rho(\bx,-t)\rhoh(\bxp,-t')
\rangle\\
&+\beta\intxp\ W(\bx-\bxp)\ \langle\rho\xn\rho\xtpn\rangle,
\end{split}
\end{equation}
or equivalently in Fourier space:
\begin{equation}\label{eqn:GC}
 {\cal G}(\bk,\omega)+ {\cal G}(\bk,-\omega)=\frac{C(\bk,\omega)}{S(\bk)}.
\end{equation}
The symmetry ${\cal U}$ is linear and thus does not mix correlators
of $n$ fields with correlators of $n+1$ fields, unlike ${\cal T}$. This makes
the identity (\ref{eqn:GC}) valid at any order in
perturbation - self-consistent or not -
as explained in section I.
In Appendix A, in order to give a concrete view of how
diagrams have to be put together to lead to (\ref{eqn:GC}),
a diagrammatic perturbative proof if given, as well as
the derivation of a similar relation between self energies:
\begin{equation}
\Sigma_{\rhoh\rhoh}(\bk,t)=-\frac{1}{S(\bk)}\Sigma_{\rhoh\rho}(\bk,t),\
\ t>0.
\label{eqn:GCself}
\end{equation}
This identity makes (\ref{eqn:GC}) compatible with the SD
equations (\ref{eqn:SD}).
Using these identities, it is possible to show that FDR is preserved
by self-consistent one-loop theory and also to all order in
self-consistent perturbation theory as explained in Appendix A.

A final remark about the quadratic case is that
from (\ref{eqn:nonerg}), it follows
$f(\bk)=0$, as there is no vertex involving
$\thetah$ and hence $\Sigma_{\thetah\thetah}$ must vanish
on general grounds.

\section{Mode-coupling approximation preserving fluctuation
  dissipation relations}

Now, we will carry out the MCA for the BDD field theory that we
studied in the previous sections.
The corresponding analysis of fluctuating nonlinear hydrodynamics
follows the same guidelines (see section \ref{sec:FNH}).
It is important to stress the difference between our and previous
approaches. We set up a self-consistent diagrammatic expansion
that preserves automatically time reversal symmetry and FDR. As a
consequence the corresponding MCA equations, or any other
approximated expression of the self-energies, will do that too.

\subsection{Mode-coupling approximation}

We first focus on the term $\int_x \thetah_x\frac{\delta
{\cal F}[\rho]}{\delta\rho_x}$. The entropic part of ${\cal F}$ gives
a contribution $\int_x \thetah_x\log
\left(1+\frac{\delta\rho_x}{\rho_0}\right)$ to this term. In order to
compute the dynamical partition function one may expand the logarithm
in powers of $\frac{\delta\rho_x}{\rho_0}$. This gives an infinite
number of vertices of the type
$T\frac{(-1)^p}{p}\thetah_x\left(\frac{\drho_x}{\rho_0}\right)^p$.
A crucial point is that, as stressed above, all
the terms in the action coming from different powers in the series
expansion are independently invariant under the transformation
${\cal U}_1$. We also can put
some couplings in front of the other vertices and carry out truncated
expansions at different orders for different vertices. However
as we focus on the one-loop theory, and for simplicity, we shall treat
these vertices as if they were all of order $T$.
Then there are two ways of dealing with the vertices arising from the
expansion of the logarithm. On one hand one can take into account all
these vertices. However this leads to the sum of an infinite number of
terms, whose meaning is not clear, due to the presence of an
infinite number of tadpoles which contribute to static vertex
renormalization. On the other hand one can truncate the series expansion
of the logarithm. However one needs to go beyond first order in order to take
into account nonlinearities. For simplicity we cut the series at
order two, but the calculation can be in principle extended to any order.

\subsection{Expression of the self-energies}

Within the approximation described in the previous paragraph, the
self-energies read:
\begin{eqnarray}
\Sigma_{\thetah\theta}(\bk,t)&=&\intq\frac{(\bk\cdot\bq)}{\rho_0^2}C_{\rho
\theta}(\bq,t) C_{\rho\rho}(\bk-\bq,t) \label{eqn:stt0}\\
\Sigma_{\rhoh\theta}(\bk,t)&=&\frac{1}{2T}\intq\Biggl\{(\bk\cdot\bq)^2
C_{\theta\theta}(\bq,t)C_{\rho\rho}(\bk-\bq,t)\label{eqn:stt01}\\
&&+(\bk\cdot\bq)[\bk\cdot(\bk-\bq)]C_{\rho
\theta}(\bq,t)C_{\rho \theta}(\bk-\bq,t)\Biggr\}\nonumber\\ \label{eqn:stt}
\Sigma_{\thetah\thetah}(\bk,t)&=&\frac{T^2}{4\rho_0^4}\intq
C_{\rho\rho}(\bq,t ) C_{\rho\rho}(\bk-\bq,t).
\end{eqnarray}
It is instructive at this stage to see how the MC equations derived by
Miyazaki
and Reichman~\cite{mya1}
in the case of the quadratic expansion can be obtained from the above
equations. The simplifications arising from the linearity of ${\cal
  U}_1$ will then become clear.
Using the method used to derive (\ref{eqn:idUquad}), one obtains
\begin{equation}\label{eqn:idUquad2}
W(\bk)\ C_{\rho \theta}(\bk,\tau)=T
C_{\theta\theta}(\bk,\tau),
\end{equation}
and (\ref{eqn:idUquad}) can be simply written as
\begin{equation}
W(\bk)\ C_{\rho\rho}(\bk,\tau)=T\Theta(\tau)\ C_{\rho
  \theta}(\bk,\tau^+)+T\Theta(-\tau)\ C_{\rho\theta}(\bk,(-\tau)^-).
\end{equation}
Using these two identities one can eliminate $C_{\rho \theta}$
and $C_{\theta\theta}$ in (\ref{eqn:SDrhr}). In addition, when the entropy is
expanded at quadratic order, all self-energies except
$\Sigma_{\rhoh\theta}$
vanish, and the equation which remains is the one derived and
discussed in~\cite{mya1}.

\subsection{Static limit}

The static equations are obtained by taking $\tau=0$ in
Eqs. (\ref{eqn:stt0}-\ref{eqn:stt}). Remarkably,
it is identical with the
one which would be obtained by doing the same MCA on the
static theory involving the density functional ${\cal F}[\rho]$.
Indeed, after truncation of the entropy at cubic order, the density
auto-correlator is
\begin{equation}\label{eqn:stat1}
C_{\rho\rho,yy'}(t=0)=\frac{\int{\cal
D}\delta\rho\,\rho(\by)\rho(\byp) e^{-\beta {\cal F}_3[\delta\rho]}}{\int{\cal
D}\delta\rho e^{-\beta {\cal F}_3[\delta\rho]}},
\end{equation}
with
\begin{equation}
{\cal F}_3[\delta\rho]=\frac{1}{6}\intx\frac{\delta\rho(\bx)^3}{\rho_0^3}
-\beta\intx\intxp W(\bx-\bxp)\delta\rho(\bx)\delta\rho(\bxp),
\end{equation}
where we have used the fact that $\intx\ W(\bx)=0$. It can be easily
verified that using the MCA as
described above for the correlator defined by (\ref{eqn:stat1}) gives
again (\ref{eqn:Ct0}).

\subsection{Equation for the non-ergodicity parameter}

Using the one-loop expression (\ref{eqn:stt}) in (\ref{eqn:nonerg}),
one gets the equation for the non-ergodicity parameter:
\begin{equation}
\frac{f(\bk)}{1-f(\bk)}=\frac{1}{2\rho_0^4}\intq
f(\bq)f(\bk-\bq)S(\bq)S(\bk-\bq)S(\bk).
\label{eqn:nonergMCA}
\end{equation}

Writing this in the form
\begin{equation}
f(\bk)=1-\frac{1}{m(\bk)},
\end{equation}
one sees that $f(\bk)=0$ and $f(\bk)=1$ are two solutions of
(\ref{eqn:nonergMCA}). However the second solution, which in addition
makes the integral in $m(\bk)$ diverge, implies
$C_{\rho\rho}(\bk,t)=1$ for all $t$ and thus has to be rejected. We
stress that in contrast with standard MCT the $q$-dependence of the
vertex (in particular, the absence of the term $(\bk\cdot\bq
c(\bq)+\bk\cdot(\bk-\bq) c(\bk-\bq))^2$) does not inforce the convergence
of the integral in (\ref{eqn:nonergMCA}).
However, the equation (\ref{eqn:Ct0}) which gives the structure factor
is also ill-defined as the integral over $\bq$ diverges too.
We have solved numerically these equations putting a cut-off for large
values of $|\bk|$. We have found a $|\bk|$ dependence of $f(\bk)$ very similar
to the usual one of MCT. However, as the cut-off
goes to infinity our numerical solution seems
to converge toward the solution $f(\bk)=1$ for all $\bk$. However,
although there is a clear {\it physical}
cut-off to the description
in terms of Langevin equations, the cut-off dependence we found
is clearly unphysical. It might be that convergence is only obtained
via a further resummation of diagrams that renormalize the vertex.
We leave this problem for a future work.
Ignoring this cut-off problem (which is in a sense
solved in the FNH case) for the time being, our MC equations have
the following properties:
\begin{itemize}
\item as in the standard MCT, one can derive a schematic theory~\cite{got} by
  assuming that the structure factor is dominated by a single mode
  $S(\bk)=1+A\delta (|\bk|-k_0)$. Then the schematic
  equation for the non-ergodicity parameter becomes identical to that
  of the schematic MCT.
\item We have also checked that the dynamical critical properties are
  the same as for standard MCT.
\item The static equations can be reduced to an equation involving
  the density field only, which is identical to that we can get from
  the equilibrium density functional by doing equivalent
  approximations. The way the theory has been written thus makes
  it possible to treat the dynamics in a similar way to the statics.
\end{itemize}
\section{Fluctuating Nonlinear Hydrodynamics\label{sec:FNH}}

In this section, we describe the case of FNH. As the
derivations follow closely those of BDD, we just give the results.

\subsection{Time-reversal symmetry and fluctuation-dissipation relations}

This time there are four response functions produced by the extra term
${\cal F}_{ext}=-\int_x(\rho_x\mu_x+{\bf g}_x\cdot{\bf p}_x)$
\begin{equation}
\begin{split}
\langle\rho(\bx,t)\rangle_\mu =&\langle\rho(\bx,t)\rangle_{\mu=0}+\int
d^3 y\int_s^t dt'' R_{\rho\rho,xx'}(t-t'')\mu(\by,t'')\\
&+\int d^3 y\int_s^t dt'' R^k_{\rho g,xx'}(t-t'')p_k(\by,t'')+
  o(\mu,{\bf p})
\end{split}
\end{equation}
\begin{equation}
\begin{split}
\langle g_i(\bx,t)\rangle_\mu=&\langle
  g_i(\bx,t)\rangle_{\mu=0}+\int d^3 y \int_s^t dt'' R^{ij}_{gg,xx'}(t-t''){
  p}_j(\by,u)\\
&+\int d^3 y\int_s^t dt'' R^i_{g\rho,xx'}(t-t'')\mu(\by,t'')+o(\mu,{\bf p}),
\end{split}
\end{equation}
which gives:
\begin{eqnarray}
R_{\rho\rho,xx'}(t-,t')
&=&\langle\rho\x\nabla\cdot(\rho\hat {\bf g})\xtp\rangle\\
R^k_{g\rho,xx'}(t-t')
&=&\langle{g}_k\x\nabla\cdot(\rho\hat{\bf g})\xtp\rangle\\
R^k_{\rho g,xx'}(t-t')
&=&\langle\rho\x\rho\xtp\nabla_k\rhoh\xtp\rangle+\langle\rho\x{g}_i\xtp\nabla_k\ \hat{g}_i\xtp\rangle\nonumber\\
&&+\langle\rho\x\nabla_i(g_k\hat{g}_i)\xtp\rangle-\langle\rho\x L_{ki}\hat{g}_i\xtp\rangle\\
R^{kl}_{gg,xx'}(t-t')
&=&\langle{g}_k\x\rho\x\nabla_l\rhoh\xtp\rangle+\langle{g}_k\x{g}_i\xtp\nabla_l\ \hat{g}_i\xtp\rangle\nonumber\\
&&+\langle{g}_k\x\nabla_i(g_l\hat{g}_i)\xtp\rangle-\langle{g}_k\x L_{li}\hat{g}_i\xtp\rangle
\end{eqnarray}
The transformation of the fields associated to time-reversal is now:
\begin{equation}
{\cal V}:\left\{\begin{array}{rl}t&\rightarrow -t\\
{\mathbf g}_x &\rightarrow -{\mathbf g}_x\\
\rhoh_x&\rightarrow-\rhoh_x+\frac{1}{T}\displaystyle{\frac{\delta {\cal F}}{\delta\rho_x}}\\
\hat{\mathbf g}_x &\rightarrow \hat{\mathbf g}_x-\frac{1}{T}\frac{\delta {\cal F}}{\delta\mathbf g_x}
\end{array}
\right.
\end{equation}
This transformations leaves the action invariant up to boundary terms:
\begin{equation}
\int_x \left[\frac{\delta {\cal F}}{\delta{\mathbf g}_x}\cdot \partial
  _t {\mathbf g}_x +\frac{\delta {\cal F}}{\delta\rho_x} \partial _t
  \rho_x+\nabla_i\left(g_{j,x}\frac{\delta{\cal
  F}}{\delta{g_{i,x}}}\frac{\delta {\cal F}}{\delta{g_{j,x}}}\right)\right].
\end{equation}
Following the procedure used for the BDD field theory one can see that
density correlation functions are invariant under time-reversal and
also derive FDR. We stress again that
the naive self-consistent perturbation
theory for the FNH violates time-reversal symmetry in the same way it does in case of the BDD.
An extension of the original model is needed providing linear time-reversal symmetry in order to satisfy
the symmetry in perturbation expansion.

\subsection{Restoration of time-reversal symmetry in perturbative expansions}

In order to make the transformation $\cal V$ linear we introduce the two additional fields
$\theta=\frac{\delta {\cal F}}{\delta\rho}$ and ${\mathbf
  v}=\frac{\delta {\cal F}}{\delta{\mathbf g}}$. We are lead to add
\begin{equation}
-\int_x \hat{\theta}_x\left[\theta_x-\frac{\delta{\cal F}}{\delta\rho_x}\right]-
\int_x\hat{\mathbf v}_x\cdot\left[\mathbf{v}_x-\frac{\delta{\cal F}}{\delta{\mathbf g}_x}\right]
\end{equation}
to the action, which becomes the integral of
\begin{eqnarray}
\label{fh-actionplus}
s_x&=&\Biggl\{-\rhoh_x\left[\partial_t\rho_x+\nabla_i\left(\rho_x v_i\right)\right]+T\hat g_{i,x} L_{ij}\hat g_{j,x}\\
&&-\hat g_{i,x}\left[\partial_t g_{i,x}+\rho_x\nabla_i\theta+\nabla_j\left(g_{i,x}v_j\right)
+g_{j,x}\nabla_i v_j+L_{ij}v_j\right]\nonumber \\
&&-\hat{\theta}_x\left[\theta_x-\frac{\delta{\cal F}}{\delta\rho_x}\right]-\hat{\mathbf v}_x\cdot\left[\mathbf{v}_x-
\frac{\delta{\cal F}}{\delta{\mathbf g}_x}\right] \Biggr\}.\nonumber
\end{eqnarray}
The corresponding linear time-reversal transformation is:
\begin{equation}
{\cal V}_1:\left\{\begin{array}{rl}t & \rightarrow -t\\
{\mathbf g}_x & \rightarrow -{\mathbf g}_x\\
{\mathbf v}_x & \rightarrow -{\mathbf v}_x\\
\rhoh_x & \rightarrow-\rhoh_x+\frac{1}{T}\theta_x\\
\hat{\mathbf g}_x & \rightarrow \hat{\mathbf g}_x-\frac{1}{T} {\mathbf v}_x\\
\hat{\theta}_x & \rightarrow \hat{\theta}_x+\frac{1}{T}\partial _t \rho_x\\
\hat{\bf v}_x & \rightarrow -\hat{\bf v}_x-\frac{1}{T}\partial _t {\mathbf g}_x
\end{array}
\right.
\end{equation}

As before, the action is conveniently split into Gaussian and
interaction parts: we expand all the
terms in powers of $\delta\rho/\rho_0$.
\begin{eqnarray}
s_{0,x}&=& -\rhoh_x(\partial_t\delta\rho_x+\rho_0\nabla\cdot{\mathbf v}_x)
-\hat g_{i,x}\left[\partial_t g_{i,x}+\rho_0\nabla_i\theta_x+L_{ij}v_{j,x}\right]\\\nonumber
&& +T\hat g_{i,x} L_{ij}\hat g_{j,x}-\thetah_x\theta_x+\thetah_x(W^{{FNH}}\star\delta\rho)_x-
\hat{\mathbf v}_x\cdot{\mathbf v}_x+\frac{1}{\rho_0}\hat{\mathbf v}_x\cdot{\mathbf g}_x\\
s_{INT,x}&=& -\rhoh_x\nabla\cdot(\delta\rho_x {\bf v}_x)-\delta\rho_x{\bf \hat g}_x\cdot\nabla\theta_x
-\hat g_{i,x}\nabla_j(g_{i,x} v_{j,x})\nonumber\\
&&-\hat g_{i,x} g_{j,x}\nabla_i v_{j,x}
-\sum_{p>1}\frac{(-1)^p}{p}\frac{T}{m}\thetah_x\frac{\delta\rho_x^p}{\rho_0^p}
\nonumber \\
&&+\sum_{n>0}(-1)^n \left[({\bf \hat v}_x\cdot{\bf
    g}_x)\delta\rho_x
+ n \thetah_x\frac{{\mathbf g_x}^2}{2}\right]\frac{\delta\rho_x^{n-1}}{\rho_0^{n+1}},
\end{eqnarray}
with
\begin{equation}
W^{{FNH}}(\bx)=\frac{T}{m}\left[\frac{1}{\rho_0}\delta(\bx)-\frac{c(\bx)}{m}\right].
\end{equation}
This expansion in powers produces two series of vertices, each being
independently invariant (also order by order) under the field
transformation ${\cal V}_1$.

Different equalities between correlators (and self-energies) result as
consequences of the use of ${\cal V}_1$ as for the BDD field theory.
In particular using that the response function is the
correlation between $\rho$ and $\hat{\theta}$ one gets FDR between
correlation and response.
The set of all relations and the simplified dynamical equations
are presented in Appendix F.

\subsection{Static limit}

The analysis of singularities of the SD equations at short time
difference gives:

\begin{eqnarray}
C_{\rho\theta}(\bk,0)&=&T\\
C_{gv}(\bk,0)&=&T\\
\label{eqn:fh-static-crr}
W(\bk)\Crr(\bk,0)&=&T+\frac{1}{T}\Sigma_{\thetah\thetah}(\bk,0)\Crr(\bk,0)\\
C_{\theta\theta}(\bk,0)&=&TW(\bk)\\
\frac{1}{\rho_0}C_{gg}(\bk,0)&=&T+\frac{1}{T}\Sigma_{\hat v\hat v}(\bk,0)C_{gg}(\bk,0)\\
C_{vv}(\bk,0)&=&\frac{T}{\rho_0}.
\end{eqnarray}

\subsection{Equation for the non-ergodicity parameter}\label{nep}

We now focus on the equation for the non-ergodicity parameter.
As for BDD, due to long time decorrelation inside ergodic components, only
$C_{\rho\rho}$, $C_{\rho g^\perp}$ and $C_{g^\perp g^\perp}$ (where
${\bf g}^\perp$ is the transverse current) can
have a non-zero limit when $\tau\to\infty$. This is confirmed by the
analysis of the dynamical equations given in Appendix F.
We however do not expect frozen currents, and thus it is quite
reasonable to asume that $C_{\rho g}(\bk,\infty)$ and
$C_{gg}(\bk,\infty)$ do vanish. Since at least one of those appear
in any diagram of
$\Sigma_{\rhoh\thetah}(\bk,\infty)$, $\Sigma_{\rhoh\hat{v}}(\bk,\infty)$,
$\Sigma_{\hat{g}\thetah}(\bk,\infty)$, $\Sigma_{\hat{g}\hat{v}}(\bk,\infty)$ and
$\Sigma_{\thetah\hat{v}}(\bk,\infty)$ those self-energies
also vanish. As a consequence one obtains the non-perturbative equation for the
non-ergodicity parameter using (\ref{eqn:non-erg-fh})
and its static limit (\ref{eqn:fh-static-crr}):
\begin{equation}
\frac{f(\bk)}{1-f(\bk)}=\frac{1}{T^2}S(\bk)\
\Sigma_{\thetah\thetah}(\bk,\infty).\label{eqn:nonergFNH}
\end{equation}
This structure is identical to the one found for BDD and is {\it an
exact result}. Any general approximation (one loop, two loops, etc) for the
self-energy on the right hand side, lead to a non linear equation
on $f (\bf k)$.
As we will discuss in detail later previous works have obtained very
different structures because in those cases time-reversal symmetry
was violated. This may be very dangerous because it can generate
spurious results as it is indeed the case for the cut-off of the
transition. For example in the analysis of \cite{das} this modifies strongly
the general structure of the Schwinger-Dyson equations and implies
that the non-ergodic parameter has to vanish. However, this has
nothing to do with the physical mechanism that cut-off the MCT
transition and is just an artefact of having violated  the time reversal symmetry.

\subsection{Mode-coupling approximation}

We restrict ourselves in this section to an approximation similar to
the one used for BDD, truncating the
vertex series in $\delta\rho/\rho_0$ at the lowest order:
\begin{eqnarray}
s_{INT,x}&=& -\rhoh_x\nabla\cdot(\delta\rho_x {\bf v}_x)-\delta\rho_x{\bf \hat g}_x\cdot\nabla\theta_x
-\hat g_{i,x}\nabla_j(g_{i,x} v_{j,x})\nonumber\\
&&-\hat g_{i,x} g_{j,x}\nabla_i v_{j,x}-{\bf \hat v}_x\cdot{\bf g}_x\frac{\delta\rho_x}{\rho_0^2}
-\thetah_x\frac{{\mathbf g_x}^2}{2\rho_0^2}-\frac{T}{2m}\thetah_x\frac{\delta\rho_x^2}{\rho_0^2}.
\end{eqnarray}
We do not write the set of all equations at the one-loop level.
We just remark that as in the BDD case the static correlation
functions can be obtained from the dynamical equations.
They coincide with the ones obtained doing the analoguous approximation
on the static theory. At the level of one-loop, the expression of
$\Sigma_{\thetah\thetah}$
reduces to the one previously obtained for BDD (up to a multiplication
by the mass). We want to stress that it is a coincidence which is
absent when higher orders schemes are considered. We believe that this is
due to the fact that the static effective free energies are different
in the two field theories.

As a consequence the MCA equation for the non-ergodic parameter is the
same as the one obtained for BDD and the previous remarks applied also
to this case. We have also checked that the critical long-time behavior
of the correlation function close to the MCT transition is the
standard MCT one and coincides with the one obtained in the BDD case.
Note that contrary to BDD there is now a cutoff
to regularize the integral over $\bq$ because FNH is not valid on
short-length scale.

\section{Relation with previous works and issues related to
  Mode-coupling Theory\label{sec:disc}}

In this section we would like to put our work in the context of
field theory derivations of MCT and discuss what can be learned
from the non-perturbative structure of the equations which we derived,
and from the resulting mode-coupling equations.

The first issue that we want to discuss is the field theory derivation
of mode-coupling equations. One can find in
 the literature~\cite{kaw,das}
different field theories derivations of the original full
$k$-dependent mode-coupling equations introduced and
studied by G{\"o}tze et al.~\cite{reviewDave,kob}. All these derivations are
not consistent because they assume some identities (related to
time-reversal symmetry) that are incompatible with the
same self-consistent equations used to derive MCT.
Indeed the Kawasaki derivation of MCT using field theory~\cite{kaw}
starts from a BDD field theory (in which the potential is replaced by
a term proportional to the static direct correlation
function). Kawasaki computed $\Sigma_{\hat\rho\hat\rho }$ for the
original (without extra fields) BDD theory and {\it assumed} that
$\Sigma_{\hat\rho\rho}$ was related to it by a symmetry transformation
similar to FDR for correlation functions:
\begin{equation}
\Sigma_{\hat\rho\rho}
(\tau)=-\frac{1}{T}\partial_\tau\Sigma_{\hat\rho\hat\rho}(\tau).
\end{equation}
Looking at the whole set of one-loop equations it is easy to see that
this relation is not verified. Actually the structure of the
Schwinger-Dyson
equations is such that if this relation was verified there would have been
a fluctuation dissipation relation between the correlation function and
$C_{\rho\hat\rho}(t)$ which {\it is not} the response function as
discussed previously.

We now focus on the Das and Mazenko~\cite{das} field theory
derivation of MCT. In the original papers and all subsequent ones
there is a linear relationship that has been assumed to hold between
$C_{\rho\rho}$ and $C_{\rho\hat\rho}$ (see eq. 6.62 of
\cite{das}). This relation, that has been used to close the equations
on $C_{\rho\rho}$, is true only in the hydrodynamic limit
(long time and length scales) or for a purely quadratic free energy
density functional. Instead, time-reversal symmetry generically
implies a more complicated identity that is the generalization of
(\ref{eqn:idU}) to the FNH case:
\begin{equation}\label{eqn:idFNH}
\beta\langle \rho(\bx,t) \displaystyle{\frac{\delta {\cal
    F}[\rho,\mathbf{g}]}{\delta\rho(\bxp,t')}}\rangle=\Theta(t-t')\langle
    \rho(\bx,t)\rhoh(\bxp,t')\rangle+\Theta(t'-t)\langle
    \rho(\bxp,t')\rhoh(\bx,t)\rangle,
\end{equation}
where $\cal F$ is the free energy functional introduced for the FNH
field theory. Forcing the relation between $C_{\rho\rho}$ and
$C_{\rho\hat\rho}$ is useful to close the equations but (1) it is
very dangerous as already discussed in \ref{nep} and as it will be
explained later and (2) it is inconsistent with time reversal symmetry
that imposes a different relation, eq. (\ref{eqn:idFNH}).

Contrary to these two approaches our derivation of MC equations
preserves time-reversal symmetry because it is constructed upon a
self-consistent expansion that preserves automatically this symmetry.
This is important for the self-consistency of the theory and it is crucial
in order to study off-equilibrium dynamics as discussed in the introduction.
At one loop, it leads to a vertex with a different $k$-dependence
than usual MCT. This is problematic for two reasons:
first the good quantitative results of MCT depend crucially on the vertex $k$-dependence and,
second, our vertex leads to a strange and cut-off dependent behavior
or the mode-coupling equations. Our conclusion is that more
refined (e.g. higher loops) self-consistent approximations have to explored is such a way
to obtain MCA that preserves time reversal symmetry and is at the same
time quantitatice succesfull as the one developed in \cite{ben}.
The investigation of this issue is left for future studies.

An interesting consequence of our derivation and our equations
concerns extended MCT and the cut-off of the transition.
Using field theory~\cite{das} and also projection operator
formalism~\cite{ben} equations beyond standard MCT have been
obtained. These are called extended MCT equations and have been conjectured to
contain the ``hopping or activated'' effects that destroy the MCT
transition. A striking result
is that these ``hopping or activated'' effects are present only if
there is a density-current coupling. As a direct consequence, they would be expected
for particles evolving under Newtonian dynamics (and thus described by
FNH) and not for particles evolving under Brownian dynamics
(and thus described by BDD). Recent numerical simulations have shown
\cite{szamel} that quite the contrary happens: the same effects
cutting off the transition are present for Brownian or Newtonian
particles. It has been also believed for some time that ideal MCT transition
might occur (i.e. is not avoided) for Brownian dynamics but it is now
clear that it
is impossible since there are mathematical theorems proving
\cite{theor} that the self-diffusion coefficient of hard spheres evolving
under Brownian dynamics never vanishes at any finite temperature or
chemical potential.

As a conclusion the Mode Coupling Transition is expected to be always
replaced by a cross-over. The crucial question is what is the physical
mechanism that cut-off the transition, how capture it in an analytical
theory and whether it is the same or not for Newtonian and Brownian dynamics.
In our derivation we find at long-times the same transition,
and at one-loop the same equations, in the Newtonian and
Brownian cases, hence there is no sign that the cut-off of the
transition is due to the presence of coupling to currents.
This conclusion remains valid at any finite order in the self
consistent perturbation theory (one loop, two loops, etc.). In
practice any approximation will provide an expression for
$\Sigma_{\thetah\thetah}(\bk,\infty)$
as a function of $f ({\bf k})$ that has to be plugged into
eq. (\ref{eqn:nonergFNH}) or its BDD counterpart.
There is no general argument implying that the resulting non-linear
equation on $f ({\bf k})$ cannot have a non vanishing solution and indeed
one expects that the transition present at one loop carries on at
higher loops. That has been indeed verified in toy models \cite{BC}
as Langevin particle inside a double well potential. Our conclusion
is therefore that the cut-off mechanism is due to
non-perturbative effects that cannot be captured at any finite order
of the self-consistent perturbation theory.

This is in clear contradiction with previous field theoretical works
that found a cut-off mechanism for one loop self-consistent expansion.
In the following we unveil, using exact results, that that mechanism
was only due to the approximations that were used and
that violated explicitely violation of time-reversal symmetry.

The evidence for an avoided transition in the field theory derivation comes mainly
from two works. Das and Mazenko~\cite{das} found that when all one-loop diagrams are considered
the transition is cut-off. However, this was due to the relation they
assumed between the two correlation functions $C_{\rho\rho}$ and
$C_{\rho\hat \rho}$.
In fact, as we have explained earlier, $C_{\rho\rhoh}$ cannot have
a plateau. Therefore by forcing this relation one kills artificially
any possibility of having a glass transition. Another  way to put it,
as discussed in \ref{nep}, is that this relationship alters
completely the non-perturbative structure of Schwinger-Dyson
equations.
Whereas it is not possible to conclude anything about the
cut-off of the transition just looking at the general equation for
the non-ergodic parameter, see section \ref{nep}, the relationship
assumed in \cite{das} between  $C_{\rho\rho}$ and
$C_{\rho\hat \rho}$ plus the general form of the Schwinger-Dyson
equations imply that the non-ergodic parameter has to be zero.
Indeed, consider one of the exact Schwinger-Dyson equations derived
in the Appendix F:
\begin{eqnarray}\label{eqa}
&&\partial_\tau C_{\rho\rho}(\bk,\tau) -i\rho_0\bk C_{v\rho}(\bk,\tau)= \frac{1}{T}\int_{0}^{\tau}dt\Sigma_{\rhoh\thetah}(\bk,\tau-t)\partial_tC_{\rho\rho}(\bk,t)\nonumber\\
 &+&\frac{1}{T}\int_{0}^{\tau}dt\Sigma_{\rhoh\hat v}(\bk,\tau-t)\partial_tC_{g\rho}(\bk,t)
 -\frac{1}{T}\int_{0}^{\tau}dt\Sigma_{\rhoh\rhoh}(\bk,\tau-t)C_{\theta\rho}(\bk,t)\nonumber\\
 &-&\frac{1}{T}\int_{0}^{\tau}dt\Sigma_{\rhoh\hat g}(\bk,\tau-t)C_{v\rho}(\bk,t)\end{eqnarray}
and the exact fluctuation dissipation relation derived in Appendix F:
\begin{equation}\label{eqb}
C_{\rho\rhoh}(\bk,\tau)=\frac{\Theta(\tau)}{T}C_{\rho\theta}(\bk,\tau).
\end{equation}
The linear relationship assumed in \cite{das} between  $C_{\rho\rho}$ and
$C_{\rho\hat \rho}$ implies that if $C_{\rho\rho}$  has an infinite
plateau at the transition so does $C_{\rho\hat \rho}$ and, using the
FDR relation (\ref{eqb}), $C_{\rho\theta}$. However, a plateau
of $C_{\rho\theta}$ is incompatible wit the exact equation
(\ref{eqa}). Indeed, integrating eq. (\ref{eqa}) over $\tau $ between zero and
infinity, we would get an infinite right hand side
and a finite right hand side \footnote{Only the third term on the
right hand side would give an infinite contribution since $C_{v\rho}$
cannot have a plateau. This can be found by
inspection studying the other equations or just on physical grounds
because at the glass transition amorphous circulating currents are not
expected.}. Therefore one would conclude that the first hypothesis,
i.e. an infinite plateau for $C_{\rho\rho}$, has to be wrong.
However, this is {\it only due} to having forced
a relation between $C_{\rho\rho}$ and $C_{\rho\hat \rho}$! Thus,
the cut-off of the transition found in \cite{das} is spurious.
The relation between $C_{\rho\rho}$ and $C_{\rho\hat \rho}$
is valid only on hydrodynamic length and timescales.

We finally remark that other arguments complementary to ours
have been given recently~\cite{cates}, in favour of
rejection of the cut-off mechanism derived within extended
of MCT provided in~\cite{das}.

The other work that is cited as a supporting evidence for the cut-off
of the transition is the one of Schmitz et al.~\cite{schmitz}.
However, they considered a purely quadratic free energy functional.
We have seen in our derivation that the transition comes from the
density nonlinearity term in the free-energy functional and therefore
it is absent in the quadratic case. In~\cite{schmitz} FDR is respected
and the corresponding equations are a particular case of the ones we
derived. The fact that there is no transition is
the natural consequence that theories with quadratic free energy functional
are {\it too simple} to lead to any MCT transition in the FNH as well as in
the BDD case. In fact, contrary to the case of non quadratic
functionals, there is no transition at any finite order in the
self-consistent perturbation theory simply because the self-energy
$\Sigma_{\hat \theta\hat \theta}$ in eq. (\ref{eqn:nonergFNH})
is identically zero!
Note that in the Brownian dynamic case this has been already found,
although in a different way, by Miyazaki and Reichman~\cite{mya1}.

\addcontentsline{toc}{section}{Conclusion}
\section*{Conclusion}
In this paper we analyzed field theories for the dynamics of glass
forming liquids focusing on
their symmetry properties. In particular we have shown that
straightforward perturbation theories generically do not preserve
time-reversal symmetry. We have found that
 time-reversal symmetry is related to a {\it nonlinear} field
transformation that leaves the action invariant. The nonlinearity
is the source of the problems. Introducing some auxiliary fields
we have shown how to to set up perturbation theories that preserve
time-reversal symmetry and hence FDR.

Furthermore we have critically revisited the field theory
derivations of MCT showing that they always assumed some
relations which are actually incompatible with the self-consistent
equations. Our derivation is completely consistent and preserves
FDR but leads to a different vertex than in the usual MCT.
This leads to similar qualitative results but clearly different
quantitative results. Furthermore strange and spurious divergences
appear if we do not put an ultraviolet cutoff on momenta integration.
We leave for future work an accurate investigation of this
puzzle that will certainly need the introduction of some kind
of diagram resummation.

We have also reconsidered the evidence for the cut-off of the MCT
transition when a coupling between density and current is present.
We have shown using exact results that
there is no obvious cut-off mechanism which
acts order by order in self-consistent perturbation theory.
>From this perspective there is no difference between Brownian and
Newtonian dynamics.  Whether or
not density-current couplings are present, we have found the same
formal structure for the equations on the non-ergodic parameter in the glass
phase. Actually at one loop the corresponding equations
for the non ergodic parameter are identical for BDD and FNH.
This structure is fundamentaly different from the one
previously obtained~\cite{das}~ that suggested a cut-off mechanism
only for Newtonian dynamics; the latter was due to the assumption of a linear
relationship between correlation functions, that although valid in the
hydrodynamic limit, is not verified in general and that, via the
general structure of the Schwinger-Dyson equations, force the
non-ergodic parameter to be zero. The correct relation,
eq. (\ref{eqn:idFNH}), has no influence on the existence of the MCT
transition. We conclude that the MCT transition has to be cut-off by
non perturbative mechanisms. A very recent work \cite{MMR} shows by a schematic
approximation that once any factorization approximation is avoided,
i.e. at a non perturbative level, the MCT transition indeed disappears.

Despite the problems - due to excessive simplicity of
approximations such as the MCA considered here - related to our vertex
we think that
our approach is promising since it automatically connects statics to
dynamics in a precise way.
It remains the issue of finding
appropriate approximations for the self energies. However, one has not
to worry without about compatibility with time-reversal symmetry,
which is guarranted by the form (\ref{eqn:SDrhr}-\ref{eqn:Ct2}) of the
Schwinger-Dyson equations. This step was missing in the previous
attemps to derive MCT equations from field theory, where
approximations were done too early.
Thus, it opens the way to dynamical equivalents of the self-consistent
schemes developed in liquid theory~\cite{Hansen,mori}. This was
already initiated in a pioneering work \cite{KW} where the
relationship between dynamic and static theory for the glass
transition was investigated.

Finally we remark that the identification of field transformations
associated to time-reversal symmetry may help developing other
non-perturbative approaches. For instance in the non-perturbative
renormalization group~\cite{ber} approach, an Ansatz for the
effective dynamical
action has to be made. The choice among different possible Ans{\"a}tze
is very wide but can be drastically reduced by the use of
symmetries. We hope that the analysis of the symmetries performed in
this work may be helpful in applying NPRG techniques to the
the glass transition problem.

{\bf Acknowledgements} We are deeply indebted to J.-P. Bouchaud for
very useful discussions, a careful reading of our manuscript and his
participation to the early stage
of this work. We also thank L. Berthier, C. Chamon, P. Calabrese,
L.F. Cugliandolo, K. Miyazaki and D. R. Reichman for discussions,
in particular K. Miyazaki and D. R. Reichman for a careful
reading of this manuscript.
A. L. acknowledges support from European training
program STIPCO under grant HPRN-CT-2002-00319. AA and GB are partially
supported by the European Community's Human Potential Program
contracts HPRN-CT-2002-00307 (DYGLAGEMEM).

\addcontentsline{toc}{section}{Appendix A. Perturbative proofs for the
quadratic theory}
\section*{Appendix A. Perturbative proofs for the quadratic theory}

\subsection*{A.1. Proof of (\ref{eqn:GC})}

In this appendix a perturbative proof of (\ref{eqn:GCself}) will be
given (in the quadratic expansion of the free-energy). The identity
(\ref{eqn:GC}) will follow from the relation
between self-energy and connected two-points functions:
\begin{equation}
G^{-1}-G_0^{-1}=\Sigma.
\end{equation}
Here the order $n$ in perturbation theory consists of diagrams with
$n$ vertices $B[\rho,\rhoh]$, the vertex including two parts $B=B_1+B_2$,
$B_1$ and $B_2$ being the noise and $W$ vertices respectively.

In this appendix, we will always have $t>0$.
As already said above, the splitting of $\rho$ into $\rho_0$
and $\delta\rho$ leads to the splitting of both noise and interaction
vertices in two parts in the dynamical action. The first one, linear
in $\rho_0$ is quadratic
and can be seen as a ``mass'' insertion, whereas the second one is
cubic and gives the vertices used in the perturbative
expansions. However it turns out that both parts (the one with
$\rho_0$ and the other) are independently invariant under the symmetry
${\cal U}$. Thus there are two strategies do carry out
perturbative expansions. One consists in keeping the quadratic term
proportional to $\rho_0$ in the bare propagator, while the other
consists of treating it as a mass insertion. For the purpose of this
appendix, both strategies are almost equivalent, and thus we will
follow the first one which generates less diagrams. In this case, the
bare propagator is given by (\ref{eqn:prop1}). At the bare level
(\ref{eqn:GC}) is verified, that is
\begin{equation}
{\cal G}_0(\bk,t)+{\cal G}_0(\bk,-t)=\beta W(\bk)\ C_0(\bk,t).
\label{eqn:GCbare}
\end{equation}
Causality implies that for $t>t'$ the terms of order $n+2$ ($n\geq 0$)
in the self
energies may be written formally as $\Sigma=\Theta_{1PI}$, with
\begin{eqnarray}
&\Theta_{\rhoh\rho}^{(n+2)}(\bk,t-t')=\intp\intpp\intq\intqp\ \delta(\bp+\bpp-\bk)\
\delta(\bq+\bqp-\bk)
\nonumber\\
&\times\Gamma(\bp,\bk,\bpp)\Biggl\{
4\Gamma(\bq,\bk,\bqp)
\langle\drho(\bp,t)\drho(\bpp,t)B[\drho,\rhoh]^n\rhoh(\bq,t')\drho(\bqp,t')\rangle
\nonumber\\
&-2T\bq\cdot\bqp
\langle\drho(\bp,t)\drho(\bpp,t)B[\drho,\rhoh]^n\rhoh(\bq,t')\rhoh(\bqp,t')\rangle
\Biggr\}\label{eqn:srhr}\\
&\Theta_{\rhoh\rhoh}^{(n+2)}(\bk,t-t')=\intp\intpp\intq\intqp\ \delta(\bp+\bpp-\bk)\
\delta(\bq+\bqp-\bk)
\nonumber\\
&\times\Gamma(\bp,\bk,\bpp)\Biggl\{ 2\Gamma(\bk,\bq,\bqp)
\langle\drho(\bp,t)\drho(\bpp,t)B[\drho,\rhoh]^n\drho(\bq,t')\drho(\bqp,t')\rangle
\nonumber\\
&-4T\bk\cdot\bq
\langle\drho(\bp,t)\drho(\bpp,t)B[\drho,\rhoh]^n\rhoh(\bq,t')\drho(\bqp,t')\rangle
\Biggr\}.\label{eqn:srhrh}
\end{eqnarray}
The subscript 1PI indicates that only
one-particle irreducible diagrams are considered. In term of diagrams
\begin{center}
\begin{tabular}{ccc}
$\Theta_{\rhoh\rho}$ &$=$&
\includegraphics[width=.4\textwidth]{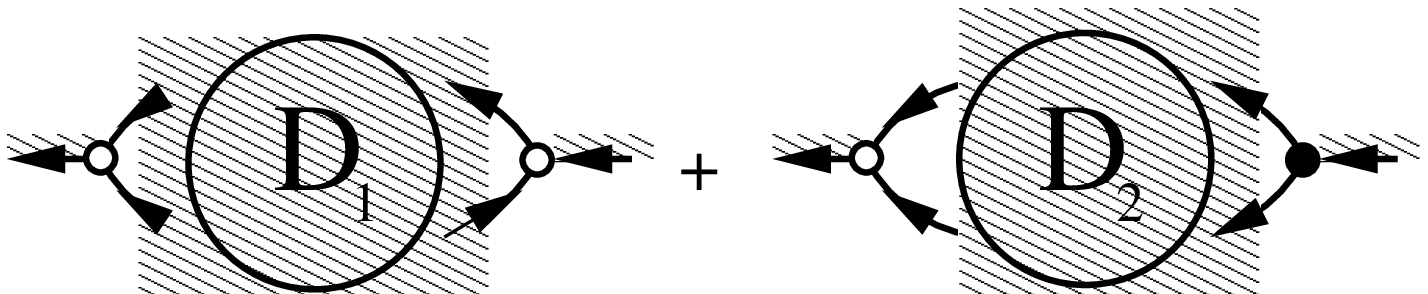}\\
$\Theta_{\rhoh\rhoh}$ &$=$&
\includegraphics[width=.4\textwidth]{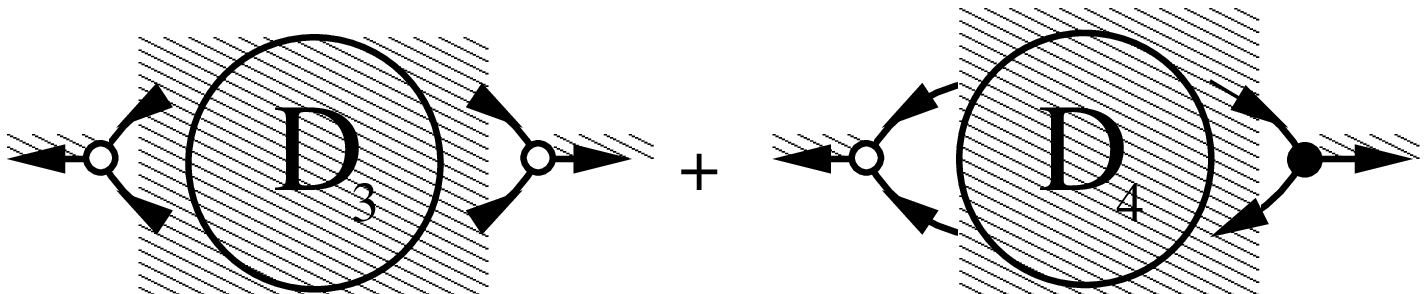}
\end{tabular}
\end{center}
where the $D_i$'s are obtained by doing all possible contractions in
the averages of (\ref{eqn:srhr}) and (\ref{eqn:srhrh}). The set of
diagrams with $m$ vertices will be noted
$\Omega^{(m)}$. The exponent $^{(m)}$ will in general design classes
of diagram of order $m$ exactly, while the subscript $_{(m)}$ will
include all orders until $m$.
It is convenient here to consider the subset of 1PI
diagrams of order $m$ as the subset of all diagrams of order $m$, minus
the classes of connected but not 1PI diagrams (subset $A^{(m)}$) and of
disconnected
diagrams (subset $B^{(m)}$). We will then proceed as follow. First we will
prove that the components of $\Theta$ verify an identity similar to
(\ref{eqn:GCself}), and that the same holds when restricted to class
$B^{(n)}$. Then we
will prove simultaneously that (\ref{eqn:GCself}) is true at order $m$
and also holds
when restricted to class $A^{(m)}$. The sum of
all diagrams belonging to class $A^{(m)}$ is noted $\Lambda^{(m)}$.

Applying ${\cal U}$ in all terms with $\rhoh$ explicitly written in
(\ref{eqn:srhr}) and (\ref{eqn:srhrh}), one gets
\begin{eqnarray}
&\Theta_{\rhoh\rho}^{(n+2)}(\bk,t-t')=\intp\intpp\intq\intqp\ \delta(\bp+\bpp-\bk)\
\delta(\bq+\bqp-\bk)
\nonumber\\
&\times\Gamma(\bp,\bk,\bpp)\langle\drho(\bp,-t)\drho(\bpp,-t)B[\drho,\rhoh]^n\drho(\bq,-t')\drho(\bqp,-t')\rangle\nonumber\\
&\left(4 \beta\
\frac{\bq\cdot\bk W(\bk)+\bq\cdot\bqp W(\bqp)}{2}
\ W(\bq)
-2\beta\bq\cdot\bqp\ W(\bq)\ W(\bqp)\right)\label{eqn:srhr1}\\
&\Theta_{\rhoh\rhoh}^{(n+2)}(\bk,t-t')=\intp\intpp\intq\intqp\ \delta(\bp+\bpp-\bk)\
\delta(\bq+\bqp-\bk)
\nonumber\\
&\times\Gamma(\bp,\bk,\bpp)\left(
2\frac{\bk\cdot\bq W(\bq)+\bk\cdot\bqp W(\bqp)}{2}\
\langle\drho(\bp,t)\drho(\bpp,t)B[\drho,\rhoh]^n\drho(\bq,t')\drho(\bqp,t')\rangle
\right.\nonumber\\
&\left.-4\bk\cdot\bq\ W(\bq)
\langle\drho(\bp,-t)\drho(\bpp,-t)B[\drho,\rhoh]^n\drho(\bq,-t')\drho(\bqp,-t')
\rangle
\right).\label{eqn:srhrh1}
\end{eqnarray}
All correlation functions involving $\rhoh(\bq,-t')$ or
$\rhoh(\bqp,-t')$ have vanished due to causality. Indeed they contribute
to
$\Theta_{\rho\rhoh}(\bk,t'-t)=\Theta_{\rhoh\rho}(\bk,t-t')$ which
vanishes for $t>t'$. Finally, using the fact that correlators
involving only $B[\drho,\rhoh]^n$ and $\rho$ are invariant under
time-reversal, the identity (\ref{eqn:GCself}) is found at order $m=n+2$.

The same steps can be follows straightforwardly for disconnected
diagrams, as it amounts to splitting $B[\drho,\rhoh]^n$ into several
powers of $B[\drho,\rhoh]$ which contribute to different connected
components. The crucial point here is the invariance of each term
$B[\drho,\rhoh]$ under ${\cal U}$.

Now we prove that diagrams of order $n$ verify the identity
(\ref{eqn:GCself}) holds at order $n$ and also within class
$A^{(n)}$. To do so, we proceed by induction.
Assuming that $\Sigma^{(p)}$ and $\Lambda^{(p)}$ verify
  (\ref{eqn:GCself}) for all $p\leq n-2$, we will show that
$\Sigma^{(n)}$ and $\Lambda^{(n)}$ verify (\ref{eqn:GCself}).

The starting point is however $n=2$. At this order:
\begin{eqnarray}
\Theta^{(2)}(\bk,t-t')&=&\Sigma^{(2)}(\bk,t-t')\\
\Lambda^{(2)}(\bk,t-t')&=&0.
\end{eqnarray}
The corresponding diagrams are:\\
\begin{center}
$\Sigma^{(2)}_{\rhoh\rho}(\bk,t-t')$ $=$
\includegraphics[width=.61\textwidth]{./figs/sh.eps} \\
$\Sigma^{(2)}_{\rhoh\rhoh}(\bk,t-t')$ $=$
\includegraphics[width=.75\textwidth]{./figs/s.eps}
\end{center}
For clarity, let us split the vertex made with $W$ into two parts:
\vskip .5cm
\begin{center}
\includegraphics[width=.3\textwidth]{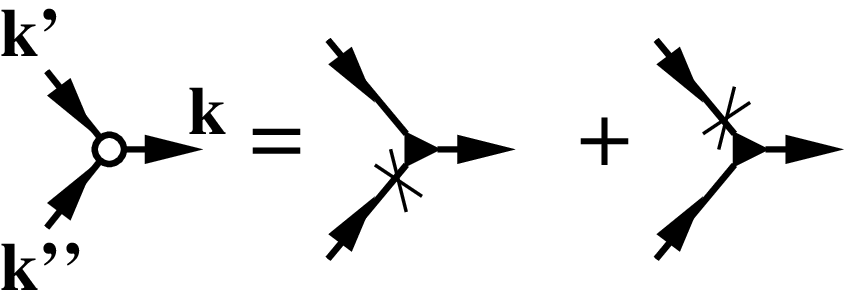}
\end{center}
The cross on a line with impulsion $\bq$ stands for $W(\bq)$ and the
black triangle stands for $\bk\cdot\bq$, $\bk$ being the momentum
of the outgoing arrow. Then we have
\begin{center}
\includegraphics[width=.81\textwidth]{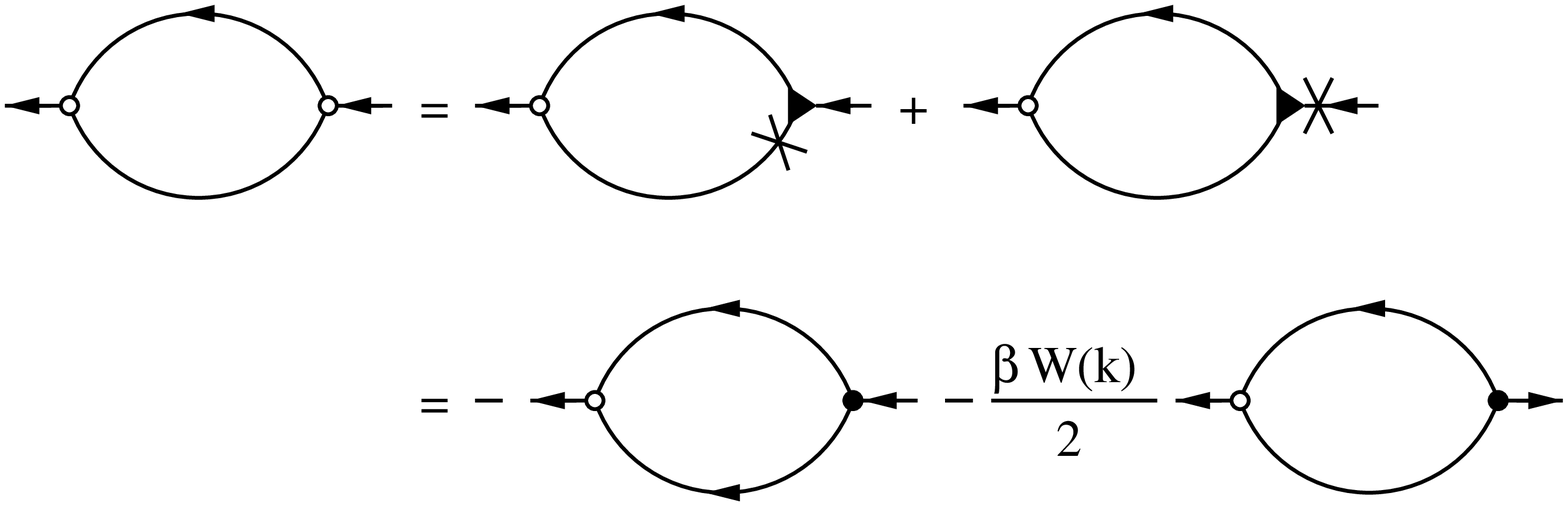}
\end{center}
and
\begin{center}
\includegraphics[width=.53\textwidth]{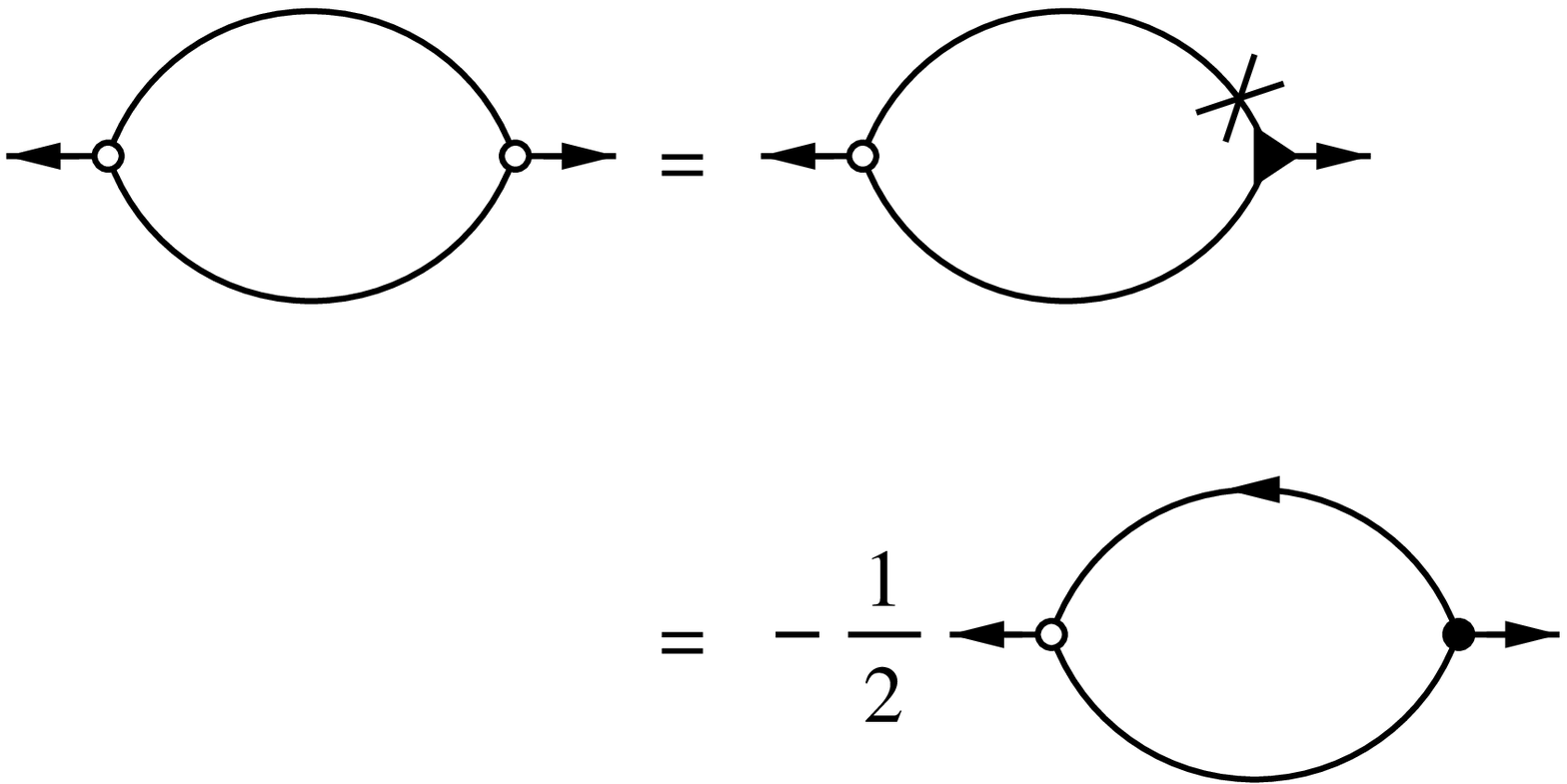}
\end{center}
Putting this altogether leads to
\begin{equation}
\Sigma_{(p)\rhoh\rho}(\bk,t-t')=-\beta\ W(\bk)\
\Sigma_{(p)\rhoh\rhoh}(\bk,t-t')
\label{eqn:GCself2}
\end{equation}
for $p=2$.

Now let us assume that $n\geq 4$ is even and (\ref{eqn:GCself2}) holds for
any even $p$ less than $n-2$.
We have in frequency:
\begin{equation}
\Lambda^{(n)}(\bk,\omega)=\sum_{q=2}^{n-2}
\left(\Sigma^{(q)}\ G_0\ \Sigma^{(n-q)}\right)(\bk,\omega)+
\left(\Sigma^{(q)}\ G_0\ \Lambda^{(n-q)}\right)(\bk,\omega),
\label{eqn:lds}
\end{equation}
where we have set $\Sigma^{(r)}=\Lambda^{(r)}=0$ for odd $r$ due to
the absence of tadpoles. Diagrammatically:
\begin{center}
\includegraphics[width=.76\textwidth]{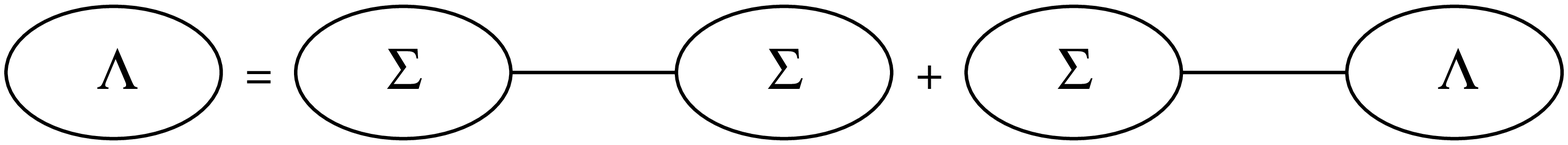}
\end{center}
Let us consider $U=A\ {\cal G}_0\ B$, where $A=\Sigma^{(q)}$ and
$B=\Sigma^{(n-q)}$ or $B=\Lambda^{(n-q)}$. We will show that $U$
verifies (\ref{eqn:GCself2}), which will be enough to show that
$\Lambda^{(n)}$ also verifies it due to (\ref{eqn:lds}).
We have:
\begin{eqnarray}
U_{\rhoh\rho}(\bk,\omega)&=&A_{\rhoh\rho}(\bk,\omega)\
R_0(\bk,\omega)\ B_{\rhoh\rho}(\bk,\omega)\nonumber\\
U_{\rhoh\rhoh}(\bk,\omega)&=&A_{\rhoh\rhoh}(\bk,\omega)\
R_0^*(\bk,\omega)\ B_{\rho\rhoh}(\bk,\omega)+A_{\rhoh\rho}(\bk,\omega)\
C_0(\bk,\omega)\ B_{\rho\rhoh}(\bk,\omega)\nonumber\\
&&+A_{\rhoh\rho}(\bk,\omega)\
R_0^*(\bk,\omega)\ B_{\rhoh\rhoh}(\bk,\omega).
\label{eqn:U}
\end{eqnarray}
However using the hypothesis made at order $n-2$, $A$ and $B$ verify
(\ref{eqn:GCself2}). In addition, ${\cal G}_0$ verifies
(\ref{eqn:GCbare}) (note the absence of the minus sign compared to
(\ref{eqn:GCself})). Thus using (\ref{eqn:GCbare}), one can express
the terms with indices $\rhoh\rhoh$ in the right hand side of
(\ref{eqn:U}) in term of those with
indices $\rhoh\rho$ and $\rho\rhoh$. Eliminating $C_0$ in profit of
$G_0$ and $G_0^*$, (\ref{eqn:U}) gives
\begin{equation}
U_{\rhoh\rhoh}(\bk,w)=-\frac{U_{\rho\rhoh}(\bk,w)+U_{\rhoh\rho}(\bk,w)}{\beta\
  W(\bk)}.
\end{equation}
Let us rewrite the above induction in such a way that the link with
similar diagrammatic proofs in the case of additive noise is clearer
(see~\cite{Ma} for a diagrammatic proof of FDT in that case).
The formal solution of (\ref{eqn:lds}) is:
\begin{equation}
\Lambda^{(n)}_{\rhoh\rho}(\bk,\omega)=\displaystyle{\sum_{r=1}^{n-2}\sum_{i_1+\cdots+i_r=n}
\Sigma^{(i_1)}_{\rhoh\rho}(\bk,\omega)\ \prod_{s=2}^r \left(R_0(\bk,\omega)\
\Sigma^{(i_s)}_{\rhoh\rho}(\bk,\omega)\right)}.
\end{equation}
Now we use the useful identity
\begin{equation*}
\Re\left(z_1 z_2\, (\cdots)\, z_n\right)=\Re(z_1) z_2^* (\cdots) z_n^*
-z_1 \Re(z_2)\ z_3^* (\cdots) z_n^*+z_1 z_2\ \Re(z_3)\ (\cdots) z_n^*-\cdots
\end{equation*}
to get the following diagrammatic identity:\\
\begin{center}
\includegraphics[width=\textwidth]{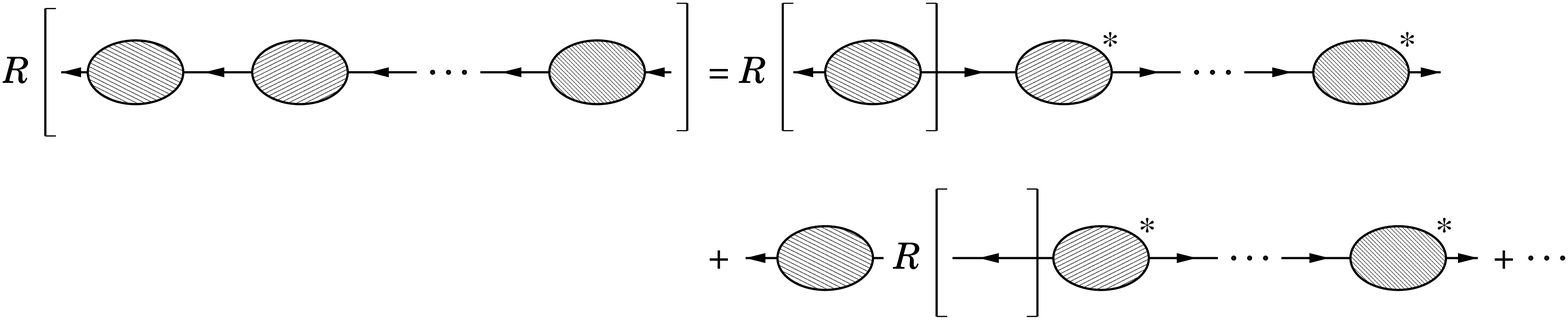}
\end{center}
Again using the hypothesis made at order $p\leq n-2$ and
(\ref{eqn:GCbare}) the result is obtained at order $n$. The importance
of the minus sign in (\ref{eqn:GCself}) compared to (\ref{eqn:GC}) is
clear here.
As said above, this establishes (\ref{eqn:GCself}) for class
$A^{(n)}$. Thus, by subtracting diagrams of classes $A^{(n)}$ and
$B^{(n)}$ from diagrams of $\Omega^{(n)}$, (\ref{eqn:GCself}) is
established at order $n$.

\subsection*{A.2. Proof of FDT in the quadratic expansion of the free-energy}

The proof will follow the lines of the previous one. We shall proceed
by induction. Here we notice that the diagrams of $\chi$ contributing
to FDT at
order $n$ are of order $n-1$. So the first diagram corresponding to
$\chi$ is for $n=2$:
\begin{center}
\includegraphics[width=.4\textwidth]{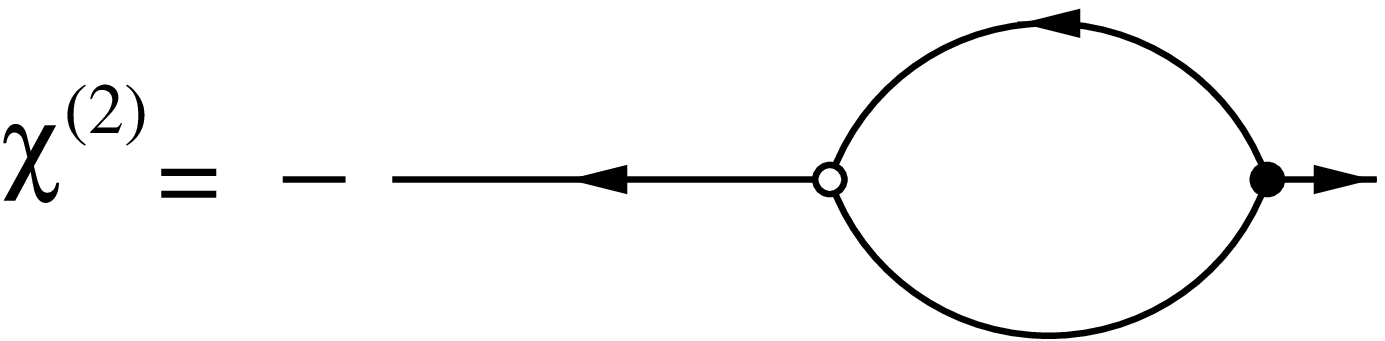}
\end{center}
Using the manipulations of the diagrams at order $2$ made in the
previous section, it is straightforward to prove FDT at this order.
Then let us assume that FDT has been proved at any order $p\leq n-2$.
The SD reads for the diagrams of order $n$ exactly:
\begin{equation}
\begin{split}
&\left(i\omega+\rho_0 T\bk^2W(\bk)\right)C^{(n)}\0=2T\bk^2\rho_0\
{\cal G}^{(n)}\0\nonumber\\
&+\sum_p\left[\{\Sigma_{\rhoh\rhoh}^{(n-p)}\0\
{\cal G}^{(p)}\0+\Sigma_{\rhoh\rho}^{(n-p)}\0\ C^{(p)}\0  \right].
\end{split}
\end{equation}
Thus, using the assumption made at order $p\leq n-2$  to
transform the self-energy part and (\ref{eqn:GC}) to transform
$ W(\bk)\ C^{(n)}\0$, we get
\begin{equation}
\begin{split}
i\omega\ C^{(n)}\0=&\frac{2 i}{\beta W(\bk)}\sum_p \Im\left(
\Sigma_{\rhoh\rho}^{(n-p)}\0\
{\cal G}^{(p)}\0\right)\\
&+2 iT\rho_0\bk^2\ \Im {\cal G}^{(n)}\0.
\label{eqn:FDTp}
\end{split}
\end{equation}
Now we remark that we have
\begin{center}
\includegraphics[width=.87\textwidth]{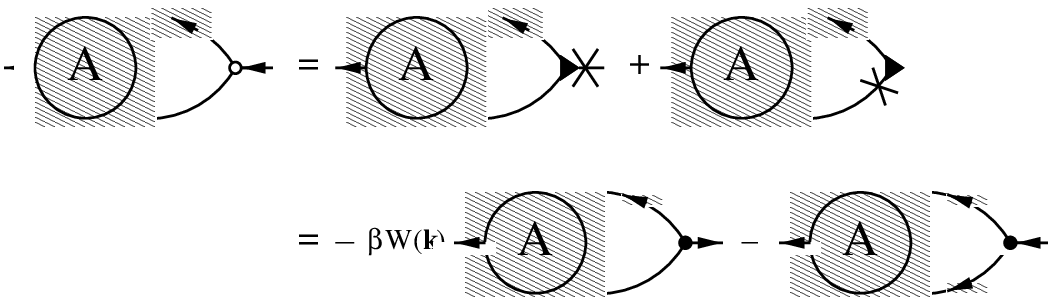}
\end{center}
which shows that the term of $\Sigma_{\rhoh\rho}\ {\cal G}$ in
(\ref{eqn:FDTp}) is nothing but the anomalous response $\chi$. Thus
the FDT is proved at order $n$.

\addcontentsline{toc}{section}{Appendix B. Causality with extra fields}
\section*{Appendix B. Causality with the extra fields}

In this appendix, we discuss briefly the difficulties in verifying
that causality is verified by the SD equations in the presence of the fields
$\theta$ and $\thetah$. Causality is mainly
insured by the form of the bare propagator and the identities
(\ref{eqn:idcorr1},\ref{eqn:idcorr2},\ref{eqn:idcorr3},\ref{eqn:idcorr4}).
However some of the diagram vanish due to the It{\^o} prescription in a
rather subtle way. We explicit this at the order of one-loop, the
generalization to higher orders being straightforward. It is not
difficult to be convinced that at one-loop the only diagram which may
eventually
cause some difficulties comes from the convolution
$\left(\Sigma_{\rho\rho}\cdot C_{\rho\rhoh}\right)(\bk,\tau)$. This
diagram, shown in Fig.~\ref{fig:one-loop} is proportional to
\begin{equation}
\intq\,\bk\cdot(\bk-\bq)\int_{-\infty}^\infty dt\, \partial_t\left[\Theta(t) C_{\rho
    \theta}(\bq,t)\right] \Theta(-t) C_{\rho\theta}(\bq-\bk,-t)
    C_{\rho\rhoh}(\bk,\tau-t).
\end{equation}
However due to the Heaviside functions the integrals can be restricted
to $[-\epsilon,\epsilon]$, with $\epsilon>0$. In addition
\begin{eqnarray}
&&\intq\,\bk\cdot(\bk-\bq)\int_{-\epsilon}^\epsilon dt\, \partial_t\left[\Theta(t) C_{\rho
    \theta}(\bq,t)\right] \Theta(-t) C_{\rho\theta}(\bq-\bk,-t)
    C_{\rho\rhoh}(\bk,\tau-t)\nonumber\\
&\approx&\intq\,\bk\cdot(\bk-\bq)\int_{-\epsilon}^{0^-} dt\, \partial_t\left[\Theta(t) C_{\rho
    \theta}(\bq,t)\right] C_{\rho\theta}(\bq-\bk,0^+)\,
    C_{\rho\rhoh}(\bk,\tau)\\ \nonumber
&=&\intq\,\bk\cdot(\bk-\bq)C_{\rho\theta}(\bq-\bk,0^+)\, C_{\rho\rhoh}(\bk,\tau)\left[\Theta(t)C_{\rho\theta}(\bq,t)\right]^{t=0-}_{t=-\epsilon} \\\nonumber
&=&0.
\end{eqnarray}

\begin{figure}[h]
\begin{center}
\caption{\label{fig:one-loop} One-loop diagram contributing to
  $\Sigma_{\rho\rho}$, which vanishing is the less obvious.}
\end{center}
\end{figure}

\addcontentsline{toc}{section}{Appendix C. Derivation of static and
  dynamical equations}
\section*{Appendix C. Derivation of dynamical and static equations}

In this appendix we sketch the derivation of equations (\ref{eqn:pp})
and (\ref{eqn:Cff0}) as an example. Other dynamical or static equations
can be obtain following the same routes.
We start from the SD equation
\begin{equation}\label{eqn:SDapp}
(G_0^{-1}\cdot G-\Sigma\cdot G)_{\rhoh\rhoh}(\bk,\tau)=\delta(\tau)
\end{equation}
for any value of $\tau$. We have
\begin{equation}
(\Sigma\cdot
G)_{\rhoh\rhoh}(\bk,\tau)=(\Sigma_{\rhoh\rho}\cdot C_{\rho\rhoh})
(\bk,\tau)+(\Sigma_{\rhoh\theta}\cdot C_{\theta\rhoh})(\bk,\tau).
\end{equation}
Indeed $C_{\rhoh\rhoh}$ and $C_{\thetah\rhoh}$ vanish by causality. We
then get:
\begin{equation}
\begin{split}
(\Sigma\cdot G)_{\rhoh\rhoh}(\bk,\tau)=&
\frac{\Theta(\tau)}{T}\int_0^\tau
dt\,\left[\partial_\tau\Sigma_{\thetah\theta}(\bk,\tau-t)C_{\rho\theta}(\bk,t)\right.\\
&\left.+\Sigma_{\rhoh\theta}(\bk,\tau-t)C_{\theta\theta}(\bk,t)\right].
\end{split}
\end{equation}
Integrating by parts, one gets
\begin{equation}
\begin{split}
&(\Sigma\cdot G)_{\rhoh\rhoh}(\bk,\tau)=\frac{\Sigma_{\thetah\theta}(\bk,\tau)}{T}
C_{\rho\theta}(\bk,t=0)\\
&+\frac{\Theta(\tau)}{T}\int_0^\tau
dt\,\left[\Sigma_{\thetah\theta}(\bk,\tau-t)\partial_t C_{\rho\theta}
(\bk,t)+\Sigma_{\rhoh\theta}(\bk,\tau-t)C_{\theta\theta}(\bk,t)\right].
\end{split}
\end{equation}
In addition we have
\begin{equation}
\begin{split}
(G_0^{-1}\cdot G)_{\rhoh\rhoh}(\bk,\tau)&\partial_\tau C_{\rho\rhoh}(\bk,\tau)+\rho_0\bk^2 C_{\theta\rhoh}(\bk,\tau)\\
&=\frac{1}{T}\partial_\tau
\left(\Theta(\tau^+)
C_{\rho\theta}(\bk,\tau)\right)+\frac{\rho_0\bk^2}{T}
C_{\theta\theta}(\bk,\tau).
\end{split}
\end{equation}
Equating the terms proportional to $\delta(\tau)$ in
(\ref{eqn:SDapp}) one gets
\begin{equation}
\frac{1}{T}C_{\rho \theta}(\bk,0)=1,
\end{equation}
and taking the limit $\tau\rightarrow 0^+$, gives
(\ref{eqn:Cff0}). Finally (\ref{eqn:pp}) is obtained by taking
$\tau>0$. All other equations for the dynamical evolutions and the
statics can be derived in the same way. When causality is not
enough to restrict explicitly time integrals between $0$ and $\tau$, one can
verify that in all cases FDT makes it possible to combine together
different contributions of the same equations to finally end up with
integrals between $0$ and $\tau$.

Let us add here that careful analysis of the self-energies shows that
$\Sigma_{\thetah\rho}$ has a tadpole contribution. However this
tadpole can be eliminated by adding a linear term $-A
\intx\,\delta\rho(\bx)$ to the entropic part of the free energy and
$A$ to the potential, with a suitable value of the constant $A$.

\addcontentsline{toc}{section}{Appendix D. Proof of the linear
  dependence of the Schwinger-Dyson equations}
\section*{Appendix D. Proof of the linear dependence of the
  Schwinger-Dyson equations}

As we have already mentioned, it may appear unnatural to
describe the evolution of $3$ correlators with $4$ dynamical
equations. A series expansions at low $\tau>0$ of these equations
makes this clearer. When expanded in series, they become a cascade of
equations for the successive derivatives of the correlators at zero
time difference. We start by expanding at first order, from which it is easy
to guess what is going on at higher orders. At order $1$ in $\tau$,
(\ref{eqn:SDrhr}), (\ref{eqn:pp}), (\ref{eqn:Ct}) and (\ref{eqn:Ct2})
read respectively:
\begin{eqnarray}
\Ct_{\rho\rho}(\bk,0)+\rho_0\bk^2C_{\rho\theta}(\bk,0)+\tau \left[
\Ctt_{\rho\rho}(\bk,0)+\rho_0\bk^2\Ct_{\rho\theta}(\bk,0)\right]=\nonumber\\
\label{eqn:o1a}
\tau\left[\Sigma_{\thetah\theta}(\bk,0^+)\Ct_{\rho\rho}(\bk,0)
+\Sigma_{\rhoh\theta}(\bk,0^+) C_{\rho\theta}(\bk,0)\right]\\
\Ct_{\rho\theta}(\bk,0)+\rho_0\bk^2C_{\theta\theta}(\bk,0)+\tau \left[
\Ctt_{\rho\theta}(\bk,0)+\rho_0\bk^2\Ct_{\theta\theta}(\bk,0)\right]\nonumber\\
\label{eqn:o1b}
-\Sigma_{\thetah\theta}(\bk,0^+)C_{\rho\theta}(\bk,0)+\tau
\left[\Sigma_{\rhoh\theta}(\bk,0) C_{\theta\theta}(\bk,0)
-\dot{\Sigma}_{\thetah\theta}(\bk,0^+)C_{\rho\theta}(\bk,0)\right]\\
 W(\bk)C_{\rho\rho}(\bk,0)-C_{\rho\theta}(\bk,0)+\tau\left[
W(\bk)\Ct_{\rho\rho}(\bk,0)-\Ct_{\rho\theta}(\bk,0)\right]=\nonumber\\
\label{eqn:o1c}
\frac{1}{T}\Sigma_{\thetah\thetah}(\bk,0)C_{\rho\rho}(\bk,0)+\tau\,
\Sigma_{\thetah\theta}(\bk,0^+)C_{\rho\theta}(\bk,0)\\
W(\bk)C_{\rho\theta}(\bk,0)-C_{\theta\theta}(\bk,0)+\tau\left[
W(\bk)\Ct_{\rho\theta}(\bk,0)-\Ct_{\theta\theta}(\bk,0)\right]=\nonumber\\
\label{eqn:o1d}
\tau\,
\left[\Sigma_{\thetah\theta}(\bk,0^+)C_{\theta\theta}(\bk,0)-\frac{1}{T}\dot{\Sigma}_{\thetah\thetah}(\bk,0)C_{\rho\theta}(\bk,0)
\right].
\end{eqnarray}
In addition, the SD equations have an apparent singularity at
$\tau=0$ which comes from the $\delta(\tau)$ in the RHS of (\ref{eqn:SD}).
This gives an initial condition: $C_{\rho\theta}(\bk,0^+)=T$. Thus there
are $5$ equations at order $0$, which fix the values of
$C_{\rho\rho}(\bk,0)$, $C_{\rho\theta}(\bk,0)$, $C_{\theta\theta}(\bk,0)$,
$\Ct_{\rho\rho}(\bk,0)$ and $\Ct_{\rho\theta}(\bk,0)$.
At order $1$, there are $4$ equations but $3$ quantities only to be
determined, namely $\Ct_{\theta\theta}(\bk,0)$,
$\Ctt_{\rho\rho}(\bk,0)$ and $\Ctt_{\rho\theta}(\bk,0)$.
Remark that the self-energies and their first derivatives appear in
the equations. However, as they can be expressed in terms of the correlators,
it can be checked that the successive derivatives of the self-energies
can be expressed in terms of the quantities already computed at
previous orders. This guarantees that at every order self-energies do
not give extra variables to be determined. Now we show that in fact
one of the equations obtained by identifying the terms of order $\tau$
is trivially satisfied by the
solution of the equations at order $0$. We
focus on the term proportional to
$\tau$ in the LHS of (\ref{eqn:o1c}). We then express this term by
using a linear combination of the terms of order $0$ of
(\ref{eqn:o1a}) and (\ref{eqn:o1b}):
\begin{eqnarray}
W(\bk) \Ct_{\rho\rho}(\bk,0)-\Ct_{\rho\theta}(\bk,0)=-\rho_0\bk^2
\left[ W(\bk) C_{\rho\rho}(\bk,0)-C_{\rho\theta}(\bk,0)\right]\\\nonumber
+\Sigma_{\thetah\theta}(\bk,0^+)C_{\theta\theta}(\bk,0).
\end{eqnarray}
>From order $0$ of (\ref{eqn:o1c}), the terms in brackets vanishes, and
then we get:
\begin{equation}
W(\bk) \Ct_{\rho\rho}(\bk,t)-\Ct_{\rho\theta}(\bk,t)\Sigma_{\thetah\theta}(\bk,0^+)C_{\rho\theta}(\bk,0),
\end{equation}
which corresponds to the terms proportional to $\tau$ in (\ref{eqn:o1c}).
Therefore the number of equations obtained at order $\tau$ is equal to
the number of variables to be determined at this order.

The non perturbative generalization of the previous approach comes from
the following remark: the SD equations form a linear system of
equations which unknown variables are the correlators and coefficients
are the components of $G_0^{-1}$ and $\Sigma$. The solution of this
system of equations is found easily using the Laplace transform.
The SD equations read in Laplace transform:
\begin{equation}
\begin{split}
\label{laplace-1}
\Crr(\bk,0^+)\left(1+\frac{\hat\Sigma_{\rhoh\thetah}(\bk,z)}{T}\right)&z\left(1+\frac{\hat\Sigma_{\rhoh\thetah}(\bk,z)}{T}\right)\hat
C_{\rho\rho}(\bk,z)\\
&+(\rho_0 \bk^2+\frac{\hat\Sigma_{\rhoh\rhoh}(\bk,z)}{T})\hat C_{\rho\theta}(\bk,z),
\end{split}
\end{equation}
\begin{equation}
\label{laplace-2}
z\left(1+\frac{\hat\Sigma_{\rhoh\thetah}(\bk,z)}{T}\right)\hat
C_{\rho\theta}(\bk,z)+(\rho_0
\bk^2+\frac{\hat\Sigma_{\rhoh\rhoh}(\bk,z)}{T})\hat
C_{\theta\theta}(\bk,z)=T
\end{equation}
\begin{eqnarray}
\label{laplace-3}
\frac{1}{T}\hat\Sigma_{\thetah\thetah}(\bk,z)\Crr(\bk,0^+)&=&-\left(1+\frac{\hat\Sigma_{\rhoh\thetah}(\bk,z)}{T}\right)
\hat C_{\rho\theta}(\bk,z)\\ \nonumber
&&+\left(W(\bk) +
\frac{z\hat\Sigma_{\thetah\thetah}(\bk,z)-\Sigma_{\thetah\thetah}(\bk,0^+)}{T}\right)\hat
C_{\rho\rho}(\bk,z)
\end{eqnarray}
\begin{equation}
\begin{split}
\label{laplace-4}
W(\bk)\ \hat
C_{\rho\theta}(\bk,z)=&\left(1+\frac{\hat\Sigma_{\rhoh\thetah}(\bk,z)}{T}\right)
\hat C_{\theta\theta}(\bk,z)
\\
&-\frac{z\hat\Sigma_{\thetah\thetah}(\bk,z)-\Sigma_{\thetah\thetah}(\bk,\tau=0)}
{T}\ \hat C_{\rho\theta}(\bk,z).
\end{split}
\end{equation}
For better clarity, we write formally this system as follows:
\begin{eqnarray}
A C_{\rho\rho}(\bk,0)&=&z A \hat C_{\rho\rho}(\bk,z)+B \hat
C_{\rho\theta}(\bk,z)\\
T&=&z A \hat C_{\rho\theta}(\bk,z)+B \hat C_{\theta\theta}(\bk,z)\\
D C_{\rho\rho}(\bk,0)&=&E \hat C_{\rho\rho}(\bk,z)-A \hat
C_{\rho\theta}(\bk,z)\\
0&=&E \hat C_{\rho\theta}(\bk,z)-A \hat C_{\theta\theta}(\bk,z).
\end{eqnarray}
The identity $E\ (RHS)_1-A\ (RHS)_2-z A\ (RHS)_3-B\
(RHS)_4=0$, where $(RHS)_i$ stands for the RHS of the $ith$ equation
above, is trivially verified. It remains to prove that the LHS are
linked by the same relation. Gathering the terms of
$E\ (LHS)_1-A\ (LHS)_2-z A\ (LHS)_3-B\
(LHS)_4=0$ (with obvious notation), one gets:
\begin{equation}
\left[W(\bk)-\frac{1}{T}\Sigma_{\thetah\thetah}(\bk,0^+)\right]C_{\rho\rho}(\bk,0)=T.
\end{equation}
This is precisely the static equation (\ref{eqn:Ct0}), and the proof
is complete.

\addcontentsline{toc}{section}{Appendix E. General fields transformation
and field theory in the case of Brownian dynamics for the
 density field}
\section*{Appendix E. General fields transformation
and field theory in the case of Brownian dynamics for the
 density field}

We have seen above (\ref{sec:morefields}) that one can make each of the two symmetries of the system linear independently. One can also render them linear simultaneously using the same method: introducing additional fields.

\begin{figure}[h]
\hspace{1.1cm}
\rotatebox{270}{
\includegraphics[height=1\textwidth]{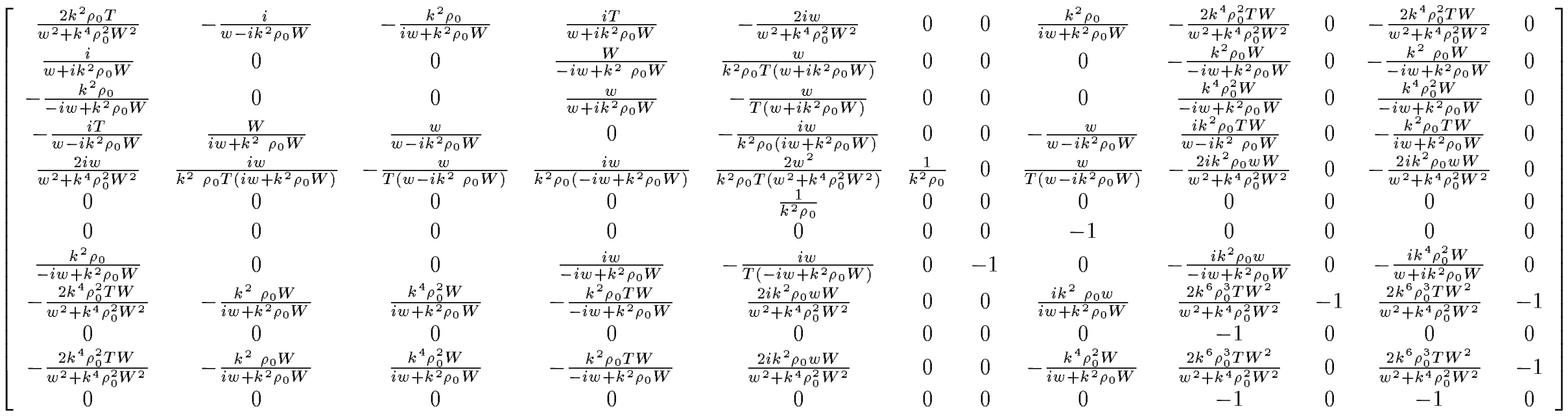}
}
\begin{center}
\caption{\label{fig:propnu} Bare propagator obtained from the action
(\ref{eqn:plusdechps}), where $W$ stands for $W(\bk)$.}
\end{center}
\end{figure}

The starting point is the action (\ref{msrd-action-1}): the field transformation is given by ${\cal T}_1$. We introduce $4$ additional fields to linearize the symmetry ${\cal U}$: $\theta=\delta{\cal F}/\delta\rho$ and $\eta=\nabla\cdot(\rho\nabla\theta)$ as well as the conjugated fields $\hat\theta$ and $\hat\eta$.
The final transformation is:

\begin{equation}
{\cal T}_1:\left\{
\begin{array}{rcl}
\rhoh_x & \rightarrow & \rhoh_x +f_x\\
\psi_x &\rightarrow& \psi_x
+\frac{1}{T}\partial_{t} \rho_x \\
 \psih_x &\rightarrow& \psih_x+T f_x\\
\fh_x &\rightarrow& -\fh_x+T f_x+\psih_x +T\rhoh_x\\
f_x&\rightarrow& -f_x
\end{array}
\right.
\end{equation}
\begin{equation}
{\cal U}_1:\left\{
\begin{array}{rcl}
\rhoh_x&\rightarrow& -\rhoh_x+\frac{1}{T}\theta_x\\
\psi_x&\rightarrow& -\psi_x+\frac{1}{T}\eta_x\\
\psih_x&\rightarrow& -\psih_x+\theta_x\\
\etah_x&\rightarrow& \etah_x+\frac{1}{T}\theta_x-\rhoh_x-\frac{1}{T}\psih_x\\
\thetah_x&\rightarrow& \thetah_x-\frac{1}{T}\partial_t\rho_x\\
\fh_x&\rightarrow& -\fh_x\\
f_x&\rightarrow& -f_x\\
\end{array}
\right.
\end{equation}
The action equals to the integral over ${\bf x},t$ of:
\begin{eqnarray}\label{eqn:plusdechps}
 &-\rhoh_x \partial_{t} \rho_x +\psi_x \theta_x-T\psi_x\rhoh_x + \fh_x
  \left(\nabla \cdot(\rho_x  \nabla f_x)+\frac{1}{T}\partial_{t}
  \rho_x \right)- \rho_x\nabla \phi_x\cdot\nabla \overline{\phi}_x\nonumber\\
&+\psih_x (\psi_x -\nabla \cdot(\rho_x \nabla \rhoh_x
  ))+\thetah_x\left( \theta_x-\frac{\delta{\cal
  F}}{\delta\rho_x}\right)
+\etah_x\left(\eta_x-\nabla\cdot(\rho_x\nabla\theta_x)\right).
\end{eqnarray}
The structure of the
two-fields correlators matrix for $\tau>0$ is schematically:
\begin{equation}\label{eqn:prop10}
\left[
\begin{array}{cccccccccc}
\Crr& \frac{\Crf}{T}& \frac{\Crrp}{T}& 0& \Crf& 0& \Crf&
-\frac{\Crrp}{T}& \Crrp& 0\\
0& 0& 0& 0& 0& 0& 0& 0& 0& \\
0& 0& 0& 0& 0& 0& 0& 0& 0& \\
\Crf& T\Cff& \frac{\Crfp}{T}& 0& T\Cff& 0& T^2\Cff& -\frac{\Crfp}{T}&
\Crfp& 0\\
\Crf& -\Cff& -\Crfp& 0& \Cff& \Cffh& -T\Cff& \frac{\Crfp}{T}& -\Crfp& 0\\
0& 0& 0& 0&  \Cffh& 0& 0& 0& 0& 0\\
\Crf& T\Cff& \frac{\Crfp}{T}& 0& T\Cff& 0& T^2\Cff& -\frac{\Crf}{T}& 0\\
0& 0& 0& 0& 0& 0& 0& 0& 0& 0\\
\Crrp& \frac{\Crfp}{T}& \frac{\Crrpp}{T}& 0& \Crfp& 0& \Crfp&
-\frac{\Crrpp}{T}& \Crrpp& C_{\eta\etah}\\
0& 0& 0& 0& 0& 0& 0& 0& C_{\eta\etah}& 0\\
\end{array}
\right]
\end{equation}
where the ordering of the fields is
$(\rho,\rhoh,\psi,\psih,f,\fh,\theta,\thetah,\eta,\etah)$. The
fermionic fields $\phi$ and $\overline{\phi}$, which correlators with
other fields vanish, have not been included. The role of these
fermionic fields will be explained later on a practical example.
For completeness, the bare propagator (still without the fermionic
fields) is given in Fig.~\ref{fig:propnu}.
Some of the correlators defined above are singular at
$\tau=0$. Indeed for $\tau\in\mathbb{R}$, we have
$C_{ff}(\bk,\tau)=a(\bk)\ \delta(\tau)+C_{ff}^{reg}(\bk,\tau)$ and
$C_{f\fh}=b(\bk)\ \delta(\tau)$, where $C_{ff}^{reg}$ is regular
(i.e. continuous by parts) and even in time. We also have
$C_{\psih\psi}(\bk,\tau)=\frac{1}{T}\partial_\tau\left(\Theta(\tau)
C_{\rho f}(\bk,\tau)\right)=-C_{\psih\thetah}(\bk,\tau)$,
$C_{\psih f}(\bk,\tau)=-C_{\rhoh f}(\bk,\tau)=c(\bk)\ \delta(\tau)+T
C^{reg}_{ff}(\bk,\tau)$ and $C_{\eta\etah}(\bk,\tau)=-\delta(\tau)$.

We conclude this appendix with a discussion about causality.
One can show using the way $\Sigma$
transforms under $\{{\cal T}_1,{\cal U}_1\}$ that
$\Sigma_{\rho\rho}$ vanishes, which guarantees causality to hold
non-perturbatively. Furthermore, it also hold perturbatively, as a
consequence of the form of the bare propagator and of the It{\^o}
discretization. We first show how it holds at one-loop order, the
generalization to higher orders being straightforward.

\begin{figure}[htb]
\begin{center}
\includegraphics[width=.65\textwidth]{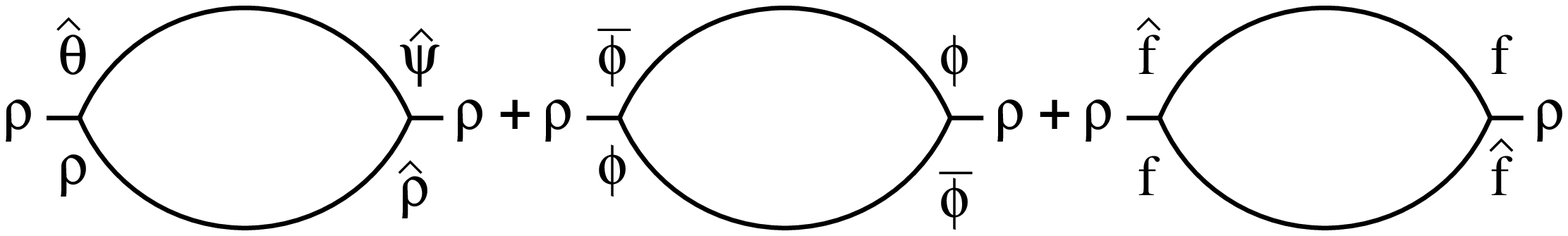}
\end{center}
\caption{Diagrams contributing to $\Sigma_{\rho\rho}$ at one-loop. The
fields involved in the Wick theorem are indicated around the vertices.
\label{fig:caus}}
\end{figure}

At one-loop, $\Sigma_{\rho\rho}$ has three non-trivially vanishing
contributions, shown in Fig.~\ref{fig:caus}.
The diagram
involving the ghost loop is
identical to the one involving the loop made with propagators
$C_{f\fh}$, with opposite sign and with $C_{\phi\overline{\phi}}$
instead of
$C_{f\fh}^0$. However the bare propagators $C_{\phi\overline{\phi}}^0$
and $C_{f\fh}^0$ are identical and the equations giving the
renormalization of $C_{\phi\overline{\phi}}$ and $C_{f\fh}$ are also
identical, thus the contributions of both loops in the self-energy -
which are divergent - cancel each other exactly. As usual in the
Fadeev-Popov method, the role of the fermionic fields is to remove the
volume of the ``gauge'' ensemble which is infinite. One can easily
check that loops of
ghost propagators annihilate at all orders with corresponding loops of
propagators $C_{f\fh}$. Following the steps of appendix B, the remaining
diagram in Fig.~\ref{fig:caus} may be shown to give a vanishing
contribution when involved in the dynamic equations.

\addcontentsline{toc}{section}{Appendix F. Dynamical equations for
 Fluctuating Nonlinear
  Hydrodynamics}
\section*{Appendix F. Dynamical equations for Fluctuating Nonlinear Hydrodynamics}

In this appendix we give the derivation of the dynamic equations for
fluctuating nonlinear
hydrodynamics. The calculus and the ideas behind are the same as the
corresponding for BDD although
somewhat more cumbersome due to a larger number of fields. We start
with the Schwinger-Dyson equations and use time-reversal to simplify them.

This time (\ref{eqn:mgm}) applied to the transformation ${\cal U}_1$
gives the following equations for correlators:
\begin{eqnarray}
C_{\rho\rhoh}(\bk,\tau)&=&\frac{\Theta(\tau)}{T}C_{\rho\theta}(\bk,\tau)\\
C_{\rho\hat g}(\bk,\tau)&=&\frac{\Theta(\tau)}{T}C_{\rho v}(\bk,\tau)\\
C_{\rho\thetah}(\bk,\tau)&=&\frac{\Theta(\tau)}{T}\partial_\tau C_{\rho\rho}(\bk,\tau)\\
C_{\rho\hat v}(\bk,\tau)&=&\frac{\Theta(\tau)}{T}\partial_\tau C_{\rho g}(\bk,\tau)
\end{eqnarray}
\begin{eqnarray}
C_{g\rhoh}(\bk,\tau)&=&\frac{\Theta(\tau)}{T}C_{g\theta}(\bk,\tau)\\
C_{g\hat g}(\bk,\tau)&=&\frac{\Theta(\tau)}{T}C_{gv}(\bk,\tau)\\
C_{g\thetah}(\bk,\tau)&=&\frac{\Theta(\tau)}{T}\partial_\tau C_{g\rho}(\bk,\tau)\\
C_{g\hat v}(\bk,\tau)&=&\frac{\Theta(\tau)}{T}\partial_\tau C_{gg}(\bk,\tau)
\end{eqnarray}
\begin{eqnarray}
C_{\theta\rhoh}(\bk,\tau)&=&\frac{\Theta(\tau)}{T}C_{\theta\theta}(\bk,\tau)\\
C_{\theta\hat g}(\bk,\tau)&=&\frac{\Theta(\tau)}{T}C_{\theta v}(\bk,\tau)\\
C_{\theta\thetah}(\bk,\tau)&=&\frac{\Theta(\tau)}{T}\partial_\tau C_{\theta\rho}(\bk,\tau)\\
C_{\theta\hat v}(\bk,\tau)&=&\frac{\Theta(\tau)}{T}\partial_\tau C_{\theta g}(\bk,\tau)
\end{eqnarray}
\begin{eqnarray}
C_{v\rhoh}(\bk,\tau)&=&\frac{\Theta(\tau)}{T}C_{v\theta}(\bk,\tau)\\
C_{v\hat g}(\bk,\tau)&=&\frac{\Theta(\tau)}{T}C_{vv}(\bk,\tau)\\
C_{v\thetah}(\bk,\tau)&=&\frac{\Theta(\tau)}{T}\partial_\tau C_{v\rho}(\bk,\tau)\\
C_{v\hat v}(\bk,\tau)&=&\frac{\Theta(\tau)}{T}\partial_\tau C_{vg}(\bk,\tau)
\end{eqnarray}
and (\ref{eqn:vdv}) yields the following identities for self-energies:
\begin{eqnarray}
\Sigma_{\rhoh\rho}(\bk,\tau)&=&\frac{1}{T}\partial_\tau\left[\Theta(\tau)\Sigma_{\rhoh\thetah}(\bk,\tau)\right]\\
\Sigma_{\rhoh g}(\bk,\tau)&=&\frac{1}{T}\partial_\tau\left[\Theta(\tau)\Sigma_{\rhoh\hat v}(\bk,\tau)\right]\\
\Sigma_{\rhoh\theta}(\bk,\tau)&=&-\frac{1}{T}\Theta(\tau)\Sigma_{\rhoh\rhoh}(\bk,\tau)\\
\Sigma_{\rhoh v}(\bk,\tau)&=&-\frac{1}{T}\Theta(\tau)\Sigma_{\rhoh\hat g}(\bk,\tau)
\end{eqnarray}
\begin{eqnarray}
\Sigma_{\hat g\rho}(\bk,\tau)&=&\frac{1}{T}\partial_\tau\left[\Theta(\tau)\Sigma_{\hat g\thetah}(\bk,\tau)\right]\\
\Sigma_{\hat g g}(\bk,\tau)&=&\frac{1}{T}\partial_\tau\left[\Theta(\tau)\Sigma_{\hat g\hat v}(\bk,\tau)\right]\\
\Sigma_{\hat g\theta}(\bk,\tau)&=&-\frac{1}{T}\Theta(\tau)\Sigma_{\hat g\rhoh}(\bk,\tau)\\
\Sigma_{\hat g v}(\bk,\tau)&=&-\frac{1}{T}\Theta(\tau)\Sigma_{\hat g\hat g}(\bk,\tau)
\end{eqnarray}
\begin{eqnarray}
\Sigma_{\thetah\rho}(\bk,\tau)&=&\frac{1}{T}\Theta(\tau)\partial_\tau\Sigma_{\thetah\thetah}(\bk,\tau)\\
\Sigma_{\thetah g}(\bk,\tau)&=&\frac{1}{T}\Theta(\tau)\partial_\tau\Sigma_{\thetah\hat v}(\bk,\tau)\\
\Sigma_{\thetah\theta}(\bk,\tau)&=&-\frac{1}{T}\Theta(\tau)\Sigma_{\thetah\rhoh}(\bk,\tau)\\
\Sigma_{\thetah v}(\bk,\tau)&=&-\frac{1}{T}\Theta(\tau)\Sigma_{\thetah\hat g}(\bk,\tau)
\end{eqnarray}
\begin{eqnarray}
\Sigma_{\hat v\rho}(\bk,\tau)&=&\frac{1}{T}\Theta(\tau)\partial_\tau\Sigma_{\hat v\thetah}(\bk,\tau)\\
\Sigma_{\hat v g}(\bk,\tau)&=&\frac{1}{T}\Theta(\tau)\partial_\tau\Sigma_{\hat v\hat v}(\bk,\tau)\\
\Sigma_{\hat v\theta}(\bk,\tau)&=&-\frac{1}{T}\Theta(\tau)\Sigma_{\hat v\rhoh}(\bk,\tau)\\
\Sigma_{\hat v v}(\bk,\tau)&=&-\frac{1}{T}\Theta(\tau)\Sigma_{\hat v\hat g}(\bk,\tau).
\end{eqnarray}
One can get some additional identities:
\begin{eqnarray}
\Sigma_{\hat g\rhoh}(\bk,\tau)&=&\Sigma_{\rhoh\hat g}(\bk,\tau)\\
\Sigma_{\thetah\rhoh}(\bk,\tau)&=&-\Sigma_{\rhoh\thetah}(\bk,\tau)\\
\Sigma_{\thetah\hat g}(\bk,\tau)&=&-\Sigma_{\hat g\thetah}(\bk,\tau)\\
\Sigma_{\hat v\rhoh}(\bk,\tau)&=&-\Sigma_{\rhoh\hat v}(\bk,\tau)\\
\Sigma_{\hat v\hat g}(\bk,\tau)&=&-\Sigma_{\hat g\hat v}(\bk,\tau)\\
\Sigma_{\hat v\thetah}(\bk,\tau)&=&\Sigma_{\hat v\thetah}(\bk,\tau),
\end{eqnarray}
and similar ones for correlators:
\begin{eqnarray}
C_{g\rho}(\bk;t,s)&=&C_{\rho g}(\bk;t,s)\\
C_{\theta\rho}(\bk;t,s)&=&C_{\rho\theta}(\bk;t,s)\\
C_{\theta g}(\bk;t,s)&=&C_{g\theta}(\bk;t,s)\\
C_{v\rho}(\bk;t,s)&=&C_{\rho v}(\bk;t,s)\\
C_{vg}(\bk;t,s)&=&C_{gv}(\bk;t,s)\\
C_{v\theta}(\bk;t,s)&=&C_{\theta v}(\bk;t,s).
\end{eqnarray}
All these identities reduce the number of independent correlators to
ten, which are
$C_{\rho\rho}$, $C_{\rho g}$, $C_{\rho\theta}$, $C_{\rho v}$, $C_{gg}$
,$C_{g\theta}$, $C_{g v}$, $C_{\theta\theta}$, $C_{\theta v}$ and $C_{vv}$.

In the case of FNH, there are in principle $64$ Schwinger-Dyson
equation. We write $16$ of these equations,
the other being trivially linear dependent on these:
\begin{eqnarray}
&&\partial_\tau C_{\rho\rho}(\bk,\tau) -i\rho_0\bk C_{v\rho}(\bk,\tau)= \frac{1}{T}\int_{0}^{\tau}dt\Sigma_{\rhoh\thetah}(\bk,\tau-t)\partial_tC_{\rho\rho}(\bk,t)\nonumber\\
 &+&\frac{1}{T}\int_{0}^{\tau}dt\Sigma_{\rhoh\hat v}(\bk,\tau-t)\partial_tC_{g\rho}(\bk,t)
 -\frac{1}{T}\int_{0}^{\tau}dt\Sigma_{\rhoh\rhoh}(\bk,\tau-t)C_{\theta\rho}(\bk,t)\nonumber\\
 &-&\frac{1}{T}\int_{0}^{\tau}dt\Sigma_{\rhoh\hat g}(\bk,\tau-t)C_{v\rho}(\bk,t)
\end{eqnarray}
\begin{eqnarray}
&&\partial_\tau C_{\rho g}(\bk,\tau) -i\rho_0\bk C_{vg}(\bk,\tau)= \frac{1}{T}\int_{0}^{\tau}dt\Sigma_{\rhoh\thetah}(\bk,\tau-t)\partial_t C_{\rho g}(\bk,t)\nonumber\\
 &+&\frac{1}{T}\int_{0}^{\tau}dt\Sigma_{\rhoh\hat v}(\bk,\tau-t)\partial_tC_{gg}(\bk,t)
 -\frac{1}{T}\int_{0}^{\tau}dt\Sigma_{\rhoh\rhoh}(\bk,\tau-t)C_{\theta g}(\bk,t)\nonumber\\
 &-&\frac{1}{T}\int_{0}^{\tau}dt\Sigma_{\rhoh\hat g}(\bk,\tau-t)C_{vg}(\bk,t)
\end{eqnarray}
\begin{eqnarray}
&&\partial_\tau C_{\rho\theta}(\bk,\tau) -i\rho_0\bk C_{v\theta}(\bk,\tau)=\Sigma_{\rhoh\hat v}(\bk,\tau)\nonumber\\
 &+&\frac{1}{T}\int_{0}^{\tau}dt\biggl[\Sigma_{\rhoh\thetah}(\bk,\tau-t)\partial_tC_{\rho\theta}(\bk,t)+\Sigma_{\rho\hat v}(\bk,\tau-t)\partial_tC_{g\theta}(\bk,t)\nonumber\\
 &-&\Sigma_{\rhoh\rhoh}(\bk,\tau-t)C_{\theta\theta}(\bk,t)-\Sigma_{\rhoh\hat g}(\bk,\tau-t)C_{v\theta}(\bk,t)\biggr]
\end{eqnarray}
\begin{eqnarray}
&&\partial_\tau C_{\rho v}(\bk,\tau) -i\rho_0\bk C_{vv}(\bk,\tau)=\Sigma_{\rhoh\hat v}(\bk,\tau)\nonumber\\
 &+&\frac{1}{T}\int_{0}^{\tau}dt\biggl[\Sigma_{\rhoh\thetah}(\bk,\tau-t)\partial_tC_{\rho v}(\bk,t)+\Sigma_{\rhoh\hat v}(\bk,\tau-t)\partial_tC_{gv}(\bk,t)\nonumber\\
 &-&\Sigma_{\rhoh\rhoh}(\bk,\tau-t)C_{\theta v}(\bk,t)-\Sigma_{\rhoh\hat g}(\bk,\tau-t)C_{vv}(\bk,t)\biggr]
\end{eqnarray}
\begin{eqnarray}
&&\partial_\tau C_{g\rho}(\bk,\tau) -i\rho_0\bk C_{\theta\rho}(\bk,\tau)+LC_{v\rho}(\bk,\tau) +\frac{1}{T}\int_{0}^{\tau}dt\Sigma_{\hat g\thetah}(\bk,\tau-t)\partial_tC_{\rho\rho}(\bk,t)\nonumber\\
 &+&\frac{1}{T}\int_{0}^{\tau}dt\Sigma_{\hat g\hat v}(\bk,\tau-t)\partial_tC_{g\rho}(\bk,t)
 -\frac{1}{T}\int_{0}^{\tau}dt\Sigma_{\hat g\rhoh}(\bk,\tau-t)C_{\theta\rho}(\bk,t)\\
 &-&\frac{1}{T}\int_{0}^{\tau}dt\Sigma_{\hat g\hat g}(\bk,\tau-t)C_{v\rho}(\bk,t)\nonumber
\end{eqnarray}
\begin{eqnarray}
&&\partial_\tau C_{gg}(\bk,\tau) -i\rho_0\bk C_{\theta g}(\bk,\tau)+LC_{vg}(\bk,\tau)=
 +\frac{1}{T}\int_{0}^{\tau}dt\Sigma_{\hat g\thetah}(\bk,\tau-t)\partial_tC_{\rho g}(\bk,t)\nonumber\\
 &+&\frac{1}{T}\int_{0}^{\tau}dt\Sigma_{\hat g\hat v}(\bk,\tau-t)\partial_tC_{gg}(\bk,t)
 -\frac{1}{T}\int_{0}^{\tau}dt\Sigma_{\hat g\rhoh}(\bk,\tau-t)C_{\theta g}(\bk,t)\nonumber\\
 &-&\frac{1}{T}\int_{0}^{\tau}dt\Sigma_{\hat g\hat g}(\bk,\tau-t)C_{vg}(\bk,t)
\end{eqnarray}
\begin{eqnarray}
&&\partial_\tau C_{g\theta}(\bk,\tau) -i\rho_0\bk
 C_{\theta\theta}(\bk,\tau)+LC_{v\theta}(\bk,\tau)=\Sigma_{\hat g\thetah}(\bk,\tau)\nonumber\\
 &+&\frac{1}{T}\int_{0}^{\tau}dt\biggl[\Sigma_{\hat g\thetah}(\bk,\tau-t)\partial_tC_{\rho\theta}(\bk,t)-\Sigma_{\hat g\hat v}(\bk,\tau-t)\partial_tC_{g\theta}(\bk,t)\\
 &-&\biggl[\Sigma_{\hat
 g\rhoh}(\bk,\tau-t)C_{\theta\theta}(\bk,t)+\Sigma_{\hat g\hat
 g}(\bk,\tau-t)C_{v\theta}(\bk,t)\biggr]
\end{eqnarray}
\begin{eqnarray}
&&\partial_\tau C_{gv}(\bk,\tau) -i\rho_0\bk
 C_{\theta v}(\bk,\tau)+LC_{vv}(\bk,\tau)=\Sigma_{\hat g\hat v}(\bk,\tau)\nonumber\\
 &+&\frac{1}{T}\int_{0}^{\tau}dt\biggl[\Sigma_{\hat g\thetah}(\bk,\tau-t)\partial_tC_{\rho v}(\bk,t)+\Sigma_{\hat g\hat v}(\bk,\tau-t)\partial_tC_{gv}(\bk,t)\\
 &-&\Sigma_{\hat g\rhoh}(\bk,\tau-t)C_{\theta v}(\bk,t)-\Sigma_{\hat g\hat g}(\bk,\tau-t)C_{vv}(\bk,t)\biggr]\nonumber
\end{eqnarray}
\begin{eqnarray}
&&C_{\theta\rho}(\bk,\tau) -W(\bk) C_{\rho\rho}(\bk,\tau)=-\frac{1}{T}\left[\Sigma_{\thetah\thetah}(\bk,0)C_{\rho\rho}(\bk,\tau)+\Sigma_{\thetah\hat v}(\bk,0)C_{g\rho}(\bk,\tau)\right]\nonumber\\
 \label{eqn:non-erg-fh}
 &+&\frac{1}{T}\int_{0}^{\tau}dt\biggl[\Sigma_{\thetah\thetah}(\bk,\tau-t)\partial_tC_{\rho\rho}(\bk,t)+\Sigma_{\thetah\hat v}(\bk,\tau-t)\partial_tC_{g\rho}(\bk,t)\\
 &-&\Sigma_{\thetah\rhoh}(\bk,\tau-t)C_{\theta\rho}(\bk,t)-\Sigma_{\thetah\hat g}(\bk,\tau-t)C_{v\rho}(\bk,t)\biggr]\nonumber
\end{eqnarray}
\begin{eqnarray}
&&C_{\theta g}(\bk,\tau) -W(\bk) C_{\rho g}(\bk,\tau)=-\frac{1}{T}\left[\Sigma_{\thetah\thetah}(\bk,0)C_{\rho g}(\bk,\tau)+\Sigma_{\thetah\hat v}(\bk,0)C_{gg}(\bk,\tau)\right]\nonumber\\
 &+&\frac{1}{T}\int_{0}^{\tau}dt\biggl[\Sigma_{\thetah\thetah}(\bk,\tau-t)\partial_tC_{\rho g}(\bk,t)+\Sigma_{\thetah\hat v}(\bk,\tau-t)\partial_tC_{gg}(\bk,t)\\
 &-&\Sigma_{\thetah\rhoh}(\bk,\tau-t)C_{\theta g}(\bk,t)-\Sigma_{\thetah\hat g}(\bk,\tau-t)C_{vg}(\bk,t)\biggr]\nonumber
\end{eqnarray}
\begin{eqnarray}
&&C_{\theta\theta}(\bk,\tau) -W(\bk) C_{\rho\theta}(\bk,\tau)=\frac{1}{T}\Sigma_{\thetah\thetah}(\bk,\tau)\nonumber\\
 &-&\frac{1}{T}\left[\Sigma_{\thetah\thetah}(\bk,0)C_{\rho\theta}(\bk,\tau)+\Sigma_{\thetah\hat v}(\bk,0)C_{g\theta}(\bk,\tau)\right]\\
 &+&\frac{1}{T}\int_{0}^{\tau}dt\biggl[\Sigma_{\thetah\thetah}(\bk,\tau-t)\partial_tC_{\rho\theta}(\bk,t)+\Sigma_{\thetah\hat v}(\bk,\tau-t)\partial_tC_{g\theta}(\bk,t)\nonumber\\
 &-&\Sigma_{\thetah\rhoh}(\bk,\tau-t)C_{\theta\theta}(\bk,t)-\Sigma_{\thetah\hat g}(\bk,\tau-t)C_{v\theta}(\bk,t)\biggr]\nonumber
\end{eqnarray}
\begin{eqnarray}
&&C_{\theta v}(\bk,\tau) -W(\bk) C_{\rho v}(\bk,\tau) \frac{1}{T}\Sigma_{\thetah\hat v}(\bk,\tau)\\
 &-&\frac{1}{T}\left[\Sigma_{\thetah\thetah}(\bk,0)C_{\rho v}(\bk,\tau)+\Sigma_{\thetah\hat v}(\bk,0)C_{gv}(\bk,\tau)\right]\nonumber\\
 &+&\frac{1}{T}\int_{0}^{\tau}dt\biggl[\Sigma_{\thetah\thetah}(\bk,\tau-t)\partial_tC_{\rho v}(\bk,t)+\Sigma_{\thetah\hat v}(\bk,\tau-t)\partial_tC_{gv}(\bk,t)\nonumber\\
 &-&\Sigma_{\thetah\rhoh}(\bk,\tau-t)C_{\theta v}(\bk,t)-\Sigma_{\thetah\hat g}(\bk,\tau-t)C_{vv}(\bk,t)\biggr]\nonumber
\end{eqnarray}
\begin{eqnarray}
&&C_{v\rho}(\bk,\tau) -\frac{1}{\rho_0} C_{g\rho}(\bk,\tau)=-\frac{1}{T}\left[\Sigma_{\hat v\thetah}(\bk,0)C_{\rho\rho}(\bk,\tau)+\Sigma_{\hat v\hat v}(\bk,0)C_{g\rho}(\bk,\tau)\right]\nonumber\\
 &+&\frac{1}{T}\int_{0}^{\tau}dt\biggl[\Sigma_{\hat v\thetah}(\bk,\tau-t)\partial_tC_{\rho\rho}(\bk,t)+\Sigma_{\hat v\hat v}(\bk,\tau-t)\partial_tC_{g\rho}(\bk,t)\\
 &-&\Sigma_{\hat v\rhoh}(\bk,\tau-t)C_{\theta\rho}(\bk,t)-\Sigma_{\hat v\hat g}(\bk,\tau-t)C_{v\rho}(\bk,t)\biggr]\nonumber
\end{eqnarray}
\begin{eqnarray}
&&C_{vg}(\bk,\tau) -\frac{1}{\rho_0} C_{gg}(\bk,\tau)=-\frac{1}{T}[\Sigma_{\hat v\thetah}(\bk,0)C_{\rho g}(\bk,\tau)+\Sigma_{\hat v\hat v}(\bk,0)C_{gg}(\bk,\tau)]\nonumber\\
 &+&\frac{1}{T}\int_{0}^{\tau}dt\biggl[\Sigma_{\hat v\thetah}(\bk,\tau-t)\partial_tC_{\rho g}(\bk,t)+\Sigma_{\hat v\hat v}(\bk,\tau-t)\partial_tC_{gg}(\bk,t)\\
 &-&\Sigma_{\hat v\rhoh}(\bk,\tau-t)C_{\theta\rho}(\bk,t)-\Sigma_{\hat v\hat g}(\bk,\tau-t)C_{vg}(\bk,t)\biggr]\nonumber
\end{eqnarray}
\begin{eqnarray}
&&C_{v\theta}(\bk,\tau) -\frac{1}{\rho_0} C_{g\theta}(\bk,\tau) +\frac{1}{T}\Sigma_{\hat v\thetah}(\bk,\tau)\nonumber\\
 &-&\frac{1}{T}\left[\Sigma_{\hat v\theta}(\bk,0)C_{\rho\thetah}(\bk,\tau)+\Sigma_{\hat v\hat v}(\bk,0)C_{g\theta}(\bk,\tau)\right]\\\nonumber
 &+&\frac{1}{T}\int_{0}^{\tau}dt\biggl[\Sigma_{\hat v\theta}(\bk,\tau-t)\partial_tC_{\rho\theta}(\bk,t)-\Sigma_{\hat v\hat v}(\bk,\tau-t)\partial_tC_{g\theta}(\bk,t)\\
 &-&\Sigma_{\hat v\rhoh}(\bk,\tau-t)C_{\theta\rho}(\bk,t)+\Sigma_{\hat v\hat g}(\bk,\tau-t)C_{v\theta}(\bk,t)\biggr]\nonumber
\end{eqnarray}
\begin{eqnarray}
&&C_{vv}(\bk,\tau) -\frac{1}{\rho_0} C_{gv}(\bk,\tau) +\frac{1}{T}\Sigma_{\hat v\hat v}(\bk,\tau)\nonumber\\
 &-&\frac{1}{T}\left[\Sigma_{\hat v\thetah}(\bk,0)C_{\rho v}(\bk,\tau)+\Sigma_{\hat v\hat v}(\bk,0)C_{gv}(\bk,\tau)\right]\\\nonumber
 &+&\frac{1}{T}\int_{0}^{\tau}dt\biggl[\Sigma_{\hat v\thetah}(\bk,\tau-t)\partial_tC_{\rho v}(\bk,t)-\Sigma_{\hat v\hat v}(\bk,\tau-t)\partial_tC_{gv}(\bk,t)\\
 &-&\Sigma_{\hat v\rhoh}(\bk,\tau-t)C_{\theta\rho}(\bk,t)+\Sigma_{\hat v\hat g}(\bk,\tau-t)C_{vv}(\bk,t)\biggr].\nonumber
\end{eqnarray}
As for BDD, the number of independent correlators is smaller than the
number of equations, and there are here $6$ redundant equations. The
extension of the proof of appendix D to the present equations is
straightforward but very painful.

{color{blue} These are the exact non-perturbative dynamical equations preserving FDT. One can then use different approximation schemes for self-energies to concretese the equations. It is worth noting that what ever the approximation is the FDT is alwas verified due to theway the equations were derived.}



\begin{thebibliography}{9}
\bibitem{ang}{C. A. Angell, {\em Science} {\bf 267}, 1924 (1995).}
\bibitem{DeBenedettiStillinger}{Recent reviews:
P.G. De Benedetti and F.H. Stillinger

{\sl Nature} {\bf  410}, (2001) 267;

M.A.Ediger Annu. Rev. Phys. Chem. {\bf 51} (2000) 99.
}
\bibitem{reviewDas}{S. P. Das
 {\em Rev. Mod. Phys.} {\bf  76} 785 (2004).}
\bibitem{reviewDave}{D. R. Reichman and P. Charbonneau,
{\em J. Stat. Mech.} (2005) P05013.}
\bibitem{kob}{W. Kob, in ``Slow relaxations and nonequilibrium dynamics
  in condensed matter'', vol. Session LXXVII of Les Houches Summer
  School (ed. J.-L. Barrat, M. Feigelman and J. Kurchan), published by
  EDP Sciences and Springer.}
\bibitem{ben}{U. Bengtzelius, W. G{\"o}tze and A. Sj{\"o}lander, {\em
    J. Phys. C} {\bf 17}, 5915 (1984).}
\bibitem{leu}{E. Leutheusser, {\em Phys. Rev. A} {\bf 29}, 2765 (1984).}
\bibitem{extendedGotze}{W. G{\"o}tze and A. Sj{\"o}lander, {\em Z. Phys. B}
{\bf  65} 415 (1987).}
\bibitem{das}{S. P. Das and G. F. Mazenko, {\em Phys. Rev. A} {\bf
    34}, 2265 (1986); S. P. Das
 {\em Phys. Rev. A} {\bf  42}, 6116 (1990).}
\bibitem{ReviewMF}{J.-P. Bouchaud, L.F. Cugliandolo, J. Kurchan and
    M. Mezard, in ``Recent progress in random magnets'', A.P. Young ed.,
    World Scientific (1997).}
\bibitem{bckm}{J.-P. Bouchaud, L.F. Cugliandolo, J. Kurchan and
    M. Mezard, Physica A {\bf 226}, 243 (1996).}
\bibitem{BerthierGarrahan}{L. Berthier, J.-P. Garrahan,
Phys. Rev. E {\bf  68}, 041201 (2003).}
\bibitem{mya1}{K. Miyazaki and D. R. Reichman, {\em J. Phys. A:
    Math. Gen.} {\bf 38}, L343-L355 (2005).}
\bibitem{KT} T.R. Kirkpatrick and D. Thirumalai
 Phys. Rev. A {\bf 37}, 4439 (1988).
\bibitem{FP}{S. Franz, G. Parisi, J. Phys.:Condens. Matter 12, 6335 (2000).
C. Donati, S. Franz, S.C. Glotzer, G. Parisi,
J. Non-Cryst. Sol., {\bf 307}, 215-224 (2002).}
\bibitem{BB}{G. Biroli and J.-P. Bouchaud, Europhys. Lett. {\bf 67} (2004) 21.}
\bibitem{TWBBB}{C. Toninelli, M. Wyart, L. Berthier, G. Biroli,
J.-P. Bouchaud, Phys. Rev. E {\bf 71}, 041505 (2005).}
\bibitem{Andersen}{H.C. Andersen, PNAS {\bf 102} 6686 (2005).}
\bibitem{latz}{A. Latz, {\em J. Phys: Condens. Matter} {\bf 12}, 6353
  (2000); A. Latz, {\em J. Stat. Phys.} {\bf 109}, 607 (2002).}
\bibitem{fuchs}{M. Fuchs ad M. E. Cates, {\em Phys. Rev. Lett.} {\bf
    89}, 248304 (2002).}
\bibitem{mya}{K. Miyazaki and D. R. Reichman, {\em Phys. Rev. E} {\bf
    66} 050501(R) (2002); K. Miyazaki, D. R. Reichman and R. Yamamoto, {\em
    Phys. Rev. E} {\bf 70}, 011501 (2004).}
\bibitem{Jarzynski}{C. Jarzynski, {\em Phys. Rev. Lett. } {\bf  78}
2690 (1997).}
\bibitem{Crooks}{G.E. Crooks, {\em J. Stat. Phys.} {\bf  90} 1481 (1998).}
\bibitem{Lubensky}{
P. M. Chaikin and T. C. Lubensky, {``The Principles of Condensed
  Matter Physics''}, Cambridge University Press (2000).}
\bibitem{dean1}{D. S. Dean, {\em J. Phys. A} {\bf 29}, L613 (1996).}
\bibitem{Pusey}{P. N. Pusey, in ``Liquids, Freezing and Glass Transition'',
J. P. Hansen, D. Levesque, J. Zinn-Justin, Eds. (North-Holland, Amsterdam, 1991), pp. 763-942.}
\bibitem{kaw1}{K. Kawasaki and T. Koga, {\em Physica A} {\bf 201}, (115).}
\bibitem{CC}. C. Chamon and L.F. Cugliandolo, unpublished, 2004.
\bibitem{RY}{T. V. Ramakrishnan and M. Youssouff, {\em Phys. Rev. B}
  {\bf 19}, 2775 (1979).}
\bibitem{zj}{J. Zinn-Justin, ``Quantum field theory and critical
  phenomena'' (Clarendon Press, Oxford, 1989).}
\bibitem{msrjd}{P. C. Martin, E. D. Siggia and H. A. Rose, {\em
    Phys. Rev. A} {\bf 8}, 423 (1973); H. K. Janssen, {\em Z. Phys. B}
    {\bf 23}, 377 (1976); C. De Dominicis, {\em J. Phys. (France),
    Colloq.} {\bf 1}, C-247 (1976).}
\bibitem{DeDom}{C. De Dominicis and
P. Martin, J. Math. Phys. {\bf 5}, 14 and 31 (1964).}
\bibitem{Ma}{S. K. Ma, {``Statistical mechanics''}, world scientific
(Singapore, 1985).; S. K. Ma, ``Modern Theory of Critical Phenomena'',
 Perseus Publishing (2000).}
\bibitem{kaw2}{K. Kawasaki, {\em Ann. Phys.} {\bf 61}, 1 (1970).}
\bibitem{cates}{M. E. Cates and S. Ramaswamy, cond-mat/0511260 (2005).}
\bibitem{schmitz}{R. Schmitz, J. W. Dufty, and P. De
 Phys. Rev. Lett. {\bf  71}, 2066 (1993).}
\bibitem{got}{W. G{\"o}tze and L. Sj{\"o}gren, {\em J. Phys.:
    Condens. Matter} {\bf 1}, 4203 (1989).}
\bibitem{got1}{W. G{\"o}tze and L. Sj{\"o}gren, {\em Rep. Prog. Phys.} {\bf
    55}, 241 (1992).}
\bibitem{kaw}{K . Kawasaki and S. Miyazima, {\em Z. Phys. B:
    Condens. Matter} {\bf 103}, 423 (1997).}
\bibitem{szamel}{G. Szamel, and E. Flenner.
{\em Europhys. Lett}. {\bf 67}, 779 (2004).}
\bibitem{theor}{H. Osada, Probab. Theory Relat. Fields {\bf  112} 53
(1998).}
\bibitem{BC} G. Biroli and A. Crisanti, unpublished.
\bibitem{Hansen}{J. P. Hansen, I. R. McDonald, ``Theory of Simple
    Liquids'', second edition, Academic Press, London (1986).}
\bibitem{mori}{T. Morita and K. Hiroike, {\em Prog. Theor. Phys.} {\bf
25}, 537 (1961).}
\bibitem{KW} T. R. Kirkpatrick, P. G. Wolynes,
 Phys. Rev. A {\bf  35}, 3072 (1987).
\bibitem{ber}{J. Berges, N. Tetradis and C. Wetterich, {\em
    Phys. Rep.} {\bf 363}, 223 (2002).}
\bibitem{MMR}  P. Mayer, K. Miyazaki, D. R. Reichman, {\it
Cooperativity Beyond Caging: Generalized Mode Coupling Theory}, cond-mat/0602248.
\end{thebibliography}
\end{document}